\documentclass[12pt]{article}

\usepackage{amsfonts}
\usepackage{cite, natbib}
\usepackage[dvips]{graphicx}
\usepackage{keyval}
\usepackage{dsfont}
\usepackage{natbib}
\usepackage{multirow}
\usepackage{rotating}
\usepackage{endnotes}
\usepackage{amssymb,amsmath,lscape, bm, amsthm}
\usepackage{amsfonts}
\usepackage{graphicx}
\usepackage{color}
\usepackage{rotating}
\usepackage{multirow}
\usepackage{caption}
\usepackage{setspace}
\usepackage{verbatim}
\usepackage{epstopdf}
\usepackage{booktabs}
\usepackage{array}
\usepackage[top=1in, bottom=1in, left=1.25in, right=1.25in]{geometry}
\usepackage{adjustbox}
\usepackage{dcolumn}
\usepackage{caption}
\usepackage{longtable}
\usepackage{mathrsfs}
\usepackage{tabularx}
\usepackage{subcaption}  
\usepackage{pdflscape}
\usepackage{subcaption}
\usepackage{booktabs}
\usepackage{adjustbox}

\usepackage[hidelinks]{hyperref}

\usepackage{bibunits}

\newtheorem*{remark}{Remark}

\newcommand{\Proj}{\mathrm{Proj}}

\usepackage{siunitx}
\newcommand{\st}[1]{\rlap{\textsuperscript{#1}}}
\makeatletter
\renewcommand{\@seccntformat}[1]{\csname the#1\endcsname.\enspace}
\makeatother

\begin{document}

\renewcommand{\baselinestretch}{1.5}
\normalsize

\renewcommand{\thefootnote}{\fnsymbol{footnote}}

\begin{center}

\noindent{\Large{\bf{A new decomposition approach to modeling financial returns: Conditioning sign on magnitude}}\footnotemark[1] }

\smallskip
\bigskip
\bigskip

\renewcommand{\baselinestretch}{1.0}
\normalsize

Ars{\`e}ne Brou and Richard Luger\footnotemark[2]

\bigskip

Universit{\'e} Laval, Canada

\bigskip
\bigskip

{\small
\emph{Author accepted manuscript.}\\
\emph{Accepted for publication in the Journal of Banking and Finance.}\\
\url{https://doi.org/10.1016/j.jbankfin.2026.107716}
}

\end{center}

\footnotetext[1]{This work draws on research supported by the Social Sciences and Humanities Research Council of Canada.}

\footnotetext[2]{Correspondence to: Department of Finance, Insurance and Real Estate, Universit{\'e} Laval, Quebec City, QC G1V  0A6, Canada.

\emph{E-mail addresses:} {kouakou-arsene.brou.1@ulaval.ca (A. Brou), richard.luger@fsa.ulaval.ca (R. Luger)}.

}

\bigskip
\bigskip

\noindent{\bf Abstract:} Changes in volatility contain valuable information about the likelihood of positive versus negative returns. We propose a new approach to modeling financial returns that exploits this insight by decomposing returns into sign and magnitude (absolute value) components, with magnitude closely related to volatility. The joint distribution used to compute expected returns combines a model for the marginal distribution of magnitude with a model for the distribution of the sign, conditional on the contemporaneous magnitude. Unlike traditional linear predictive regressions, this decomposition framework captures nonlinear predictability in return dynamics. An out-of-sample forecasting evaluation using monthly U.S. stock market excess returns demonstrates substantial statistical and economic gains relative to linear regression and complete subset regression, while delivering performance that is competitive with copula-based return-decomposition methods and other nonlinear benchmarks.



\thispagestyle{empty}

\newpage

\renewcommand{\baselinestretch}{1.75}
\normalsize

\pagenumbering{arabic}

\renewcommand{\thefootnote}{\arabic{footnote}}
\setcounter{footnote}{0}

\section{Introduction}

Traditional methods for investigating return predictability typically rely on ordinary least squares (OLS) regressions in which the conditional expectation of returns is modeled as a linear function of lagged economic variables. Yet a large body of evidence shows that such linear predictive regressions deliver poor out-of-sample performance and often fail to outperform the historical average benchmark that treats returns as essentially unpredictable \citep[see, e.g.,][]{Breen-Glosten-Jagannathan:1989,Bossaerts-Hillion:1999,Butler-Grullon-Weston:2005,Welch-Goyal:2008,Rapach-Zhou:2013,Borup-Eriksen-Kjaer-Thyrsgaard:2024}. This has motivated alternative approaches that aim to better exploit the information content of existing predictors rather than introducing new ones.

One prominent approach that addresses model uncertainty in return prediction is the complete subset regression (CSR) method of \citet{Elliott-Gargano-Timmermann:2013}. CSR averages forecasts across all linear models formed from fixed-size subsets of the available predictors. This is particularly well suited to settings where many predictors are only weakly informative and where conventional OLS forecasts are unstable. By systematically averaging across many small models, CSR mitigates model selection risk and implicitly regularizes estimation without imposing strong parametric structure. \citet{Elliott-Gargano-Timmermann:2013} show that CSR can outperform several common benchmarks, and \citet{Bianchi-Rubesam-Tamoni:2025} document that CSR remains competitive with a wide range of modern statistical learning methods in out-of-sample equity premium forecasting.

In contrast to CSR, which remains within the linear OLS framework, \citet[][hereafter AG]{Anatolyev-Gospodinov:2010} propose a decomposition-based approach that targets nonlinear return dynamics by expressing returns as the product of a sign and a magnitude (absolute value) component. 
The magnitude is specified via a multiplicative error model, the sign via a binary choice model, and the two components are then coupled with a bivariate copula.
The resulting joint distribution is used to compute expected returns. This framework accommodates nonlinear structure that may be missed by linear regressions, and it has motivated a growing literature; see, among others, 
\citet{Liu:2015}, \citet{Liu:2017}, \citet{Liu-Luger:2015}, 
\citet{Frazier-Liu:2016}, \citet{Anatolyev-Gospodinov-Jamali-Liu:2017}, 
and \citet{Anatolyev-Gospodinov:2019}.

In this paper, we introduce a new decomposition strategy that we call the conditioning sign on magnitude (CSM) approach. Like AG, CSM decomposes returns into sign and magnitude components. However, rather than specifying their dependence through a copula, CSM models the magnitude using a multiplicative error model and the sign using a binary choice model that conditions explicitly on the contemporaneous magnitude. This structure avoids copula selection and estimation while still allowing the predictive content of volatility states to influence expected returns.

The economic motivation for conditioning sign on magnitude stems from the \sloppy well-documented persistence of volatility. When the magnitude component is large, volatility tends to remain high, and this can shift the probability of positive versus negative returns. As shown by \citet{Christoffersen-Diebold:2006}, volatility dynamics can therefore induce sign predictability. The CSM approach exploits this channel by letting the sign model incorporate information embedded in the contemporaneous magnitude, which in turn improves the implied conditional mean forecast.

To benchmark the CSM approach against other economically interpretable nonlinear specifications, we also consider two widely used alternatives. First, we estimate a GARCH-in-mean model \citep{Engle-Lilien-Robins:1987}, in which conditional volatility enters the return equation directly. Second, we estimate a Markov-switching predictive regression with regime-dependent coefficients and variances, following the regime-switching literature \citep[e.g.,][]{Hamilton:1989,Ang-Bekaert:2002,Ang-Bekaert:2004,Henkel-Martin-Nardari:2011}.

Our empirical results deliver three main messages. 
First, forecasting performance is typically strongest at moderate predictor dimensions. Decomposition-based models tend to produce their most reliable out-of-sample gains when the information set is neither too small nor too large: improvements are generally more stable under absolute loss for small predictor sets, while under squared loss the gains concentrate at intermediate dimensions and can dissipate as the predictor set becomes large.

Second, formal model comparisons broadly reinforce this pattern. Across both loss functions, decomposition-based specifications are typically difficult to distinguish from the best-performing models, whereas the regime-switching benchmark is more often dominated. Overall, the evidence suggests that return decomposition captures salient nonlinearities that are missed by linear predictive regressions and by forecast averaging within the linear class.

Third, while copula-based and CSM specifications often exhibit similar statistical forecast accuracy, differences become more visible once performance is assessed through market-timing outcomes and investor utility. The CSM approach frequently delivers large economic gains relative to standard benchmarks and remains highly competitive across predictor dimensions. Copula-based strategies can occasionally match or slightly exceed CSM in some configurations, but these incremental utility gains are typically modest, suggesting that conditioning sign on magnitude provides a practical alternative to copula-based dependence modeling.

Our empirical analysis should be viewed as a small-scale application: we work with a deliberately moderate set of candidate predictors and focus on fixed-size subsets within this set. Recent studies such as \citet{Bianchi-Rubesam-Tamoni:2025} and \citet{Goyal-Welch-Zafirov:2024} consider substantially larger predictor sets. Moreover, \citet{Bianchi-Rubesam-Tamoni:2025} emphasize that the optimal model size can be time-varying, whereas the CSR benchmark implemented here assumes a constant subset size throughout. Accordingly, our results demonstrate the incremental value of decomposition-based forecasting given our moderate predictor set and fixed subset size, rather than providing level comparisons to studies using larger predictor sets or time-varying model selection.

The remainder of the paper proceeds as follows. Section~2 presents the 
statistical framework, including the benchmark 
specifications and our return decomposition approach. Section~3 presents 
the empirical evaluation, comparing out-of-sample forecasting performance 
and economic value against the historical average, linear predictive 
regression, CSR, and nonlinear benchmarks. Section~4 concludes. Technical 
details related to the copulas are provided in the Appendix and further 
empirical results are reported in the Supplementary material.

\section{Statistical framework}

We begin by reviewing the OLS predictive regression and CSR as linear 
benchmarks, then summarize the GARCH-in-mean and Markov-switching 
nonlinear benchmarks. Next, we introduce the sign-magnitude decomposition 
and a motivating example illustrating why it can capture nonlinear 
predictive structure missed by linear projections. We then present the 
AG copula-based approach and the CSM approach proposed in this paper.

Consider an asset return $R_{t}$ at time $t$ and a vector of variables $\boldsymbol X_{t-1} = (X_{1,t-1},\ldots, X_{K, t-1})^{\prime}$ observed at time $t-1$ that could potentially predict $R_t$. We treat $R_t$ as a continuous random variable that can take both positive and negative values.
The traditional method for investigating stock return predictability uses OLS to fit the linear predictive regression model
\begin{equation}
R_{t} = \beta_0 + \beta_{1}  X_{1,t-1} + \ldots + \beta_{K}  X_{K,t-1} + \varepsilon_{t},
\label{OLSregression}
\end{equation}
where the error term is assumed to satisfy $\mathbb{E}(\varepsilon_{t} \mid \boldsymbol X_{t-1}) = 0$, and $\boldsymbol{\beta} = (\beta_1,\ldots,\beta_K)'$ collects the slope coefficients.
Let $\Proj(R_t \mid \boldsymbol{X}_{t-1})$ denote the linear projection of $R_t$ onto the space spanned by the constant and the components of $\boldsymbol X_{t-1}$.
If the model in \eqref{OLSregression} is correctly specified, then
$ \mathbb{E}(R_t \mid \boldsymbol X_{t-1}) = \Proj(R_t \mid \boldsymbol X_{t-1}) = \beta_0 + \boldsymbol\beta' \boldsymbol X_{t-1},$
and hence, for $\boldsymbol X_{t-1}=\boldsymbol x_{t-1}$, we have
$\mathbb{E}(R_t \mid \boldsymbol x_{t-1}) = \big[\Proj(R_t \mid \boldsymbol X_{t-1})\big]_{\boldsymbol X_{t-1}=\boldsymbol x_{t-1}} = \beta_0 + \boldsymbol\beta' \boldsymbol x_{t-1}.$
If $\beta_i \neq 0$, then $X_{i,t-1}$ is said to have predictive ability for the mean of $R_t$. This predictive ability can be evaluated using in-sample or out-of-sample performance. When $\beta_1 = \ldots = \beta_K = 0$, the model reduces to the historical average model, in which the conditional expectation becomes constant: $\mathbb{E}(R_t \mid \boldsymbol{X}_{t-1}) = \beta_0$, and hence $\mathbb{E}(R_t) = \beta_0$.

While conceptually straightforward, the OLS framework can perform poorly in empirical applications with many potential predictors. In such settings, forecasts may suffer from instability and weak out-of-sample performance due to estimation noise and model selection risk. To mitigate these issues, \citet{Elliott-Gargano-Timmermann:2013} introduce the CSR method, which averages forecasts across all linear models based on fixed-size subsets of $k \leq K$ predictors. Formally, the CSR forecast is given by
\[
\Proj_{\text{CSR}}(R_{t} \mid \boldsymbol{x}_{t-1})
= \frac{1}{n_{k, K}} \sum_{i=1}^{n_{k, K}}
\Big[ \Proj \big(R_{t}\mid S_i \boldsymbol{X}_{t-1}\big) \Big]_{\boldsymbol{X}_{t-1}=\boldsymbol{x}_{t-1}},
\]
where $n_{k, K} = \binom{K}{k}$, and $S_i$ is a $K \times K$ diagonal selection matrix with ones on the diagonal entries corresponding to the included variables (and zeros otherwise).
In practice, the population projections are estimated by OLS on each subset, and the sample CSR forecast averages these OLS forecasts.
This averaging scheme implicitly regularizes estimation and accounts for model uncertainty. When $ k = 1 $, CSR reduces to the forecast combination method of \citet{Rapach-Strauss-Zhou:2010}; when $ k = K $, it becomes the full linear predictive regression model in \eqref{OLSregression}.

As in  \citet{Elliott-Gargano-Timmermann:2013}, we implement CSR for a fixed subset size over the out-of-sample period. This differs from \citet{Bianchi-Rubesam-Tamoni:2025}, who emphasize that the optimal model size may vary over time and consider adaptive choices of the subset size. 
We keep the subset size fixed to maintain comparability across models and to isolate the incremental role of return decomposition.

We next describe the two benchmark nonlinear specifications used in the empirical comparison.
The GARCH-in-mean (GARCH-M) model \citep{Engle-Lilien-Robins:1987} allows expected returns to depend on conditional volatility. Specifically,
\begin{equation*}
R_t = \beta_0 + \boldsymbol{\beta}' \boldsymbol{X}_{t-1} + \lambda \, \sigma_t + \varepsilon_t, 
\quad \varepsilon_t \sim N(0, \sigma_t^2),
\end{equation*}
where $\sigma_t^2$ follows a standard GARCH(1,1) recursion, $\sigma_t^2 = \omega + \alpha \varepsilon_{t-1}^2 + \beta_g \sigma_{t-1}^2$, with $\omega > 0$, $\alpha \geq 0$, $\beta_g \geq 0$, and $\alpha+\beta_g<1$. The parameter $\lambda$ captures the sensitivity of expected returns to time-varying risk. This specification aligns with the view emphasized by \citet{Cederburg-Johnson-Doherty:2023} that volatility is an important state variable for equity premium predictability and the economic value of forecasting. The model is estimated via quasi-maximum likelihood.

To capture shifts in return dynamics, we also estimate a two-state Markov-switching (MS) predictive regression in the spirit of 
\citet{Hamilton:1989}, \citet{Ang-Bekaert:2002,Ang-Bekaert:2004}, and \citet{Henkel-Martin-Nardari:2011}. Returns depend on an unobserved state variable $c_t \in \{1,2\}$ that evolves as a first-order Markov chain. We specify
\begin{equation*}
R_t = \beta_{0,c_t} + \boldsymbol{\beta}_{c_t}' \boldsymbol{X}_{t-1} + \varepsilon_t, 
\quad \varepsilon_t \sim N(0, \sigma_{c_t}^2),
\end{equation*}
where $\beta_{0,c_t}$, $\boldsymbol{\beta}_{c_t} = (\beta_{1,c_t},\ldots,\beta_{K,c_t})'$, and $\sigma_{c_t}^2$ are regime-specific parameters. The transition probabilities $\Pr(c_t=j \mid c_{t-1}=i)=p_{ij}$, $i,j\in\{1,2\}$, govern movements between regimes. This specification allows the intercept, slope coefficients, and error variance to vary across regimes. We estimate the model by maximum likelihood using the expectation-maximization (EM) algorithm.\footnote{In our empirical application, we implement the MS model using the R package \texttt{MSwM} \citep{SanchezEspigares-LopezMoreno:2021}.}

The return decomposition approach offers an alternative nonlinear perspective to \eqref{OLSregression} by first separating $R_t$ into distinct components. This approach is based on the mathematical identity
\begin{equation}\label{DecomFond}
R_{t} = \text{sign}(R_{t}) \vert R_{t} \vert,
\end{equation}
where $\text{sign}(R_{t})$ and $\vert R_{t} \vert$ are the sign and magnitude (or absolute value) components of $R_{t}$, respectively. Note that there is no loss of information in this decomposition. Following AG, we apply this decomposition directly to raw returns rather than mean-adjusted returns; \citet{Christoffersen-Diebold:2006} show that removing unconditional means before decomposing can disguise sign predictability, providing the rationale for this choice. 

The conditional mean of $R_t$, given that $ \boldsymbol X_{t-1} = \boldsymbol x_{t-1} $, can then be expressed as
\[
 \mathbb{E}(R_{t} \mid \boldsymbol x_{t-1}) = \mathbb{E}\big( \text{sign}(R_{t}) \, \vert R_{t} \vert \,\big|\, \boldsymbol x_{t-1} \big).
\]
The objective of decomposing $R_t$ as in (\ref{DecomFond}) is to obtain $\mathbb{E} (R_{t} \mid \boldsymbol x_{t-1})$ by modeling the joint conditional distribution of $\text{sign}(R_{t})$ and $\vert R_{t} \vert$, given the predictors. 
Next, we present a simple example that illustrates why such a decomposition-based approach can be advantageous compared to OLS and its extensions, including CSR, which relies on linear projections as its foundational building blocks.

\subsection{Motivating example}\label{example}

This toy example illustrates how linear predictive regression models can fail to capture fundamental nonlinear dynamics, such as those arising from the decomposition in \eqref{DecomFond}. While this identity holds for any real-valued random variable, linear models may fail to exploit the structure it reveals, thereby missing essential aspects of the data-generating process.

Assume that the asset return is related to two predictor variables via the nonlinear relationship
\begin{equation*}
R_{t} = \text{sign}(X_{1,t-1}) \, \vert X_{2,t-1} \vert,
\end{equation*}
where $X_{1,t-1}$ and $X_{2,t-1}$ are independent random variables, each following an $N(0,1)$ distribution. This independence assumption is purely for expositional simplicity in this toy example; the CSM framework developed below does not require the predictors to be independent.

Economically, this stylized structure can be viewed as separating direction and magnitude in return dynamics.
The variable $X_{1,t-1}$ summarizes information that tilts the market toward “good” or “bad” states, and its sign determines whether the return is positive or negative.
In contrast, $X_{2,t-1}$ can be interpreted as a magnitude (or ``volatility-state'') signal: $|X_{2,t-1}|$ determines how large the return move is.
The realized return is then the product of a directional signal and a magnitude signal.

The return is fully predictable because the population conditional expectation satisfies
\begin{equation*}
\mathbb{E}( R_{t} \mid X_{1,t-1}, X_{2,t-1} ) = \text{sign}(X_{1,t-1}) \, \vert X_{2,t-1} \vert,
\end{equation*}
which equals $R_t$ with probability one, since $R_t$ is a deterministic function of its predictors. This confirms that both $X_{1,t-1}$ and $X_{2,t-1}$ play an essential role in predicting $R_t$.

If, however, one assumes the linear relationship
\begin{equation*}
R_{t} = \beta_{1} X_{1,t-1} + \beta_{2} X_{2,t-1} + \varepsilon_{t}
\end{equation*}
with $\mathbb{E}( \varepsilon_{t} \mid X_{1,t-1}, X_{2,t-1} ) = 0$ and fits the model using OLS, then only $X_{1,t-1}$ 
will have a nonzero coefficient in the population regression.
Indeed, the population OLS coefficient $\beta_1$ is given by $\beta_1 = \mathbb{E}(R_t X_{1,t-1})$. Substituting the definition of $R_t$, we obtain
\[
\beta_1 = \mathbb{E} \left[ \text{sign}(X_{1,t-1}) \, |X_{2,t-1}| \, X_{1,t-1} \right].
\]
Observe that $\text{sign}(X_{1,t-1})  X_{1,t-1} = |X_{1,t-1}|$, so the expression simplifies to
\[
\beta_1 = \mathbb{E} \left[ |X_{1,t-1}| \, |X_{2,t-1}| \right].
\]
Because $X_{1,t-1}$ and $X_{2,t-1}$ are independent standard normal variables, the expectation factorizes:
\[
\beta_1 = \mathbb{E}(|X_{1,t-1}|)  \mathbb{E}(|X_{2,t-1}|) = \left( \sqrt{\frac{2}{\pi}} \right)^2 = \frac{2}{\pi}.
\]
On the other hand, the contribution of $X_{2,t-1}$ is overlooked, since the population OLS coefficient $\beta_{2}$ is
\begin{equation*}
\beta_{2} = \mathbb{E}( R_{t} X_{2,t-1} ) =  \mathbb{E}\big(\text{sign}(X_{1,t-1}) \big)  \mathbb{E}\big( \vert X_{2,t-1} \vert X_{2,t-1} \big) = 0,
\end{equation*}
due to the independence of $X_{1,t-1}$ and $X_{2,t-1}$ and the symmetry of $X_{1,t-1}$, which implies $\mathbb{E}\big (\mathrm{sign}(X_{1,t-1}) \big)=0$.

To quantify the consequences of this misspecification, we compare the population mean squared error (MSE) of the OLS predictor to that of the true conditional expectation. Since $R_t = \text{sign}(X_{1,t-1}) \, |X_{2,t-1}|$ by construction, the oracle predictor $\mathbb{E}(R_t \mid X_{1,t-1}, X_{2,t-1})$ perfectly recovers $R_t$, resulting in $\text{MSE}_{\text{oracle}} = 0$. In contrast, the population OLS predictor corresponds to the best linear approximation in mean square, given by
\[
\Proj(R_t \mid X_{1,t-1}, X_{2,t-1}) = \frac{2}{\pi} X_{1,t-1},
\]
which fails to capture the nonlinear interaction between the sign and magnitude components. The resulting population MSE is strictly positive:
\[
\text{MSE}_{\text{OLS}} = \mathbb{E}\left[\left(R_t - \frac{2}{\pi} X_{1,t-1} \right)^2\right] = 1 - \frac{4}{\pi^2} \approx 0.595.
\]
This demonstrates that although OLS correctly identifies $X_{1,t-1}$ as having predictive power, it fails to recover the true nonlinear structure and overlooks the essential role of $X_{2,t-1}$, resulting in a substantial loss of predictive accuracy.

This example highlights a key limitation of linear predictive models: their inability to capture interactions between sign and magnitude that are central to return dynamics. While the CSR method improves upon OLS by averaging over many low-dimensional linear models to reduce estimation risk, it remains confined to the linear span of predictors and cannot recover the nonlinear structure illustrated here. More generally, when return dynamics involve distinct directional and scale components, purely linear predictive models may miss economically meaningful structure. This limitation motivates the use of decomposition-based models, notably the proposed CSM approach; see \citet{Anatolyev-Gospodinov-Jamali-Liu:2017} for another illustrative example.

To preview the empirical relevance of this decomposition, we turn to the data used in our application. Table~\ref{correlations} reports simple correlations between lagged predictors and the realized return $r_t$, as well as with its sign and magnitude components, denoted $s_t$ and $m_t$, respectively. Although these unconditional correlations are generally modest, several predictors exhibit stronger associations with one of the two components than with $r_t$ itself. For example, the short-term interest rate $tbl_{t-1}$ is more strongly correlated with the sign component ($-0.143$) than with returns ($-0.097$), whereas credit-related variables such as $dfy_{t-1}$ and $dfr_{t-1}$ display the strongest correlations with the magnitude component ($0.166$ and $-0.146$, respectively). These patterns suggest that modeling directional and scale information jointly, rather than forcing both channels through a single linear conditional-mean relation, may capture economically meaningful structure.

\subsection{Copula-based decomposition approach}

In this subsection, we present the AG copula-based decomposition approach. 
This method  builds upon  (\ref{DecomFond}) by first rewriting it as
\begin{equation}\label{DecomCop}
R_{t} = M_t(2 S_t - 1), 
\end{equation}
where  $M_{t} = |R_t|$ is a non-negative random variable and $S_{t} =  \frac{1}{2} \bigl(\text{sign}(R_{t}) + 1 \bigr) = \mathds{1} \{ R_{t} > 0 \} $ is a Bernoulli random variable.
Here $\mathds{1} \{ \cdot \} $ is the indicator function. In the decomposition approach, $M_t$ and $S_t$ are treated as the dependent variables. 
AG specify a multiplicative error model (MEM) for the ``volatility'' component $M_t$ (since under a scale model $R_t=\sigma_{t}  Z_t$, we have  $\vert R_t \vert = \sigma_t \vert Z_t \vert$, so the absolute return is proportional to volatility; \citealt{Granger-Sin:2000}), a  binary choice model for the direction-of-change component $S_t$, and then construct their joint distribution by using a bivariate copula that links the two marginal distributions.

\subsubsection{Marginal distributions}

Following the framework of \citet{Engle:2002}, AG specify the dynamics of $M_t \ge  0$ to be an MEM model with a Weibull distribution for the error term. 
In the present context with $\boldsymbol X_{t-1}$ comprising the driving variables, the model for the magnitude component in (\ref{DecomCop}) becomes
\begin{equation}\label{MEM1}
\begin{split}
M_{t} & = \psi_{t}  \eta_{t},   \\ 
 \log(\psi_{t})  & = w_{v} +   \boldsymbol \delta^{\prime}_{v} \boldsymbol{x}_{t-1},   
 \end{split}
\end{equation}
where $\psi_{t}  = \mathbb{E}( M_{t} \, \vert \, \boldsymbol x_{t-1}  ) $ and the error term $\eta_{t} \ge  0$ follows  a Weibull distribution, scaled to ensure $\mathbb{E}( \eta_{t} \, \vert \, \boldsymbol x_{t-1}  )= 1 $.
AG consider a richer specification that includes dynamic autoregressive terms in addition to the predictor variables. However, for our comparison with the traditional OLS predictive regression, we focus on specifications that use only the same predictor variables as in (\ref{OLSregression}).
Under (\ref{MEM1}), the cumulative distribution function (CDF) of $M_t$ becomes
\begin{equation}
F_{M_{t} \vert \boldsymbol X_{t-1} }(m \, \vert \, \boldsymbol x_{t-1}) = 1- \exp \bigg( - \Big( \frac{m}{\psi_{t}} \Gamma \big( 1 + \kappa^{-1} \big) \Big)^{\kappa} \bigg),
\label{magnitude_cdf}
\end{equation}
with corresponding probability density function 
\begin{equation}\label{magnitude_pdf}
f_{M_t \vert \boldsymbol X_{t-1}}(m \, \vert \, \boldsymbol x_{t-1}) = \kappa \psi_{t}^{-\kappa} \Gamma\left(1 + \kappa^{-1} \right)^{\kappa} m^{\kappa-1} \exp \bigg( - \Big(\frac{m}{\psi_{t}} \Gamma (1 + \kappa^{-1} ) \Big)^{\kappa} \bigg),
\end{equation}
where $\Gamma(\cdot)$ is the gamma function and $\kappa > 0 $ is a shape parameter.
We stack the parameters on which (\ref{magnitude_cdf}) and (\ref{magnitude_pdf}) depend into the vector $\boldsymbol \theta_v = ( w_{v}, \boldsymbol \delta^{\prime}_v, \kappa)^{\prime}.$

The conditional distribution of ${S}_{t}$ given $\boldsymbol x_{t-1}$ is of course Bernoulli  $\mathcal{B}(p_{t} )$ with probability mass function (PMF) $f_{S_t \vert \boldsymbol X_{t-1}}(v \, \vert \, \boldsymbol x_{t-1}) = p_t^v (1-p_t)^{1-v},$ $ v \in \{0,1\}$ and cumulative mass function  $F_{S_t \vert \boldsymbol X_{t-1}} = 1- p_t(1-v), $ where $p_{t} = \Pr(R_t > 0 \, \vert \, \boldsymbol x_{t-1})$.
Following \citet{Nyberg:2011} and \citet{Ponka:2017}, we parameterize $p_{t}$ as a predictive probit model of the form
\begin{equation}
\begin{split}
     p_{t} & =  \Phi(\theta_{t}), \\      
\theta_{t} & =   w_{d} +   \boldsymbol \delta^{\prime}_{d} \boldsymbol{x}_{t-1},   
\end{split} \label{probit1}
\end{equation}
where $\Phi(\cdot)$ is the CDF of the standard normal distribution, and we let $\boldsymbol \theta_d = ( w_{d}, \boldsymbol \delta^{\prime}_d)^{\prime}$ regroup the associated parameters.\footnote{We prefer probit over logit as it is conformable with the Gaussian copula \citep{Anatolyev-Gospodinov:2019}.}
AG consider a more general binary choice model that allows for own dynamics. 
As already mentioned, our focus here is on specifications that use only the predictor variables appearing in (\ref{OLSregression}), as we aim to investigate the channels through which their predictive ability operates.

\subsubsection{Joint distribution}

In order to construct the bivariate distribution of $Y_t=(M_t, S_t)^{\prime},$ the AG approach appeals to the theory of copulas. Specifically,  a conditional meta-distribution is created as
\begin{equation}
 F_{Y_t \vert \boldsymbol X_{t-1}} (u, v \, \vert \, \boldsymbol x_{t-1} ) = C \big( F_{M_t \vert \boldsymbol X_{t-1}} (u \, \vert \,\boldsymbol x_{t-1} ) , F_{S_t \vert \boldsymbol X_{t-1} }( v \, \vert \, \boldsymbol x_{t-1}  )\big),  
 \label{copula1}
\end{equation}
where $C(w_1, w_2) $ is a  copula distribution function defined on $[0,1]^2$, with dependency parameter $\theta_c.$
Upon differentiation of (\ref{copula1}), AG obtain the joint conditional density/mass function of $M_t$ and $S_t$ as
\begin{equation}
f_{Y_{t}  \vert \boldsymbol X_{t-1}} (u,v \, \vert \, \boldsymbol x_{t-1} ) = \varrho_{t} \Big(F_{M_t \vert \boldsymbol X_{t-1} }(u \, \vert \, \boldsymbol x_{t-1} ) \Big)^{v} \Big(1-\varrho_{t} \big(F_{M_t \vert \boldsymbol X_{t-1}}(u \, \vert \, \boldsymbol x_{t-1}) \big) \Big)^{1-v}
     f_{M_t \vert \boldsymbol X_{t-1}}(u \, \vert \, \boldsymbol x_{t-1}),
\label{copulajoint}
\end{equation}
where $ \varrho_{t}(z) = 1 - \partial C(z, 1-p_{t}  )/\partial w_{1} $ represents a ``deformed'' probability mass of $S_t$ whose probability of success is given by 
$\varrho_{t}\big(F_{M_t \vert \boldsymbol X_{t-1} }(u \, \vert \, \boldsymbol x_{t-1} ) \big)$.

The bivariate copulas used in the empirical application are the Gaussian, Frank, Clayton, and Farlie-Gumbel-Morgenstern (FGM) copulas, following AG and \citet{Anatolyev-Gospodinov:2019}. Since these copulas are well known, the Appendix provides only their definitions and the corresponding expressions for the deformed probability of success, $\varrho_t(z)$; see \citet{Trivedi-Zimmer:2007} for more detailed discussions.

\subsubsection{Conditional mean prediction and parameter estimation}

The decomposition in  (\ref{DecomCop}) implies that the conditional mean of $R_t$ can be written as
\[
 \mathbb{E} (R_t \mid \boldsymbol x_{t-1}) = 2 \xi_t -  \psi_{t},
\]
where $\xi_t = \mathbb{E}(  M_t S_t \, \vert \, \boldsymbol x_{t-1})$ is the expected cross-product of  $M_t$ and $S_t$, and  $ \psi_{t} = \mathbb{E}( M_{t} \, \vert \, \boldsymbol x_{t-1}  )  $ is given in (\ref{MEM1}). The cross-product term is given by AG as
\begin{equation}
\xi_t = \int_{0}^{1} F^{-1}_{M_t \vert \boldsymbol X_{t-1}}(z \, \vert \, \boldsymbol x_{t-1})  \varrho_t (z) dz, 
\label{xi}
\end{equation}
where $ F^{-1}_{M_t \vert \boldsymbol X_{t-1}}(z \, \vert \, \boldsymbol x_{t-1}) =  \psi_{t} \Gamma(1 + \kappa^{-1})^{-1} \big( - \log(1-z) \big)^{1/\kappa}$ is the conditional quantile function of $M_t$.  
For practical applications, we evaluate the above integral by Monte Carlo integration \citep[][Ch.~3]{Robert-Casella:2010}.

The complete vector of model parameters is $\boldsymbol \theta = ( \boldsymbol \theta_v^{\prime}, \boldsymbol \theta_d^{\prime},  \theta_c)^{\prime}$,
comprising: (i) $\boldsymbol \theta_v = ( w_{v}, \boldsymbol \delta^{\prime}_v, \kappa)^{\prime}$,  the parameters of the marginal distribution of $M_t$ in (\ref{MEM1}); (ii) $\boldsymbol \theta_d = ( w_{d}, \boldsymbol \delta^{\prime}_d)^{\prime}$, the parameters of the marginal distribution of $S_t$ in (\ref{probit1}); and (iii) $\theta_c$, the parameter of the copula distribution in (\ref{copula1}).
Given the sample of returns $r_1,\ldots, r_T,$ which upon decomposition yields the realizations $(m_1, s_1), \ldots, (m_T, s_T)$, along with the realized values of the predictor variables,  the full log-likelihood function can be computed as 
\begin{equation}
L(\boldsymbol \theta) = L_v(\boldsymbol \theta_v) + L_c(\boldsymbol \theta_v, \boldsymbol \theta_d, \theta_c) 
\label{fullMLE}
\end{equation}
with 
\[
L_v(\boldsymbol \theta_v)  = \sum_{t=1}^{T} \log  f_{M_t \vert \boldsymbol X_{t-1}}(m_t \, \vert \, \boldsymbol x_{t-1} ) 
\]
and
\[
L_c(\boldsymbol \theta_v, \boldsymbol \theta_d, \theta_c)   = \sum_{t=1}^{T}  s_t \log \varrho_{t} ( u_t )  
  + (1-s_t) \log \big( 1-\varrho_{t} ( u_t ) \big), 
\]
where $u_t = F_{M_{t} \vert \boldsymbol X_{t-1}} (m_t \, \vert \, \boldsymbol x_{t-1} ).$
The maximum likelihood estimates (MLE) are then obtained as
\[
\hat{\boldsymbol \theta} =  \arg \max_{\boldsymbol \theta_v, \boldsymbol \theta_d,  \theta_c }  L(\boldsymbol \theta),
\]
using standard numerical optimization methods \citep[][Ch.~4]{Judd:1998}.

The estimation can be simplified by employing the method of inference for margins (IFM) of \citet{Shih-Louis:1995} and \citet{Joe-Xu:1996}.
The IFM method in the present context consists of first estimating $\boldsymbol \theta_v$ by computing 
$\hat{\boldsymbol \theta}_v =  \arg \max_{\boldsymbol \theta_v}  L_v(\boldsymbol \theta_v),$ and then obtaining estimates of the remaining parameters as
\[
(\hat{\boldsymbol \theta}_d, \hat \theta_c) =  \arg \max_{\boldsymbol \theta_d,  \theta_c }  L_c(\hat{\boldsymbol \theta}_v, \boldsymbol \theta_d,  \theta_c ),
\]
in which the first-step estimates $\hat{\boldsymbol \theta}_v$ are taken as given. This procedure is computationally simpler than estimating all the parameters in $\boldsymbol \theta$ simultaneously. The drawback of the IFM approach is that there is some loss of statistical efficiency relative to the full MLE because the first step ignores the dependency of $L_c(\boldsymbol \theta_v, \boldsymbol \theta_d, \theta_c) $ on $\boldsymbol \theta_v $ when finding $\hat{\boldsymbol \theta}_v$.

\subsection{CSM decomposition approach}

\subsubsection{Factorization and its motivation}

The CSM approach constructs the joint density of $Y_t=(M_t, S_t)^{\prime}$ by conditioning the distribution of $S_t$ on both $\boldsymbol x_{t-1}$ and the contemporaneous value of the return magnitude.
Specifically, the CSM approach proceeds via the factorization
\begin{equation}
f_{Y_t \vert  \boldsymbol X_{t-1}}(m_t, s_t \, \vert \, \boldsymbol x_{t-1}) = f_{S_t \vert M_t, \boldsymbol X_{t-1}}( s_t \, \vert \, m_t, \boldsymbol x_{t-1}) f_{M_t \vert \boldsymbol X_{t-1}}(m_t \, \vert \, \boldsymbol x_{t-1}),
\label{cond}
\end{equation}
where $f_{M_t \vert \boldsymbol X_{t-1}}(m_t \, \vert \, \boldsymbol x_{t-1})$ is given by (\ref{magnitude_pdf}), and the distribution $f_{S_t \vert M_t, \boldsymbol X_{t-1}}( s_t \, \vert \, m_t, \boldsymbol x_{t-1})$ is conditional on $M_t= m_t$.
This way of expressing a joint distribution of two random variables as the product of a conditional distribution and a marginal distribution is a standard result in probability theory; it also implicitly defines a copula linking the two components \citep{Sklar:1959}.
Since the magnitude component is modeled identically in both (\ref{copulajoint}) and (\ref{cond}), any performance disparities between 
the copula-based and CSM approaches can be attributed solely to differences in how the sign-magnitude dependence structure is modeled.

Why should we prefer conditioning the sign  on  the magnitude as in (\ref{cond}), instead of the magnitude  on the sign?
There are at least three reasons.
First, the magnitude is more likely to be predictive of the sign. Indeed, there is overwhelming evidence of GARCH-type volatility clustering in financial markets, where periods of high volatility (large $m_t$) are followed by more high-volatility periods. \citet{Christoffersen-Diebold:2006} demonstrate that these volatility dynamics can lead to predictability in the sign of returns, even in the absence of mean predictability. This means that as $m_t$ changes, so too does the probability of observing a positive return.

Second, behavioral finance insights support this approach. Due to differences in the timing of information receipt, information processing, behavioral biases, and feedback trading, \citet{Treynor-Ferguson:1985},
\citet{Brown-Jennings:1989},
\citet{Hong-Stein:1999},
\citet{Cespa-Vives:2011}, and
\citet{Edmans-Goldstein-Jiang:2015}, among others, show 
that past stock prices can lead to predictable patterns in returns. 
For instance, large values of $m_t$ can trigger specific investor behaviors: significant negative returns might lead to panic selling, increasing the likelihood of subsequent negative returns, while large positive returns might lead to profit-taking or increased buying pressure, influencing the sign of the next return.

Third, in econometrics, conditioning on a variable that explains more variation in the dependent variable usually leads to better model performance. Since the magnitude exerts  a greater influence on the sign of the return than vice versa, modeling $S_t$ given $m_t$ is likely to improve the accuracy and reliability of our predictions.   
Related empirical evidence supporting the usefulness of return volatility in forecasting the direction of asset returns is reported in  \citet{Christoffersen-Diebold-Mariano-Tay-Tse:2007}, \citet{Bekiros-Georgoutous:2008}, \citet{Chevapatrakul:2013}, \citet{Algieri-Leccadito:2019}, and \citet{Campisi-Muzzioli-DeBaets:2024}.

\subsubsection{CSM model}

In the CSM approach, the specification for the magnitude component $M_t$ remains as described in (\ref{MEM1}), with the corresponding density function $f_{M_t \vert \boldsymbol X_{t-1}}(m_t \, \vert \, \boldsymbol x_{t-1}) $ given in (\ref{magnitude_pdf}).
Importantly, the conditional distribution of the sign component $S_t$  becomes Bernoulli $\mathcal B(p_t^{\ast})$, where 
$p^{\ast}_{t} = \Pr( R_{t} > 0 \, \vert \, m_{t}, \boldsymbol{x}_{t-1}  )$ now conditions on $m_t$ in addition to $\boldsymbol{x}_{t-1}$.
We specify $p^{\ast}_{t}$ as
\begin{equation}
\begin{split}
     p^{\ast}_{t} & =  \Phi(\theta^{\ast}_{t}), \\      
\theta^{\ast}_{t} & =   w_{d} +   \boldsymbol \delta^{\prime}_{d} \boldsymbol{x}_{t-1} + \beta m_t,   
\end{split} \label{probit2}
\end{equation}
so that $\beta$ captures the effect of the realized magnitude $m_t$ on $\text{sign}(R_t).$
The corresponding PMF is 
$f_{S_t \vert M_t, \boldsymbol X_{t-1}}(v \, \vert \, m_t, \boldsymbol x_{t-1}) = p_t^{\ast v} (1-p^{\ast}_t)^{1-v},$ for $ v \in \{0,1\}$.

With this approach, the conditional mean of $R_t$ given that $ \boldsymbol X_{t-1} = \boldsymbol x_{t-1} $ becomes
\begin{equation}
 \mathbb{E} (R_t \, \vert \, \boldsymbol x_{t-1}) = 2 \xi_t^{\ast} -  \psi_{t}, \label{ER}
\end{equation}
where $\psi_{t} = \mathbb{E}( M_{t} \, \vert \, \boldsymbol x_{t-1}  )  $ is given in (\ref{MEM1}), and
the term $\xi_t^{\ast} = \mathbb{E}( M_t S_t \, \vert \, \boldsymbol x_{t-1})$ is obtained via iterated expectations as
\begin{equation*}
\begin{split}
     \mathbb{E}( M_t S_t  \, \vert \, \boldsymbol x_{t-1})  & = \mathbb{E} \big( M_{t} \mathbb{E} ( S_t \, \vert \,  M_t, \boldsymbol x_{t-1} )       \, \big \vert \, \boldsymbol x_{t-1} \big) \\[1.0ex]
                                                           & = \int_0^{+\infty} m  \Phi(w_{d} +   \boldsymbol \delta^{\prime}_{d} \boldsymbol{x}_{t-1} + \beta m) f_{M_t \vert \boldsymbol X_{t-1}}(m \, \vert \, \boldsymbol x_{t-1}) d m, 
\end{split}     
\end{equation*}
which must be evaluated numerically given the absence of a closed-form solution.
Upon the change of variable $ \upsilon = F_{M_{t} \vert \boldsymbol X_{t-1} }(m \, \vert \, \boldsymbol x_{t-1}), $ this integral can be rewritten as 
\begin{equation}
\xi^{\ast}_{t}  = \int_{0}^{1} q_t(\upsilon) \Phi \big(w_{d} +   \boldsymbol \delta^{\prime}_{d} \boldsymbol{x}_{t-1} + \beta q_t(\upsilon) \big) d \upsilon, \label{xi2}
\end{equation}
where $q_t(\upsilon) = F^{-1}_{M_{t} \vert \boldsymbol X_{t-1} }(\upsilon \, \vert \, \boldsymbol x_{t-1})$ is the conditional quantile function of $M_t$ appearing in (\ref{xi}).
Here again we use Monte Carlo integration to evaluate (\ref{xi2}), which then serves to compute (\ref{ER}).

The parameters entering the CSM model specification are $\boldsymbol \theta^{\ast} = ( \boldsymbol \theta_v^{\prime}, \boldsymbol \theta_d^{\ast \prime})^{\prime}$,
where $\boldsymbol \theta_d^{\ast} = ( w_{d}, \boldsymbol \delta_d^{\prime}, \beta)^{\prime}$.
In light of  (\ref{cond}), the sample log-likelihood function is expressed as
\begin{equation}
L(\boldsymbol \theta^{\ast}) = L_v(\boldsymbol \theta_v) + L_d(\boldsymbol \theta_d^{\ast}), 
\label{fullMLE2}
\end{equation}
where $L_v(\boldsymbol \theta_v) $ is given in (\ref{fullMLE}) and $L_d(\boldsymbol \theta_d^{\ast})   = \sum_{t=1}^{T}  s_t \log p_t^{\ast}  + (1-s_t) \log ( 1- p_t^{\ast} ). $
The parameter estimates needed to make the CSM model operational can be computed as
$\hat{\boldsymbol \theta}^{\ast} =  \arg \max_{\boldsymbol \theta^{\ast} }   L (\boldsymbol \theta^{\ast}).$
Since $L_v(\boldsymbol \theta_v)$ and $L_d(\boldsymbol \theta_d^{\ast})$ have no common parameters, 
 the values $\hat{\boldsymbol \theta}^{\ast} = ( \hat{\boldsymbol \theta}_v^{\prime}, \hat{\boldsymbol \theta}_d^{\ast \prime})^{\prime}$ that maximize (\ref{fullMLE2}) can be equivalently found by
computing $\hat{\boldsymbol \theta}_v =  \arg \max_{\boldsymbol \theta_v}  L_v(\boldsymbol \theta_v)$ and 
$\hat{\boldsymbol \theta}_d^{\ast} =  \arg \max_{\boldsymbol \theta_d^{\ast}}  L_d(\boldsymbol \theta_d^{\ast})$, separately, with no loss in statistical efficiency.

We also note that the maximization of $L_d(\boldsymbol \theta_d^{\ast})$ is not sensitive to the choice of starting values. For probit models with a log-likelihood function of that form, the Hessian matrix is negative definite for all values of $\boldsymbol \theta_d^{\ast}$, ensuring global concavity and a unique maximum \citep[see, e.g.,][p.~26]{Maddala:1983}. As a result, a Newton-Raphson procedure will typically converge to the unique maximum likelihood estimator from any reasonable starting point.

The baseline specification in (\ref{probit2}) can be extended to allow for more flexible modeling, including interaction terms between $m_t$ and elements of $ \boldsymbol{x}_{t-1} $, as well as nonlinear effects of $m_t$ itself. To illustrate the latter, we consider a polynomial specification in which the index function
\[
\theta^{\ast}_{t} = w_{d} + \boldsymbol{\delta}^{\prime}_{d} \boldsymbol{x}_{t-1} + \beta m_t + \beta_{2} m_t^2 + \beta_{3} m_t^3
\]
includes higher-order terms in $m_t$. The additional components $\beta_2 m_t^2$ and $\beta_3 m_t^3$ introduce curvature into the probit index, allowing the probability of a positive return to respond nonlinearly to changes in return magnitude. This extension increases the model’s flexibility while maintaining parsimony. We refer to this nonlinear index specification as the CSM (Poly) model, whereas the version in which $\theta_t^{\ast}$ is linear in $(\boldsymbol{x}_{t-1}, m_t)$ is referred to as the CSM (Baseline) model.

While more flexible functional forms, such as splines or kernel-based methods, could model nonlinearities in the sign-magnitude relationship, they typically require tuning and regularization to prevent overfitting. In contrast, the polynomial expansion used in the CSM (Poly) model offers a simple, interpretable alternative that captures moderate nonlinearities without the computational burden of nonparametric methods. Finally, one could also augment  (\ref{MEM1}) and (\ref{probit2}) with lagged values of $m_{t}$, $\psi_{t} $, $ s_t$, and $\theta^{\ast}_{t}$, to get  dynamic autoregressive specifications; see \citet{Kauppi-Saikkonen:2008}, 
\citet{Anatolyev-Gospodinov:2010}, \citet{Nyberg:2011}, \citet{Liu-Luger:2015}, \citet{Anatolyev-Gospodinov-Jamali-Liu:2017}, and \citet{Ponka:2017} for examples. 
We explored such extensions but found that none consistently outperformed the baseline CSM model, in line with AG's finding that forecasting gains stem primarily from predictor variables rather than autoregressive structures (p.~240).

\begin{remark}
\textup{(CSM vs.\ copula-based decomposition: Two non-nested models). Although both the CSM and copula-based decomposition models build on the decomposition in \eqref{DecomFond}, they represent distinct and, under the parametric specifications we consider, non-nested approaches. The copula-based framework models the marginals of $M_t$ and $S_t$ separately (each conditional on past information) and then captures their contemporaneous dependence via a bivariate copula. 
Because the copula operates on the marginal distributions, any dependence of the sign on the magnitude is induced implicitly through the joint distribution, rather than through explicit conditioning on $m_t$ in the sign equation.
In contrast, the CSM model embeds dependence by conditioning the sign directly on the contemporaneous magnitude, without requiring a copula specification for the joint distribution. The two approaches coincide only under conditional independence: in the copula framework, this corresponds to using the independence copula; in CSM, this corresponds to excluding $m_t$ and any function of it from the probit equation.}
\end{remark}

\section{Empirical application}

In this section, we examine the predictability of monthly excess returns $r_t$ on the S\&P 500 value-weighted index. We begin with ten financial predictors that are widely used in the return predictability literature, including several proposed by \citet{Welch-Goyal:2008}: the log dividend-price ratio ($dp$), the log earnings-price ratio ($ep$), the book-to-market ratio ($btm$), the default yield spread ($dfy$), the term spread ($tms$), the short-term interest rate ($tbl$), the long-term government bond return ($ltr$), the default return spread ($dfr$), net equity expansion ($ntis$), and inflation ($infl$). These variables are a subset of those in \citet{Welch-Goyal:2008}, updated through 2021. The monthly data span January 1948 to December 2021 and are obtained from Amit Goyal's website.

To avoid multicollinearity and improve model stability, we exclude two predictors ($ep$ and $btm$) due to their high correlation (above 0.80) with other variables. The final predictor set thus includes eight variables.\footnote{We also consider an extended specification that adds a time-series momentum indicator following \citet{Moskowitz-Ooi-Pedersen:2012}. Section~D of the Supplementary material shows that the main conclusions are unchanged.} We deliberately work with a moderate set of predictors to facilitate transparent subset selection and interpretation. This design differs from studies using substantially larger predictor sets \citep[e.g.,][]{Bianchi-Rubesam-Tamoni:2025,Goyal-Welch-Zafirov:2024}, which can change the absolute magnitude of economic performance measures and may increase the scope for gains from flexible nonlinear methods.

Table~\ref{correlations} reports sample correlations between lagged predictors and excess returns, as well as the sign and magnitude components of returns. Panel A shows that correlations between $r_t$ and the lagged predictors are generally weak, with the largest absolute correlation occurring for the short-term interest rate $tbl_{t-1}$ ($-0.097$). Panel B shows that several predictors exhibit stronger associations with individual components than with returns themselves. In particular, $tbl_{t-1}$ exhibits the strongest correlation with the sign component $s_t$ ($-0.143$), while the default yield spread $dfy_{t-1}$ shows the strongest correlation with the magnitude component $m_t$ ($0.166$), followed closely by $dfr_{t-1}$ ($-0.146$). Panel C reports correlations among predictors within the final eight-variable set; the largest absolute correlation is about $0.50$, suggesting no evidence of severe multicollinearity in the specification used in the empirical analysis.

To ensure a fair comparison across subset sizes $k \in \{1,\ldots,8\}$, 
we fix the number of predictors at $k$ for all methods and, for methods 
that involve subset selection, use a common in-sample selection criterion. 
We split the sample into two parts after aligning lagged predictors and returns: 
the first 400 monthly return observations (February 1948 to May 1981) are used for 
in-sample predictor selection (when applicable) and model estimation, consistent with AG's rolling-window design, and the remaining 
487 observations are reserved for out-of-sample forecast evaluation.

For each $k$, we enumerate all $n_{k,8}=\binom{8}{k}$ candidate predictor combinations 
using the first 400 observations and generate a sequence of one-step-ahead in-sample 
forecasts for each combination. 
Methods that select a best $k$-predictor subset (the linear model, 
GARCH-M, MS, the AG copula-based models, and the CSM models) retain the 
single $k$-predictor combination that maximizes directional accuracy.
Directional accuracy is measured by the 
area under the (correct classification) curve (AUC), following \citet{Jorda-Taylor:2011,Jorda-Taylor:2012}. 
This choice is motivated by the fact that the limited success in return predictability is largely 
attributable to the difficulty of forecasting the sign of returns, rather than their magnitude. 
Since the AUC captures the ability of forecasts to discriminate between positive- and 
negative-return periods, it provides a natural and informative criterion for predictor 
selection in this context. In contrast, the CSR approach does not select 
a single best subset; for each $k$, it forms forecasts by averaging across all linear regressions 
that use exactly $k$ predictors drawn from the full set of eight candidates \citep{Elliott-Gargano-Timmermann:2013}. 
The historical average serves as an additional benchmark.

Let $\hat r_t$ denote the one-step-ahead in-sample forecast of return $r_t$. Define \sloppy 
$N_n = \sum_{t=1}^{400} \mathds{1}\{\text{sign}(r_t) = -1\}$, and let $v_j$ be the value of 
$\hat r_t$ when $\text{sign}(r_t) = -1$, for $j = 1, \ldots, N_n$. Similarly, define 
$N_p = \sum_{t=1}^{400} \mathds{1}\{\text{sign}(r_t) = 1\}$, and let $u_i$ be the value of 
$\hat r_t$ when $\text{sign}(r_t) = 1$, for $i = 1, \ldots, N_p$. The AUC is then computed as
\[
\mathrm{AUC} = \frac{1}{N_n N_p} \sum_{j=1}^{N_n} \sum_{i=1}^{N_p} \left( \mathds{1}\{v_j < u_i\} 
+ \frac{1}{2} \mathds{1}\{v_j = u_i\} \right),
\]
where the second term serves to break ties. The AUC estimates $\Pr(v < u)$, which equals 
0.5 when the forecasts are uninformative and approaches 1 when they assign systematically 
higher values to positive-return periods than to negative-return periods. 
See \citet{Jorda-Taylor:2011} for further discussion.

As a robustness check, we also perform an alternative best $k$-predictor subset selection using the MSE criterion. Although the resulting subsets differ from those based on AUC, they overlap substantially for most models and subset sizes. Out-of-sample evaluation shows that MSE-based selection underperforms relative to AUC-based selection, particularly for the linear model. We therefore proceed with the AUC-based subsets for the remainder of the analysis. Additional details and results for the MSE-based selection are provided in Section A of the Supplementary material.

Table~\ref{Bestsubset_AUC} reports the variables selected in the best subset for each value of $k$. Panel A shows the linear and GARCH-M models, which select identical subsets at all values of $k$. Panel B shows the MS model, and Panel C shows the decomposition-based models (CSM and copula-based), for which selection results are identical across all copula specifications.

Across models, there is substantial overlap in variable selection. The short-term interest rate ($tbl_{t-1}$) is consistently chosen at $k = 1$ in all specifications, reflecting its strong predictive power for return direction. This variable remains in every selected subset for the linear and decomposition-based models from $k = 1$ to $k = 8$. In the MS model, $tbl_{t-1}$ is selected for $k=1$--$3$ and again from $k=6$ onward.

Moreover, the linear and decomposition-based models produce identical subsets for most values of $k$, differing only at $k = 2$, $k = 5$, and $k = 6$, where a single variable accounts for the discrepancy. Notably, $dp_{t-1}$ enters the decomposition-based models earlier (at $k = 5$) than in the linear model (at $k = 7$), which is consistent with decomposition-based models exploiting predictive content from valuation ratios at smaller subset sizes. The MS model exhibits greater variation at small $k$, with inflation ($infl_{t-1}$) entering already at $k=2$. All models converge at $k = 8$, where all predictors are included by construction. Overall, the consistency in selected variables across specifications enhances the credibility of the comparative analysis by ensuring that performance differences are driven primarily by model structure, rather than predictor choice.

\subsection{Forecasting performance}

We assess out-of-sample forecasting performance using a one-step-ahead rolling-window scheme with a fixed length of $L = 400$, consistent with AG. This window provides the initial sample for model estimation and yields 487 out-of-sample return forecasts. At each step, the oldest observation is dropped, the next available data point is added, and the model is re-estimated. This procedure continues until the end of the sample, allowing forecasts to adapt to new information. By re-estimating the model each month, we accommodate potential changes in the data-generating process and mitigate parameter instability in a simple but practical manner.

Let $\hat{r}_{t+1}=\widehat{\mathbb{E}}(R_{t+1}\mid \boldsymbol{x}_t)$ denote the forecast obtained from a given conditional model, and let $\bar{r}_{t+1}$ denote the historical average estimate of $\mathbb{E}(R_{t+1})$, both computed using the $L$ observations in the rolling window.
It is worth noting that all historical-average forecasts were positive.

To assess the performance of the predictive methods relative to the historical average, we follow AG and consider the out-of-sample $ R^{2} $ statistic \citep{Campbell-Thompson:2008} computed as
\begin{equation}\label{Rsquare}
R^{2}_{\text{OOS}} = 1 - \frac{\sum_{t = L}^{ T-1} \mathcal L  ( r_{t+1} - \hat{r}_{t+1} )}{ \sum_{t = L}^{T-1} \mathcal L ( r_{t+1} -  \bar{r}_{t+1}  ) },
\end{equation}
where the loss function is  $\mathcal L ( e) = e^2$ when using squared forecast errors and $\mathcal L ( e) = |e|$ when using absolute forecast errors.
A positive $ R^2_{\text{OOS}} $ indicates that the conditional model outperforms the historical average, while a negative value suggests the opposite. Additionally, we report the results of the well-known \citet{Diebold-Mariano:1995} (DM) test of equal predictive ability. 
The null hypothesis under test is $H_0 : \mathbb{E}( \Delta \mathcal L_{t+1}) = 0$, where $\Delta \mathcal L_{t+1} =  \mathcal L  ( r_{t+1} - \hat{r}_{t+1} ) - \mathcal L ( r_{t+1} -  \bar{r}_{t+1}  )$ is the loss function differential (using either squared or absolute loss functions). This differential is considered statistically significant if the DM  $t$-ratio statistic exceeds the critical values from the standard normal distribution.

\begin{figure}
      \caption{Out-of-sample $R^2$ as a function of the number of predictors ($k$)}
      \label{fig:ROOS dynamics}
      \vspace{0.5 cm}
   \begin{minipage}[c]{0.5\linewidth}
     \centering
     (a) Squared errors 
    \vspace{2.5 cm}
     \includegraphics[width=\textwidth, height=\textheight, keepaspectratio]{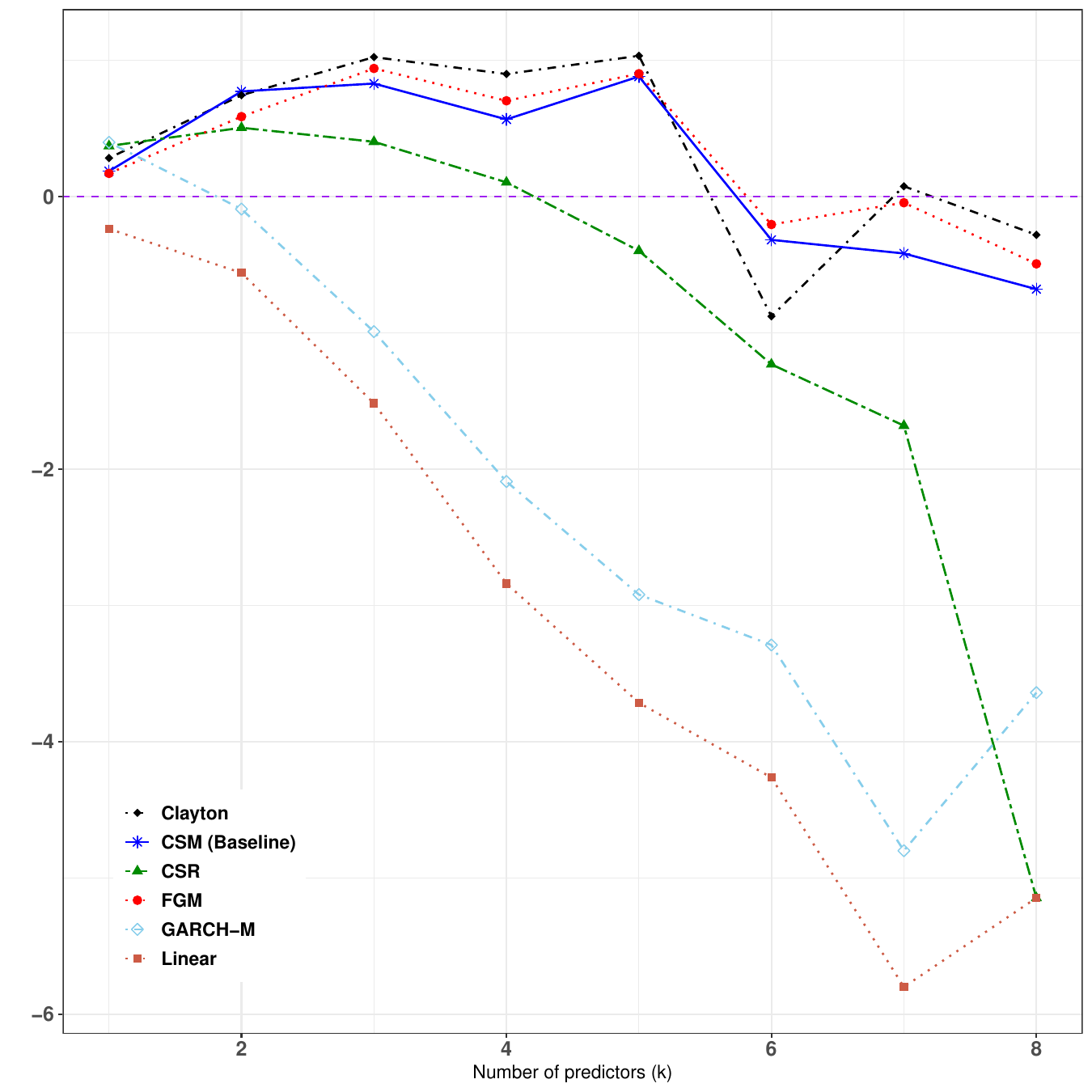}
  \end{minipage}\hspace{1 cm}
  \begin{minipage}[c]{0.5\linewidth}
     \centering
     (b) Absolute errors
    \vspace{2.5 cm}
     \includegraphics[width=\textwidth, height=\textheight, keepaspectratio]{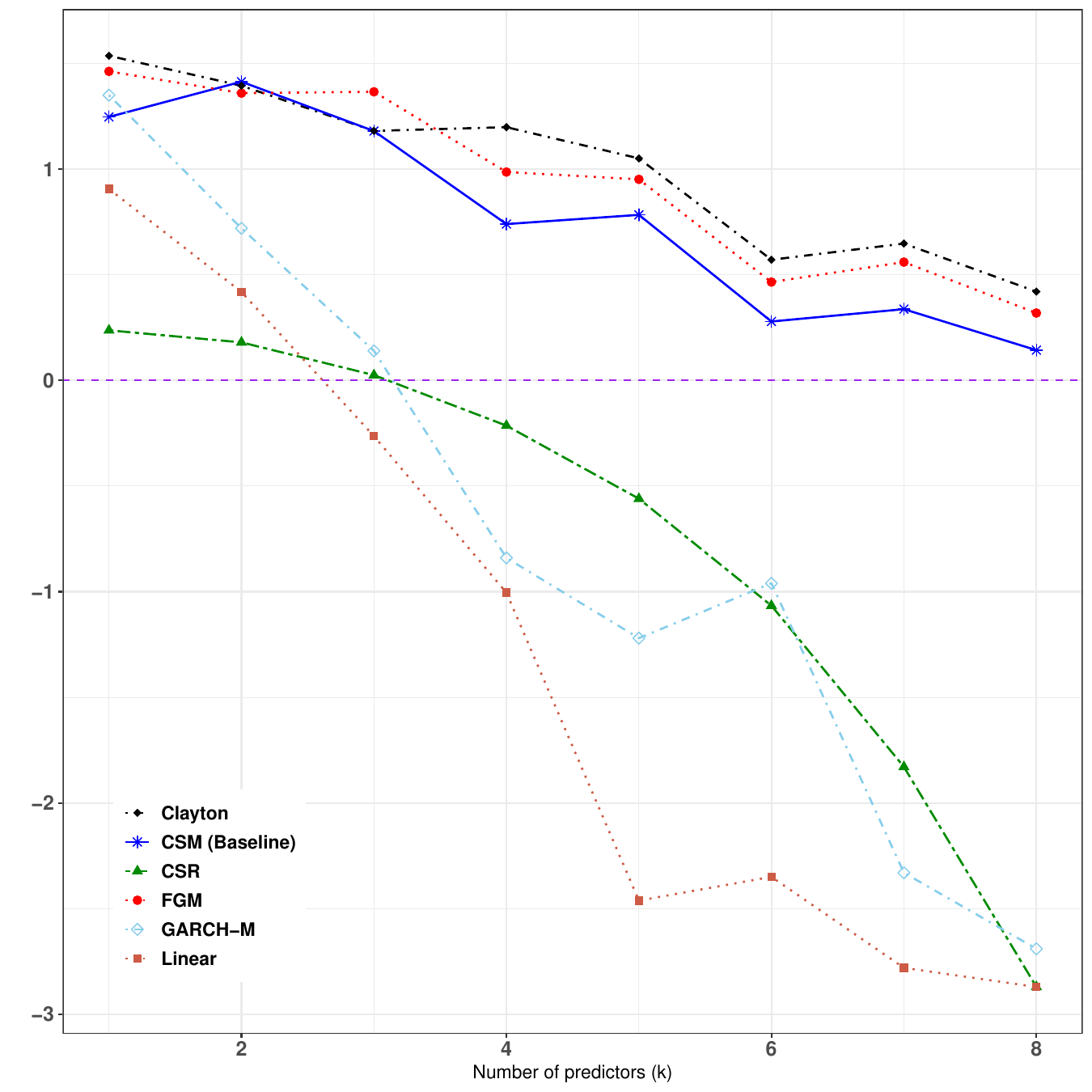}
  \end{minipage}
 \vspace{1.5 cm}
 \end{figure}

Table~\ref{ROOS_all} reports the $R^2_{\text{OOS}}$ values (in \%) across models and subset sizes $k$, under both squared and absolute loss functions. 
We consider the linear predictive regression, the CSR approach, the two nonlinear benchmarks (MS and GARCH-M), and the decomposition-based models (copula-based and CSM). 
Figure~\ref{fig:ROOS dynamics} complements the table by plotting the $R^2_{\text{OOS}}$ profiles for a selected subset of models (Linear, CSR, GARCH-M, Clayton and FGM copulas, and CSM Baseline), facilitating a visual comparison of how performance changes with $k$.

Under squared loss, all decomposition-based models deliver positive $R^2_{\text{OOS}}$ values for $k \le 5$, with the strongest gains concentrated at intermediate subset sizes (roughly $k=3$--$5$); see Table~\ref{ROOS_all}. Figure~\ref{fig:ROOS dynamics}(a) illustrates this pattern for the decomposition-based 
specifications included in the plot (Clayton, FGM, and CSM Baseline). The CSR approach performs reasonably well at low dimensionality (positive for $k \le 4$) 
but deteriorates once additional predictors are included. In contrast, the linear predictive regression and the MS model yield negative 
$R^2_{\text{OOS}}$ values across all $k$, indicating persistent underperformance relative to the historical average benchmark; this is consistent with evidence that 
regime-switching models often deliver strong in-sample fit but weaker out-of-sample performance due to uncertainty around regime forecasts \citep{Boot-Pick:2018}. 
The GARCH-M model delivers gains only at very low dimensionality (notably $k=1$), and its performance declines steadily as $k$ increases. For larger subset sizes ($k \ge 6$), squared-loss performance weakens across specifications, including the decomposition-based class, yet these models remain the most competitive overall in this metric.

Under absolute loss, a similar ranking emerges, but performance is more robust for the decomposition-based models. 
In particular, all decomposition-based models deliver positive $R^2_{\text{OOS}}$ values 
across all subset sizes and thus consistently outperform the historical average under 
absolute loss; see Table~\ref{ROOS_all}. Figure~\ref{fig:ROOS dynamics}(b) shows that the plotted decomposition-based models (Clayton, FGM, and CSM Baseline) stay above zero throughout, while CSR and the linear model deliver improvements only for very small subset sizes (CSR for $k \le 3$, with values close to zero at $k=3$, and the linear model for $k \le 2$) before turning clearly negative. The GARCH-M model also produces positive gains only for small $k$ (up to $k=3$) and then deteriorates, and the MS model continues to perform poorly with large negative values (Table~\ref{ROOS_all}).

Taken together, these results indicate that the CSM and copula-based decomposition models provide the most robust out-of-sample performance across loss functions and subset sizes. 
More broadly, the evidence highlights that increasing the number of predictors does not necessarily improve forecast accuracy: for the benchmark methods (linear, CSR, GARCH-M, and MS), performance is maximized at small subset sizes, reinforcing the role of parsimony emphasized by \citet{Elliott-Gargano-Timmermann:2013}. 
The linear model delivers negative $R^2_{\text{OOS}}$ values under squared loss for all $k$, and under absolute loss it improves on the historical average only at very small subset sizes ($k \le 2$), consistent with evidence that linear predictive regressions often fail to outperform the historical average in equity premium forecasting \citep{Welch-Goyal:2008,Campbell-Thompson:2008}.

The results so far indicate that the decomposition-based approaches (both copula-based and CSM) generally outperform the historical average benchmark as well as the linear predictive regression and the MS specification. 
However, an important question remains: do the decomposition-based models differ significantly in performance, and if so, which specification performs best?

To address this question, we adopt the model confidence set (MCS) methodology of \citet{Hansen-Lunde-Nason:2011}, which provides a joint comparison across many competing models and avoids the limitations of relying on numerous pairwise tests. For a given subset size $k$, exhaustively comparing the 11 candidate models would require $\binom{11}{2}=55$ pairwise tests, raising multiple-comparison concerns and inflating the scope for spurious rejections; see \citet{Harvey-Liu-Saretto:2020} for a discussion of multiple testing in finance.

The MCS procedure circumvents these limitations by identifying a subset of models that are statistically indistinguishable in terms of predictive performance. 
For each model, we compute the sequence of out-of-sample losses under a pre-specified loss function and construct test statistics to compare predictive performance. 
The weakest model is iteratively removed, and elimination continues until the remaining set contains only models that cannot be statistically differentiated in terms of predictive ability.\footnote{We base the MCS on the $T_{\max,\mathcal{M}}$ statistic of \citet[][Section~3.1.2]{Hansen-Lunde-Nason:2011} and implement it using the \texttt{MCS} package in R \citep{Bernardi-Catania:2018}.}

Tables \ref{MCS squared} and \ref{MCS absolute} report the results of the 80\% MCS ($\text{MCS}_{80\%}$) procedure, based on squared and absolute loss functions, respectively. The 80\% confidence level applies a more aggressive elimination criterion than higher levels, so $\text{MCS}_{80\%} \subseteq \text{MCS}_{90\%} \subseteq \text{MCS}_{95\%}$, and a model belongs to $\text{MCS}_{80\%}$ if its MCS $p$-value is at least $0.20$. For each subset size $k$, the tables report the average loss (MSE or 
MAE, scaled by 100) together with the corresponding MCS $p$-values. Under squared loss (Table~\ref{MCS squared}), the MS model is excluded from the MCS 
for every $k$ (MCS $p$-values equal to zero), and the linear model is excluded for 
several subset sizes, most notably for $k=3$--$6$ and $k=8$.
The GARCH-M specification remains competitive only at small subset sizes, but is excluded at $k=4$--$6$ and receives comparatively low MCS $p$-values at larger $k$.
The CSR approach is retained for most values of $k$, but is excluded at $k=8$, consistent with its deterioration at high dimensionality.
Overall, under squared loss, differences in average MSE are relatively small, which limits the extent to which the MCS can separate closely performing models.

Under absolute loss (Table~\ref{MCS absolute}), the MCS delivers sharper differentiation among the benchmark specifications. The linear and MS models are excluded for several intermediate and large subset sizes (notably for $k=3$--$5$ and $k=8$). The CSR approach is retained for $k=1$--$7$ but is excluded at $k=8$, and its MCS $p$-values are sometimes close to the 0.20 cutoff (e.g., $k=1$ and $k=5$). The GARCH-M benchmark is competitive at low dimensions (included with MCS $p$-values equal to one at $k=1$ and $k=2$, and still retained at $k=3$), suggesting that incorporating volatility feedback may yield modest gains when the predictor dimension is limited; however, it is excluded at $k=4$, $k=5$, and $k=8$, indicating sensitivity to $k$.

In contrast, the decomposition-based models (the copula-based specifications and the CSM specifications) are retained in the MCS across all values of $k$ under both loss functions, with MCS $p$-values typically close to one.
This shows that these models are consistently among the statistically best-performing specifications out of sample.
Empirically, the MCS does not meaningfully distinguish between the CSM models and the copula-based decomposition models, consistent with the view in AG that the dependence between the sign and magnitude of monthly returns is weak, so that modeling this dependence flexibly yields limited incremental gains in overall loss.

Finally, while the MCS results suggest broadly similar performance within the decomposition class under standard loss functions, the CSM framework is designed to target directional predictability more directly by modeling the return sign conditional on predictors and contemporaneous magnitude. The next section therefore turns to the economic value of these directional forecasts.

\subsection{Economic value}

What matters to investors is not necessarily predictive accuracy as measured by statistical criteria \citep{Jorda-Taylor:2011}, but rather the ability to correctly anticipate whether returns will be positive or negative, especially when significant profits or losses are at stake. 
\citet{Merton:1980} likewise argues that fund managers place greater emphasis on the direction of returns than on their magnitude. 
In this section, we therefore assess the economic significance of the models through their ability to predict return direction. Specifically, we examine their value for market timing by comparing the returns from a passive buy-and-hold strategy, under which the risky asset is held throughout the sample period, with those from market timing strategies constructed from forecasts generated by the various model specifications.

Following \citet{Breen-Glosten-Jagannathan:1989}, \citet{Pesaran-Timmermann:1995}, \citet{Guo:2006}, and AG, among others, we use a simple trading rule known as a switching strategy, which only requires the direction of the forecasts. This rule involves investing the current wealth in the stock market by buying shares of the index if the predicted excess return is positive, or investing the current wealth in the risk-free asset if the predicted excess return is negative. Once the investor makes a decision following this trading strategy, the funds remain invested until the end of the month, before a new decision is made at the beginning of the next month. The trading begins with an initial wealth of \$1, and the portfolio value is recalculated and reinvested every month.

To make the economic evaluation more realistic, we introduce proportional transaction costs of $c=10$ basis points of the portfolio's value when the investor makes a switch.\footnote{\citet{French:2008} documents a sharp decline in trading costs over this period, providing the basis for our choice of 10 basis points as a conservative estimate.} Let $r_{p,t+1}$ denote the out-of-sample portfolio return realized at time $t+1$, and let $R_{p,t+1}$ denote the corresponding random portfolio return. The investor's wealth then evolves from month $t$ to month $t+1$ according to $W_{t+1}=W_t(1+r_{p, t+1})(1-c),$ when a switch is made, and 
$W_{t+1}=W_t(1+r_{p, t+1}),$ otherwise. The portfolios formed using the switching rule based on forecasts from the different candidate models are compared to the benchmark buy-and-hold strategy. We note that the historical average model resulted in a strategy that is equivalent to the buy-and-hold strategy, as all the historical-average forecasts turned out positive.

The active strategies compared with the buy-and-hold benchmark include those based on the previously discussed forecasting models. In addition, we consider momentum-based switching strategies, where the forecasted sign of next month’s return is taken to be the sign of the cumulative return over a recent historical period. These backward-looking rules offer a useful contrast to the forward-looking model-based strategies. We report results for momentum strategies based on cumulative returns over the past 3, 6, and 12 months.

For each portfolio strategy, the following out-of-sample performance metrics are computed: (i) TW, the terminal wealth in dollars at the end of the trading period; (ii) AV, the average return (annualized by multiplying by 12) in percent; (iii) SD, the standard deviation of returns (annualized by multiplying by the square root of 12) in percent; (iv) SR, the Sharpe ratio; and (v) MDD, the maximum drawdown in percent over the trading period. To assess statistical significance, we test whether the Sharpe ratio of each strategy differs from that of the buy-and-hold benchmark using the two-sided $p$-value from the prewhitened $\mathrm{HAC_{PW}}$ method of \citet{Ledoit-Wolf:2008}, which tests the null hypothesis of equal Sharpe ratios.

We evaluate out-of-sample portfolio performance across predictor subset sizes $k=1$ to $8$. 
Table~\ref{tab:performance_matrix1} focuses on four informative cases ($k=1, 2, 3, 7$) that capture early-stage gains and behavior at higher dimensions, with the remaining results ($k=4, 5, 6, 8$) reported in the Supplementary material (Table~B1).
Over the out-of-sample trading period (May 1981--December 2021), the buy-and-hold benchmark delivers solid performance (annualized return 12.65\%, standard deviation 15.00\%, Sharpe ratio 0.17, terminal wealth \$104.63), despite encompassing major crisis episodes.\footnote{We conduct a dedicated sub-sample analysis of the Global Financial Crisis and the COVID-19 crisis to assess model performance during turbulent periods. Results are reported in Section C of the Supplementary material.}

\begin{figure}[hp]

\caption{Wealth growth for the  CSM (Baseline, $k = 3$), buy-and-hold, and momentum  strategies}

  \label{fig:Wealth dynamics k3Mom}
  \centering
  \includegraphics[width=\textwidth, height=\textheight, keepaspectratio]{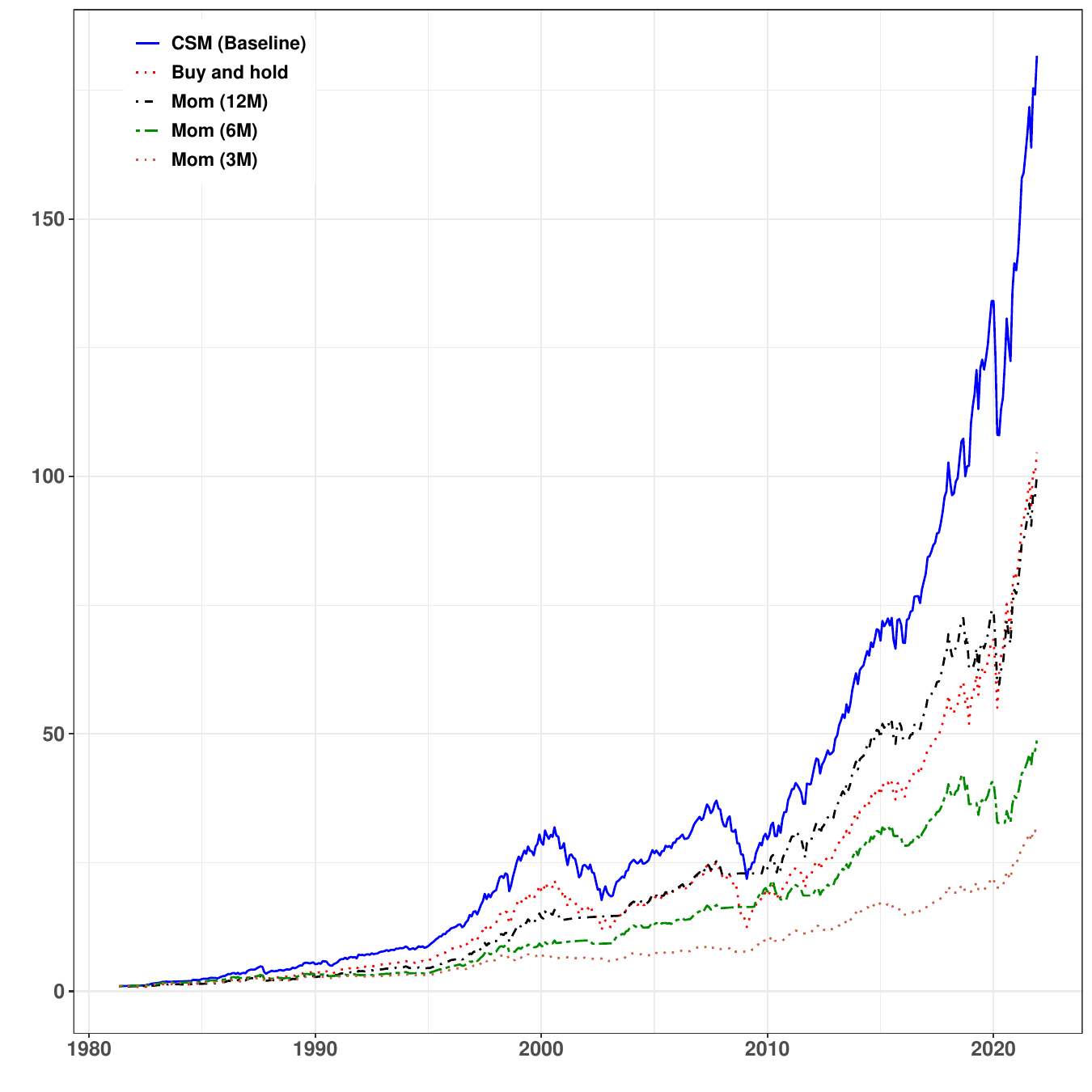}
\end{figure}

Momentum strategies show mixed performance: while 12-month momentum proves competitive (TW \$100.21, SR 0.20), short-horizon variants (3-month and 6-month) substantially underperform, with terminal wealth below \$50. 
Figure~\ref{fig:Wealth dynamics k3Mom} plots wealth growth for CSM (Baseline) at $k=3$ alongside buy-and-hold and momentum-based switching strategies, showing that CSM accumulates wealth more rapidly over time and ends above both buy-and-hold and all momentum horizons, with the gap widening in the later part of the sample.

The linear predictive regression exhibits mixed performance at very small subset sizes but deteriorates sharply as predictor dimension increases (terminal wealth \$71.58 at $k=4$ and \$62.31 at $k=5$; Supplementary material, Table~B1), highlighting the instability of traditional linear models with many predictors.
The CSR approach partially stabilizes performance relative to the individual linear model at moderate $k$ (reaching \$126.80 at $k=5$; Supplementary material, Table~B1), but it does not dominate uniformly and coincides with the linear model at $k=8$ by construction.
The nonlinear benchmarks are also sensitive to predictor dimension: GARCH-M performs best at low to moderate $k$ (TW \$151.01 at $k=2$ and \$147.63 at $k=3$), whereas the MS strategy collapses at $k=3$ (TW \$25.65).

\begin{figure}[hp]
  \caption{Wealth growth for the CSM (Baseline), copula-based, GARCH-M, CSR, and linear strategies, each with $k = 3$}
  \label{fig:Wealth dynamics k3}
  \centering
  \includegraphics[width=\textwidth, height=\textheight, keepaspectratio]{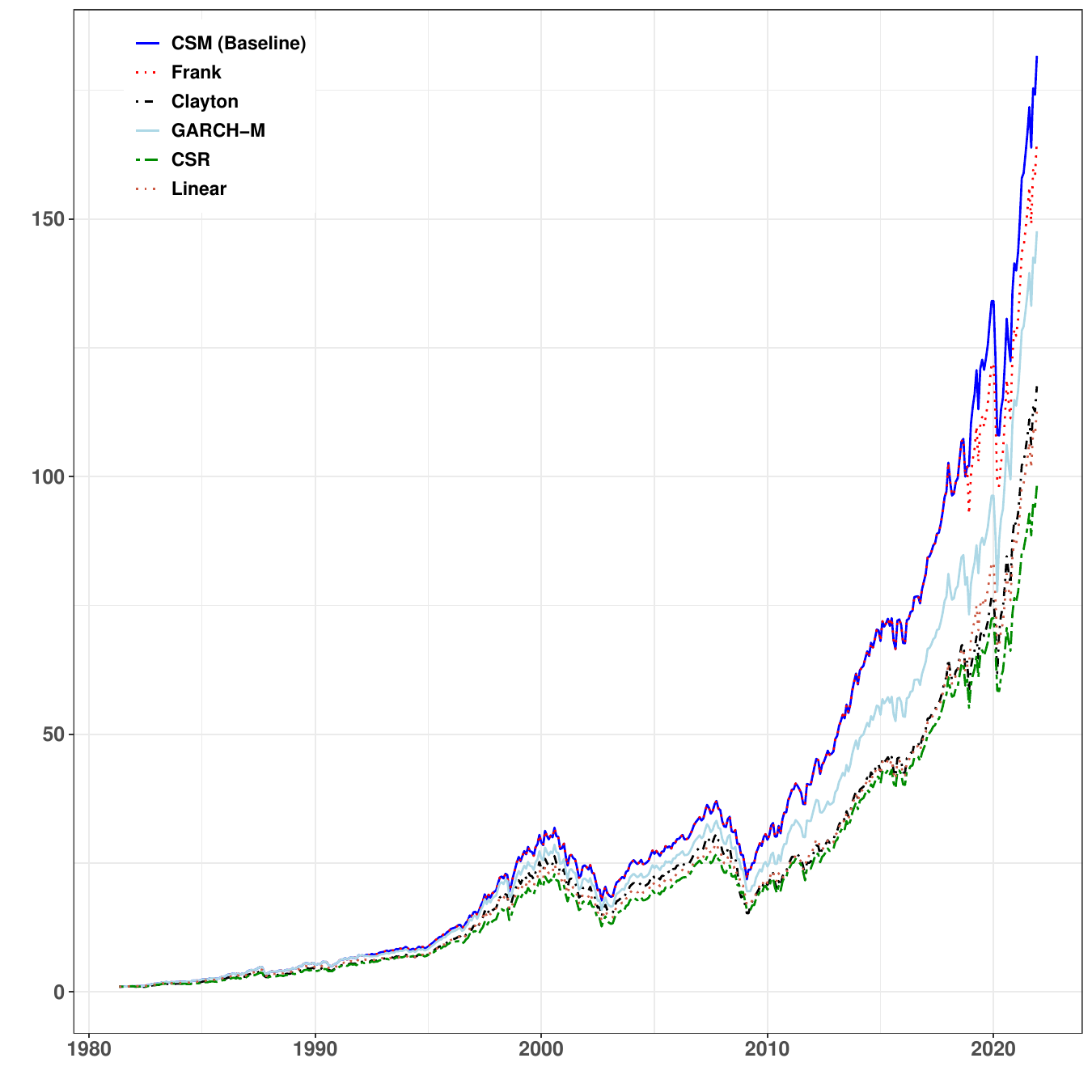}
\end{figure}

Overall, the decomposition-based strategies deliver the strongest economic performance, particularly at intermediate and larger subset sizes.
At $k=3$, the copula-based strategies (Gaussian/Frank/FGM) reach TW \$165.23, well above CSR (\$98.22) and the linear model (\$112.79), while the proposed CSM strategy (Baseline and Poly coincide at this dimension) attains the highest terminal wealth (TW \$181.68) with a statistically significant Sharpe-ratio improvement over buy-and-hold (SR 0.21, 5\% level).
Figure~\ref{fig:Wealth dynamics k3} shows that CSM and the Frank copula track each other closely for much of the sample, with CSM pulling ahead in the later years; by contrast, CSR and the linear strategy lag behind after major drawdowns, resulting in substantially lower terminal wealth.
This temporal perspective indicates that the superior performance of decomposition-based strategies reflects sustained accumulation rather than isolated episodes of outperformance.

At higher dimensions, decomposition-based strategies continue to perform strongly. At $k=7$, copula-based strategies deliver terminal wealth of \$176--\$178 with Sharpe ratios of 0.22 (statistically significant at the 10\% level), slightly outperforming CSM (Baseline) at this dimension (TW \$168.05, SR 0.22). Within the cases reported in Table~\ref{tab:performance_matrix1}, statistically significant Sharpe-ratio improvements at the 10\% level are also achieved by multiple strategies at $k=2$ (GARCH-M, Frank copula, CSM Baseline) and at $k=3$ (Gaussian, Frank, and FGM copulas). Complementary evidence for the remaining subset sizes, including significance at $k=4$ for CSM (Poly), is reported in the Supplementary material (Table~B1).

Figure~\ref{fig:CSM vs. Linear} illustrates how terminal wealth varies with the number of predictors across selected strategies, revealing starkly different dimensional-stability properties. 
CSM (Baseline) is clearly non-monotone in $k$, peaking at $k=3$ and remaining highly competitive at $k=7$--$8$ while being weaker at $k=4$--$6$. 
In contrast, the linear model exhibits severe instability (e.g., dropping from \$112.79 at $k=3$ to \$62.31 at $k=5$), while CSR offers only partial stabilization relative to the linear benchmark. 
The Clayton copula also displays pronounced non-monotonicity, with strong performance at $k=7$--$8$.

We further gauge the economic value of the models by calculating their certainty equivalent returns (CER), as in \citet{Campbell-Thompson:2008}, 
\citet{Cenesizoglu-Timmermann:2012}, \citet{Ciner:2022}, \citet{Christoffersen-Langlois:2013}, and \citet{Borup-Eriksen-Kjaer-Thyrsgaard:2024}, among many others. The CER is the guaranteed return that the investor would accept instead of a risky investment, such that the utility derived from the CER equals the expected utility of the risky position. Typically, the average utility is used as a consistent estimate of the expected utility.

We compute the CER associated with each strategy by assuming two different types of investor preferences. 
First, we consider mean-variance preferences summarized by the mean--variance criterion
\begin{equation*}
    \mathbb{E}(R_{p,t+1})-\frac{1}{2}\gamma \mathrm{Var}(R_{p,t+1}),
\end{equation*}
where $\gamma$ is the risk aversion parameter. The associated CER is computed as 
\begin{equation*}
   \mathrm{CER} = \hat{\mu}_{p} - \frac{1}{2} \gamma \hat{\sigma}^{2}_{p}, 
\end{equation*}
where $\hat{\mu}_{p}$ and $\hat{\sigma}^{2}_{p}$ 
are the sample mean and variance, respectively, of the realized portfolio returns $\{r_{p,t+1}\}_{t=L+1}^{T}$ over the out-of-sample evaluation period.

Second, we consider constant relative risk aversion (CRRA) preferences,
\begin{equation*}
    U(R_{p,t+1})=\frac{(1+R_{p,t+1})^{1-\gamma}}{1-\gamma}.
\end{equation*}
The advantage of CRRA utility is that it incorporates preferences toward higher-order
moments \citep{Brandt-SantaClara-Valkanov:2009}.
In this case, the average utility over the out-of-sample period is
\[
\bar U = \frac{1}{T-L}\sum_{t=L}^{T-1}\frac{(1+r_{p,t+1})^{1-\gamma}}{1-\gamma}.
\]
By solving for CER in
$(1+\text{CER})^{1-\gamma} = (T-L)^{-1}\sum_{t=L}^{T-1}(1+r_{p,t+1})^{1-\gamma},$
we obtain the expression
\[
\text{CER}
=
\Biggl[
\frac{1}{T-L}\sum_{t=L}^{T-1}(1+r_{p,t+1})^{1-\gamma}
\Biggr]^{\frac{1}{1-\gamma}} - 1.
\]
The CER gain is defined as the difference between the CER for the predictive methods and the CER for the historical average benchmark. We annualize the CER gain so that it can be interpreted as the annual portfolio management fee that an investor would be willing to pay to access the information provided by the predictive methods compared to simply using the historical average.

The results are summarized in Table~\ref{CER_k1237}, which reports the CER gains relative to the historical average benchmark for an investor with moderate risk aversion ($\gamma = 5$; results for $\gamma \in \{3, 10\}$ were qualitatively similar to those discussed here).
For brevity, we present results for $k = 1, 2, 3, 7$; the remaining cases are provided in Section~B of the Supplementary material. 
Panel~A reports results for a mean-variance investor, while Panel~B corresponds to CRRA preferences. 
In both cases, the CSM (Baseline) model delivers strong CER gains at $k=2$--$3$ and remains competitive at larger subset sizes (e.g., $k=7$, and also $k=8$ in Supplementary Table~B2), though performance is non-monotone in $k$ with weaker gains at intermediate dimensions (e.g., $k=4$ and $k=6$; Supplementary Table~B2). 
For example, at $k = 2$, it yields CER gains of 1.122 (Panel~A) and 1.108 (Panel~B), outperforming the CSR approach (0.596 and 0.303, respectively). 
A similar pattern holds at $k = 3$, where the CSM (Baseline) achieves 1.878 and 1.939, which are higher than most competing models. 
At $k = 7$, the CSM (Baseline) continues to perform strongly with gains of 2.302 and 2.612 under mean-variance and CRRA preferences, respectively.

The copula-based models deliver competitive results and frequently outperform the linear and CSR benchmarks, though their relative performance varies with $k$ and the specific copula specification. 
For instance, under mean-variance preferences and $k = 2$, the Frank copula achieves the highest CER gain among the copulas (1.122), while at $k = 7$, the Gaussian copula attains 2.442, slightly exceeding the CSM (Baseline) model (2.302). 
Under CRRA preferences, copula-based strategies remain competitive at larger $k$ values, with the Gaussian copula reaching 2.748 at $k = 7$ (versus 2.612 for CSM Baseline). 
Momentum strategies based on 12-month signals perform consistently well, while 3-month momentum substantially underperforms (CER gains approximately $-0.71\%$ and $-0.56\%$ under mean-variance and CRRA preferences, respectively) and 6-month momentum delivers only modest positive gains. 
The linear model exhibits non-monotone performance in $k$: it delivers positive CER gains at small subset sizes and again at larger sizes (e.g., $k=7$--$8$), but turns negative at intermediate dimensions (e.g., $-0.408$ at $k=5$ under mean-variance preferences; Supplementary Table~B2).

The nonlinear benchmark models exhibit more mixed performance. 
GARCH-M performs well at smaller subset sizes, for example yielding CER gains of 1.075 (Panel~A) and 1.062 (Panel~B) at $k = 2$, and remains positive at $k = 7$ with 0.412 and 0.675, respectively. 
The MS model performs strongly only at $k = 1$, with CER gains above 1.6 in both panels, but its performance declines rapidly thereafter, becoming negative by $k = 3$ and remaining negative at $k = 7$. 
Overall, while some copula specifications deliver the highest CER gains at particular $k$, the CSM framework delivers substantial economic value and remains consistently competitive across subset sizes.

Nevertheless, a natural question arises as to whether an investor stands to benefit from transitioning away from the CSM (Baseline) model in favor of forecasts generated by alternative copula-based decomposition strategies. Table~\ref{CER_CSM_k1237} addresses this question directly by reporting CER gains computed relative to the CSM (Baseline) benchmark, using the same setup and investor preferences as in Table~\ref{CER_k1237}. For brevity, we report results for $k = 1,2,3,7$; the remaining cases are provided in Section~B of the Supplementary material.

Across both preferences, copula-based strategies generally offer no improvement in utility over the CSM (Baseline) model when the number of predictors is relatively low ($k \le 3$). For example, under mean-variance preferences, adopting a Gaussian copula instead of the CSM (Baseline) implies annual CER losses of 8.7, 50.4, and 28.8 basis points for $k=1,2,3$, respectively. Under CRRA preferences, the corresponding losses are 6.4, 45.0, and 29.6 basis points. These negative CER differentials indicate that, for modest predictor specifications, the CSM (Baseline) model remains difficult to beat (at best, the Frank copula essentially ties the baseline at $k=2$).

At $k=7$, some copula specifications modestly improve upon the baseline: under mean-variance preferences, the Gaussian and Clayton copulas deliver gains of 14.0 and 11.0 basis points relative to the baseline, respectively, while under CRRA preferences they deliver 13.6 and 10.7 basis points. At $k=8$ (Supplementary material, Table~B2), these differentials remain small overall: Gaussian/Frank/FGM are essentially neutral (or slightly negative), while the Clayton copula continues to provide a modest positive gain of about 9 basis points. At intermediate dimensions, however, the scope for improvements over CSM (Baseline) can be economically meaningful: for example, at $k=4$ and $k=6$, several copula specifications and, in particular, the CSM (Poly) variant deliver sizable positive CER differentials relative to the baseline (Supplementary material, Table~B2). Overall, these results underscore that CSM (Baseline) is difficult to beat for low-dimensional specifications ($k \le 3$), while at larger dimensions ($k=7$--$8$) copula-based alternatives offer at most modest incremental utility, and the Poly variant provides an effective way to strengthen robustness at intermediate subset sizes.

Finally, we stress that the levels of certainty equivalents reported here are not meant to be directly compared to large-scale applications with more extensive predictor sets, such as those in \citet{Bianchi-Rubesam-Tamoni:2025} and \citet{Goyal-Welch-Zafirov:2024}. Our focus is on relative performance across methods within our moderate predictor set and under the same portfolio construction and evaluation framework.

\section{Conclusion}

In this paper, we introduced the CSM decomposition approach, a new method for return prediction that builds on the AG framework by decomposing returns into sign and magnitude components. 
The expected return is modeled using the joint distribution of these components, combining a multiplicative error model for the magnitude with a binary choice model for the sign conditional on the contemporaneous magnitude. 
By avoiding the need to explicitly specify or estimate a copula function, the CSM framework simplifies implementation while preserving the ability to capture nonlinear return dynamics. 
This structure exploits the well-documented relationship between return magnitude (i.e., volatility) and direction, as emphasized by \citet{Christoffersen-Diebold:2006}.

In our empirical application, return decomposition-based strategies often outperform traditional benchmarks such as the historical average and the linear predictive regression, as well as the CSR forecast-combination method. Relative to economically motivated nonlinear benchmarks that incorporate time-varying risk and regime dependence (GARCH-in-mean and Markov-switching specifications), the CSM approach remains highly competitive. Compared with copula-based decomposition methods, CSM provides a direct way to exploit sign-magnitude dependence, and the CSM (Baseline) specification delivers consistently strong CER gains across investor preferences and predictor dimensions. Copula-based methods are also competitive, with performance that varies across copula choices and subset sizes.

Economically, these findings translate into high terminal wealth outcomes and robust utility gains in the market-timing exercise, with CSM typically among the best-performing specifications even as performance differences narrow at larger predictor dimensions. 
Overall, the results highlight the value of modeling direction and magnitude jointly and position CSM as a practical and effective tool for return forecasting and portfolio allocation.

\newpage
\bibliographystyle{chicago}
\bibliography{CSM-References.bib}

\newpage
\clearpage

\begin{table}
\begin{center}
\caption{Sample correlations}
\label{correlations}

\begin{tabular*}{\textwidth}{@{\extracolsep{\fill}}l*{10}{S[table-format=1.3]}}
\toprule

\multicolumn{11}{l}{Panel A: Between $r_t$ and $\boldsymbol x_{t-1}$ } \\

& {} 
& \multicolumn{1}{c}{$r_t$}
& \multicolumn{1}{c}{$dp_{t-1}$}
& \multicolumn{1}{c}{$dfy_{t-1}$}
& \multicolumn{1}{c}{$tms_{t-1}$}
& \multicolumn{1}{c}{$tbl_{t-1}$}
& \multicolumn{1}{c}{$ltr_{t-1}$}
& \multicolumn{1}{c}{$dfr_{t-1}$}
& \multicolumn{1}{c}{$ntis_{t-1}$}
& \multicolumn{1}{c}{$infl_{t-1}$} \\

$r_t$ 
& {} & 1.000 & 0.059 & 0.022 & 0.058 & -0.097 & 0.045 & 0.038 & -0.018 & -0.069 \\

\midrule

\multicolumn{11}{l}{Panel B: Between $s_t$, $m_t$, and $\boldsymbol x_{t-1}$ } \\

& \multicolumn{1}{c}{$s_t$}
& \multicolumn{1}{c}{$m_t$}
& \multicolumn{1}{c}{$dp_{t-1}$}
& \multicolumn{1}{c}{$dfy_{t-1}$}
& \multicolumn{1}{c}{$tms_{t-1}$}
& \multicolumn{1}{c}{$tbl_{t-1}$}
& \multicolumn{1}{c}{$ltr_{t-1}$}
& \multicolumn{1}{c}{$dfr_{t-1}$}
& \multicolumn{1}{c}{$ntis_{t-1}$}
& \multicolumn{1}{c}{$infl_{t-1}$} \\

$s_t$ 
& 1.000 & -0.011 & 0.006 & -0.025 & 0.059 & -0.143 & 0.026 & 0.049 & -0.046 & -0.118 \\

$m_t$ 
& {}    &  1.000 & 0.029 & 0.166  & -0.015 & 0.044 & 0.093 & -0.146 & -0.062 & 0.024 \\

\midrule

\multicolumn{11}{l}{Panel C: Among $\boldsymbol x_{t-1}$ } \\

& {} & {}
& \multicolumn{1}{c}{$dp$}
& \multicolumn{1}{c}{$dfy$}
& \multicolumn{1}{c}{$tms$}
& \multicolumn{1}{c}{$tbl$}
& \multicolumn{1}{c}{$ltr$}
& \multicolumn{1}{c}{$dfr$}
& \multicolumn{1}{c}{$ntis$}
& \multicolumn{1}{c}{$infl$} \\

$dp$   
& {} & {} & 1.000 & 0.125 & 0.322 & 0.344 & -0.006 & 0.389 & 0.169 & 0.168 \\

$dfy$  
& {} & {} & {}    & 1.000 & 0.260 & 0.139 & 0.067  & -0.361 & 0.069 & 0.063 \\

$tms$  
& {} & {} & {}    & {}    & 1.000 & -0.374 & 0.076  & -0.261 & -0.197 & -0.198 \\

$tbl$  
& {} & {} & {}    & {}    & {}    & 1.000 & -0.045 & -0.085 & 0.407 & 0.499 \\

$ltr$  
& {} & {} & {}    & {}    & {}    & {}    & 1.000 & -0.457 & 0.078 & 0.102 \\

$dfr$  
& {} & {} & {}    & {}    & {}    & {}    & {}    & 1.000 & 0.003 & -0.004 \\

$ntis$ 
& {} & {} & {}    & {}    & {}    & {}    & {}    & {}    & 1.000 & 0.102 \\

$infl$ 
& {} & {} & {}    & {}    & {}    & {}    & {}    & {}    & {}    & 1.000 \\

\bottomrule
\end{tabular*}

\end{center}

{\small \textit{Notes:} This table reports sample correlations between the variables used in the analysis over the full sample from January 1948 to December 2021. 
Panel A shows correlations between the current excess return ($r_t$) and the lagged predictors: $dp_{t-1}$ (dividend-price ratio), $dfy_{t-1}$ (default yield spread), $tms_{t-1}$ (term spread), $tbl_{t-1}$ (short-term interest rate), $ltr_{t-1}$ (long-term return), $dfr_{t-1}$ (default return spread), $ntis_{t-1}$ (net equity expansion), and $infl_{t-1}$ (inflation). Panel B presents correlations between the sign ($s_t$) and magnitude ($m_t$) components of the excess return and the lagged predictors. Panel C reports the correlations among the lagged predictors themselves.}

\end{table}

\newpage
\clearpage

\begin{table}[p]
\begin{center}
\begin{minipage}{0.96\textwidth} 
\caption{Best subset selection based on AUC criterion}
\label{Bestsubset_AUC}
\centering
\begin{tabular}{@{\extracolsep{1.5em}}l l *{8}{c}}
\toprule
      & const & dp & dfy & tms & tbl & ltr & dfr & ntis & infl \\
\midrule
\multicolumn{10}{l}{Panel A: Linear and GARCH-M models} \\
$k=1$ & *     &    &     &     & *   &     &     &      &      \\
$k=2$ & *     &    &     &     & *   &     &     & *    &      \\
$k=3$ & *     &    &     &     & *   &     & *   & *    &      \\
$k=4$ & *     &    & *   & *   & *   &     &     & *    &      \\
$k=5$ & *     &    & *   & *   & *   &     & *   & *    &      \\
$k=6$ & *     &    & *   & *   & *   &     & *   & *    & *    \\
$k=7$ & *     & *  & *   & *   & *   &     & *   & *    & *    \\
$k=8$ & *     & *  & *   & *   & *   & *   & *   & *    & *    \\
\midrule
\multicolumn{10}{l}{Panel B: MS model} \\
$k=1$ & *     &    &     &     & *   &     &     &      &      \\
$k=2$ & *     &    &     &     & *   &     &     &      & *    \\
$k=3$ & *     &    & *   & *   & *   &     &     &      &      \\
$k=4$ & *     & *  & *   & *   &     &     & *   &      &      \\
$k=5$ & *     & *  & *   & *   &     &     & *   & *    &      \\
$k=6$ & *     &    & *   & *   & *   &     & *   & *    & *    \\
$k=7$ & *     &    & *   & *   & *   & *   & *   & *    & *    \\
$k=8$ & *     & *  & *   & *   & *   & *   & *   & *    & *    \\
\midrule
\multicolumn{10}{l}{Panel C: Decomposition-based models (CSM and copula-based)} \\
$k=1$ & *     &    &     &     & *   &     &     &      &      \\
$k=2$ & *     &    &     &     & *   &     & *   &      &      \\
$k=3$ & *     &    &     &     & *   &     & *   & *    &      \\
$k=4$ & *     &    & *   & *   & *   &     &     & *    &      \\
$k=5$ & *     & *  & *   & *   & *   &     &     & *    &      \\
$k=6$ & *     & *  & *   & *   & *   &     &     & *    & *    \\
$k=7$ & *     & *  & *   & *   & *   &     & *   & *    & *    \\
$k=8$ & *     & *  & *   & *   & *   & *   & *   & *    & *    \\
\bottomrule
\end{tabular}
\\[0.5em]
\small
\parbox{1.0\textwidth}{\textit{Notes:} A star ($\ast$) indicates that the corresponding variable is included in the best subset of size $k$. The subsets are selected based on the AUC criterion using one-step-ahead in-sample forecasts. An empty cell indicates exclusion of the variable from the selected subset. Panel A shows the selected subsets for the linear and GARCH-M models, which select identical variables for all values of $k$. Panel B shows the selected subsets for the MS model. Panel C shows the selected subsets for the decomposition-based models (CSM and copula-based); results are identical across all copula specifications. At $k = 8$, all predictors are included by construction in all models, so no subset selection is required.}
\end{minipage}
\end{center}
\end{table}

\clearpage

\newpage

\begin{landscape}

\begin{table}
\vspace*{-1.3 cm}
\begin{center}
\caption{Out-of-sample $R^2$ values by subset size ($k$) and loss function}
\label{ROOS_all}
\begin{minipage}{1.225\textwidth}

\begin{tabular}{l *{8}{S[table-format=-2.2]}}
\toprule
& \multicolumn{2}{c}{$k=1$} & \multicolumn{2}{c}{$k=2$} & \multicolumn{2}{c}{$k=3$} & \multicolumn{2}{c}{$k=4$} \\
\cmidrule(lr){2-3} \cmidrule(lr){4-5} \cmidrule(lr){6-7} \cmidrule(lr){8-9}
& {Squared} & {Absolute} & {Squared} & {Absolute} & {Squared} & {Absolute} & {Squared} & {Absolute} \\
\midrule
Linear model            & -0.24 &  0.91 & -0.56 &  0.42 & -1.52 & -0.26 & -2.28 & -1.00 \\
CSR approach            &  0.37 &  0.24 &  0.51 &  0.18 &  0.40 &  0.02 &  0.11 & -0.21 \\
GARCH-M model           &  0.40 &  1.35\st{**} & -0.09 &  0.72 & -0.99 &  0.14 & -2.09 & -0.84 \\
MS model                & -5.15\st{*}  & -0.93 & -4.49 & -0.24 & -7.60\st{***} & -2.68 & -9.61\st{***} & -4.26\st{***} \\
Copula-based approach   & \multicolumn{8}{c}{} \\
\quad Gaussian          &  0.20 &  1.39\st{**} &  0.47 &  1.27\st{*} &  0.97 &  1.34 &  0.75 &  0.95 \\
\quad Frank             &  0.18 &  1.47\st{**} &  0.47 &  1.55\st{*} &  0.94 &  1.37 &  0.71 &  0.99 \\
\quad Clayton           &  0.28 &  1.54\st{**} &  0.75 &  1.39\st{*} &  1.03 &  1.18 &  0.90 &  1.20 \\
\quad FGM               &  0.17 &  1.46\st{**} &  0.59 &  1.36\st{*} &  0.94 &  1.37 &  0.70 &  0.99 \\
CSM approach            & \multicolumn{8}{c}{} \\
\quad Baseline          &  0.19 &  1.25\st{**} &  0.77 &  1.41\st{*} &  0.83 &  1.18 &  0.57 &  0.74 \\
\quad Poly              &  0.04 &  1.32\st{**} &  0.23 &  1.16 &  0.67 &  1.19 &  0.28 &  0.73 \\
\midrule
\end{tabular}

\vspace{1.5ex}

\begin{tabular}{l *{8}{S[table-format=-2.2]}}
& \multicolumn{2}{c}{$k=5$} & \multicolumn{2}{c}{$k=6$} & \multicolumn{2}{c}{$k=7$} & \multicolumn{2}{c}{$k=8$} \\
\cmidrule(lr){2-3} \cmidrule(lr){4-5} \cmidrule(lr){6-7} \cmidrule(lr){8-9}
& {Squared} & {Absolute} & {Squared} & {Absolute} & {Squared} & {Absolute} & {Squared} & {Absolute} \\
\midrule
Linear model            & -3.52 & -1.57 & -3.76 & -1.03 & -5.80\st{*} & -2.78 & -5.15 & -2.87 \\
CSR approach            & -0.40 & -0.56 & -1.23 & -1.07 & -2.68 & -1.83 & -5.15 & -2.87 \\
GARCH-M model           & -2.92 & -1.22 & -3.29 & -0.96 & -4.80 & -2.33 & -3.64 & -2.69 \\
MS model                & -10.69\st{***} & -5.08\st{***} & -7.30\st{*} & -2.23 & -7.77\st{*} & -2.87 & -16.93\st{***} & -7.21\st{***} \\
Copula-based approach   & \multicolumn{8}{c}{} \\
\quad Gaussian          &  0.93 &  0.92 & -0.34 &  0.37 & -0.11 &  0.52 & -0.39 &  0.35 \\
\quad Frank             &  1.35 &  1.51 & -0.27 &  0.46 & -0.04 &  0.56 & -0.40 &  0.34 \\
\quad Clayton           &  1.04 &  1.05 & -0.09 &  0.57 &  0.08 &  0.65 & -0.28 &  0.42 \\
\quad FGM               &  0.90 &  0.95 & -0.20 &  0.47 & -0.04 &  0.56 & -0.49 &  0.32 \\
CSM approach            & \multicolumn{8}{c}{} \\
\quad Baseline          &  0.88 &  0.78 & -0.38 &  0.26 & -0.42 &  0.34 & -0.68 &  0.14 \\
\quad Poly              &  1.04 &  0.94 &  0.72 &  0.80 & -0.96 &  0.47 & -0.99 &  0.35 \\
\bottomrule
\end{tabular}
\\[0.5em]
{\small \textit{Notes:} This table reports $R^2_{\text{OOS}}$ statistics (in \%) for squared and absolute loss functions, relative to the historical-average benchmark.
Positive values indicate improvement over the historical average, while negative values indicate deterioration.
Stars indicate rejection of equal predictive accuracy (DM test) against the historical average benchmark: $^{\ast}$, $^{\ast\ast}$, and $^{\ast\ast\ast}$ denote significance at the 10\%, 5\%, and 1\% levels, respectively.}

\end{minipage}
\end{center}
\end{table}

\end{landscape}

\begin{landscape}

\begin{table}
\vspace*{-1.25 cm}
\begin{center}

\caption{Out-of-sample model performance comparison using MCS procedure with squared loss}\label{MCS squared}

\begin{minipage}{1.27\textwidth}
\begin{tabular}{lcccccccc}
\toprule
& \multicolumn{2}{c}{$k=1$} & \multicolumn{2}{c}{$k=2$} & \multicolumn{2}{c}{$k=3$} & \multicolumn{2}{c}{$k=4$} \\
\cmidrule(lr){2-3} \cmidrule(lr){4-5} \cmidrule(lr){6-7} \cmidrule(lr){8-9}
                       & MSE & MCS $p$-value & MSE & MCS $p$-value & MSE & MCS $p$-value & MSE & MCS $p$-value \\
\midrule
Historical average     & 0.189 & 0.999 & 0.189 & 0.970 & 0.189 & 0.953 & 0.189 & 0.946 \\
Linear model           & 0.189 & 0.292 & 0.189 & 0.593 & 0.191 & 0.000 & 0.193 & 0.000 \\
CSR approach           & 0.188 & 1.000 & 0.188 & 1.000 & 0.188 & 1.000 & 0.188 & 0.959 \\
GARCH-M model          & 0.188 & 1.000 & 0.188 & 0.938 & 0.190 & 0.222 & 0.193 & 0.000 \\
MS model               & 0.198 & 0.000 & 0.197 & 0.000 & 0.203 & 0.000 & 0.206 & 0.000 \\
Copula-based approach  &       &       &       &       &       &       &       &       \\
\quad Gaussian         & 0.188 & 1.000 & 0.187 & 1.000 & 0.187 & 1.000 & 0.187 & 1.000 \\
\quad Frank            & 0.188 & 1.000 & 0.187 & 1.000 & 0.187 & 1.000 & 0.187 & 1.000 \\
\quad Clayton          & 0.188 & 1.000 & 0.187 & 1.000 & 0.186 & 1.000 & 0.187 & 1.000 \\
\quad FGM              & 0.188 & 1.000 & 0.188 & 1.000 & 0.187 & 1.000 & 0.187 & 1.000 \\
CSM approach           &       &       &       &       &       &       &       &       \\
\quad Baseline         & 0.188 & 1.000 & 0.187 & 1.000 & 0.187 & 1.000 & 0.188 & 1.000 \\
\quad Poly             & 0.189 & 0.952 & 0.188 & 1.000 & 0.187 & 1.000 & 0.188 & 0.891 \\
\midrule
\end{tabular}
\vspace{1.5ex}
\begin{tabular}{lcccccccc}
& \multicolumn{2}{c}{$k=5$} & \multicolumn{2}{c}{$k=6$} & \multicolumn{2}{c}{$k=7$} & \multicolumn{2}{c}{$k=8$} \\
\cmidrule(lr){2-3} \cmidrule(lr){4-5} \cmidrule(lr){6-7} \cmidrule(lr){8-9}
                       & MSE   & MCS $p$-value & MSE & MCS $p$-value & MSE & MCS $p$-value & MSE & MCS $p$-value \\
\midrule
Historical average     & 0.189 & 0.895 & 0.189 & 1.000 & 0.189 & 1.000 & 0.189 & 1.000 \\
Linear model           & 0.196 & 0.000 & 0.196 & 0.000 & 0.195 & 0.665 & 0.198 & 0.000 \\
CSR approach           & 0.189 & 0.743 & 0.191 & 0.825 & 0.194 & 0.506 & 0.198 & 0.000 \\
GARCH-M model          & 0.194 & 0.000 & 0.195 & 0.000 & 0.198 & 0.320 & 0.195 & 0.243 \\
MS model               & 0.208 & 0.000 & 0.201 & 0.000 & 0.204 & 0.000 & 0.220 & 0.000 \\
Copula-based approach  &       &       &       &       &       &       &       &       \\
\quad Gaussian         & 0.187 & 1.000 & 0.189 & 0.995 & 0.189 & 1.000 & 0.189 & 1.000 \\
\quad Frank            & 0.187 & 1.000 & 0.189 & 1.000 & 0.189 & 1.000 & 0.189 & 0.829 \\
\quad Clayton          & 0.187 & 1.000 & 0.189 & 1.000 & 0.188 & 1.000 & 0.189 & 1.000 \\
\quad FGM              & 0.187 & 1.000 & 0.189 & 1.000 & 0.189 & 1.000 & 0.190 & 0.829 \\
CSM approach           &       &       &       &       &       &       &       &       \\
\quad Baseline         & 0.187 & 1.000 & 0.189 & 0.982 & 0.189 & 1.000 & 0.189 & 1.000 \\
\quad Poly             & 0.187 & 1.000 & 0.187 & 1.000 & 0.190 & 1.000 & 0.190 & 0.829 \\
\bottomrule
\end{tabular}
{\small \textit{Notes:} This table reports MSE values multiplied by 100, along with the associated MCS $p$-values. The MCS $p$-value indicates the lowest confidence level at which a given model would still be included in the MCS. In other words, it reflects whether the model is among those not significantly worse than the best model in terms of forecasting accuracy. A $p$-value of zero indicates that the model has been excluded from the MCS.
} 
\end{minipage}
\end{center}
\end{table}

\end{landscape}

\clearpage
\newpage

\begin{landscape}
\begin{table}
\vspace*{-1 cm}
\begin{center}
\caption{Out-of-sample model performance comparison using MCS procedure with absolute loss}
\label{MCS absolute}
\begin{minipage}{1.27\textwidth}
\begin{tabular}{lcccccccc}
\toprule
& \multicolumn{2}{c}{$k=1$} & \multicolumn{2}{c}{$k=2$} & \multicolumn{2}{c}{$k=3$} & \multicolumn{2}{c}{$k=4$} \\
\cmidrule(lr){2-3} \cmidrule(lr){4-5} \cmidrule(lr){6-7} \cmidrule(lr){8-9}
                       & MAE & MCS $p$-value & MAE & MCS $p$-value & MAE & MCS $p$-value & MAE & MCS $p$-value \\
\midrule
Historical average     & 3.269 & 0.000 & 3.269 & 0.559 & 3.268 & 0.322 & 3.268 & 0.731 \\
Linear model           & 3.239 & 0.805 & 3.255 & 0.878 & 3.277 & 0.000 & 3.301 & 0.000 \\
CSR approach           & 3.261 & 0.225 & 3.262 & 0.543 & 3.268 & 0.388 & 3.276 & 0.296 \\
GARCH-M model          & 3.224 & 1.000 & 3.247 & 1.000 & 3.264 & 0.472 & 3.296 & 0.000 \\
MS model               & 3.299 & 0.000 & 3.276 & 0.500 & 3.356 & 0.000 & 3.402 & 0.000 \\
Copula-based approach  &       &       &       &       &       &       &       &       \\
\quad Gaussian         & 3.223 & 1.000 & 3.227 & 1.000 & 3.224 & 1.000 & 3.237 & 1.000 \\
\quad Frank            & 3.221 & 1.000 & 3.218 & 1.000 & 3.224 & 1.000 & 3.234 & 1.000 \\
\quad Clayton          & 3.218 & 1.000 & 3.224 & 1.000 & 3.230 & 1.000 & 3.229 & 1.000 \\
\quad FGM              & 3.220 & 1.000 & 3.222 & 1.000 & 3.224 & 1.000 & 3.236 & 1.000 \\
CSM approach           &       &       &       &       &       &       &       &       \\
\quad Baseline         & 3.228 & 1.000 & 3.222 & 1.000 & 3.230 & 1.000 & 3.244 & 1.000 \\
\quad Poly             & 3.237 & 0.982 & 3.231 & 1.000 & 3.230 & 1.000 & 3.245 & 1.000 \\
\midrule
\end{tabular}
\vspace{1.5ex}
\begin{tabular}{lcccccccc}
& \multicolumn{2}{c}{$k=5$} & \multicolumn{2}{c}{$k=6$} & \multicolumn{2}{c}{$k=7$} & \multicolumn{2}{c}{$k=8$} \\
\cmidrule(lr){2-3} \cmidrule(lr){4-5} \cmidrule(lr){6-7} \cmidrule(lr){8-9}
                       & MAE & MCS $p$-value & MAE & MCS $p$-value & MAE & MCS $p$-value & MAE & MCS $p$-value \\
\midrule
Historical average     & 3.267 & 0.800 & 3.269 & 1.000 & 3.269 & 0.984 & 3.269 & 0.952 \\
Linear model           & 3.320 & 0.000 & 3.302 & 0.649 & 3.301 & 0.924 & 3.362 & 0.000 \\
CSR approach           & 3.287 & 0.247 & 3.303 & 0.753 & 3.328 & 0.421 & 3.362 & 0.000 \\
GARCH-M model          & 3.308 & 0.000 & 3.300 & 0.552 & 3.345 & 0.445 & 3.356 & 0.000 \\
MS model               & 3.434 & 0.000 & 3.328 & 0.503 & 3.362 & 0.326 & 3.504 & 0.000 \\
Copula-based approach  &       &       &       &       &       &       &       &       \\
\quad Gaussian         & 3.239 & 1.000 & 3.256 & 1.000 & 3.252 & 1.000 & 3.253 & 1.000 \\
\quad Frank            & 3.219 & 1.000 & 3.253 & 1.000 & 3.250 & 1.000 & 3.250 & 1.000 \\
\quad Clayton          & 3.234 & 1.000 & 3.250 & 1.000 & 3.248 & 1.000 & 3.250 & 1.000 \\
\quad FGM              & 3.237 & 1.000 & 3.253 & 1.000 & 3.250 & 1.000 & 3.259 & 1.000 \\
CSM approach           &       &       &       &       &       &       &       &       \\
\quad Baseline         & 3.234 & 1.000 & 3.260 & 1.000 & 3.258 & 1.000 & 3.248 & 0.659 \\
\quad Poly             & 3.243 & 1.000 & 3.242 & 1.000 & 3.253 & 1.000 & 3.257 & 1.000 \\
\bottomrule
\end{tabular}
{\small \textit{Notes:} This table reports mean absolute error (MAE) values multiplied by 100, along with the associated MCS $p$-values. 
See Table~\ref{MCS squared} for more details.} 
\end{minipage}
\end{center}
\end{table}

\end{landscape}

\newpage

\begin{table}[htbp]
\vspace*{-1 cm}
\centering
\caption{Out-of-sample performance statistics for the trading strategies for $k = 1,2,3,7$}
\label{tab:performance_matrix1}
\resizebox{\textwidth}{!}{%
\begin{tabular}{l
S[table-format=3.2] S[table-format=2.2] S[table-format=2.2] S[table-format=1.2] S[table-format=1.2]
S[table-format=3.2] S[table-format=2.2] S[table-format=2.2] S[table-format=1.2] S[table-format=1.2]}
\toprule
& \multicolumn{5}{c}{$k=1$} & \multicolumn{5}{c}{$k=2$} \\
\cmidrule(lr){2-6} \cmidrule(lr){7-11}
& \multicolumn{1}{c}{TW} & \multicolumn{1}{c}{AV} & \multicolumn{1}{c}{SD} & \multicolumn{1}{c}{SR} & \multicolumn{1}{c}{MDD}
& \multicolumn{1}{c}{TW} & \multicolumn{1}{c}{AV} & \multicolumn{1}{c}{SD} & \multicolumn{1}{c}{SR} & \multicolumn{1}{c}{MDD} \\
\midrule
Buy and hold           & 104.63 & 12.65 & 15.00 & 0.17 & 0.50 & 104.63 & 12.65 & 15.00 & 0.17 & 0.50 \\[1.0ex]
Momentum               & {}     & {}    & {}    & {}   & {}   & {}     & {}    & {}    & {}   & {}   \\
\quad 3-month          &  32.39 &  9.18 & 10.69 & 0.15 & 0.23 &  32.39 &  9.18 & 10.69 & 0.15 & 0.23 \\
\quad 6-month          &  48.71 & 10.26 & 11.30 & 0.17 & 0.23 &  48.71 & 10.26 & 11.30 & 0.17 & 0.23 \\
\quad 12-month         & 100.21 & 12.15 & 12.15 & 0.20 & 0.30 & 100.21 & 12.15 & 12.15 & 0.20 & 0.30 \\[1.0ex]
Linear model           &  93.90 & 12.29 & 14.33 & 0.17 & 0.50 & 109.38 & 12.69 & 14.50 & 0.18 & 0.50 \\[1.0ex]
CSR approach           & 104.63 & 12.65 & 15.00 & 0.17 & 0.50 & 119.54 & 12.91 & 14.54 & 0.18 & 0.44 \\[1.0ex]
GARCH-M model          & 112.08 & 12.73 & 14.40 & 0.18 & 0.50 & 151.01 & 13.52 & 14.72 & 0.19\st{*} & 0.50 \\[1.0ex]
MS model               & 116.96 & 12.59 & 12.56 & 0.20 & 0.29 &  59.60 & 10.92 & 12.58 & 0.17 & 0.48 \\[1.0ex]
Copula-based approach  & {}     & {}    & {}    & {}   & {}   & {}     & {}    & {}    & {}   & {}   \\
\quad Gaussian         & 120.26 & 12.93 & 14.57 & 0.18 & 0.50 & 105.28 & 12.51 & 13.96 & 0.18 & 0.44 \\
\quad Frank            & 122.55 & 12.98 & 14.57 & 0.18 & 0.50 & 154.85 & 13.59 & 14.75 & 0.19\st{*} & 0.50 \\
\quad Clayton          & 118.17 & 12.87 & 14.44 & 0.18 & 0.50 &  89.80 & 12.09 & 13.78 & 0.18 & 0.44 \\
\quad FGM              & 122.55 & 12.98 & 14.57 & 0.18 & 0.50 & 111.75 & 12.66 & 13.97 & 0.19 & 0.44 \\[1.0ex]
CSM approach           & {}     & {}    & {}    & {}   & {}   & {}     & {}    & {}    & {}   & {}   \\
\quad Baseline         & 123.55 & 13.00 & 14.57 & 0.18 & 0.50 & 154.85 & 13.59 & 14.75 & 0.19\st{*} & 0.50 \\
\quad Poly             & 122.55 & 12.98 & 14.57 & 0.18 & 0.50 &  96.89 & 12.27 & 13.71 & 0.18 & 0.44 \\[1.0ex]
\midrule
& \multicolumn{5}{c}{$k=3$} & \multicolumn{5}{c}{$k=7$} \\
\cmidrule(lr){2-6} \cmidrule(lr){7-11}
& \multicolumn{1}{c}{TW} & \multicolumn{1}{c}{AV} & \multicolumn{1}{c}{SD} & \multicolumn{1}{c}{SR} & \multicolumn{1}{c}{MDD}
& \multicolumn{1}{c}{TW} & \multicolumn{1}{c}{AV} & \multicolumn{1}{c}{SD} & \multicolumn{1}{c}{SR} & \multicolumn{1}{c}{MDD} \\
\midrule
Buy and hold           & 104.63 & 12.65 & 15.00 & 0.17 & 0.50          & 104.63 & 12.65 & 15.00 & 0.17 & 0.50 \\[1.0ex]
Momentum               & {}     & {}    & {}    & {}   & {}            & {}     & {}    & {}    & {}   & {}   \\
\quad 3-month          &  32.39 &  9.18 & 10.69 & 0.15 & 0.23          &  32.39 &  9.18 & 10.69 & 0.15 & 0.23 \\
\quad 6-month          &  48.71 & 10.26 & 11.30 & 0.17 & 0.23          &  48.71 & 10.26 & 11.30 & 0.17 & 0.23 \\
\quad 12-month         & 100.21 & 12.15 & 12.15 & 0.20 & 0.30          & 100.21 & 12.15 & 12.15 & 0.20 & 0.30 \\[1.0ex]
Linear model           & 112.79 & 12.67 & 13.84 & 0.19 & 0.44          & 101.16 & 12.17 & 12.27 & 0.20 & 0.44 \\[1.0ex]
CSR approach           &  98.22 & 12.39 & 14.32 & 0.18 & 0.44          &  62.91 & 11.00 & 12.27 & 0.17 & 0.45 \\[1.0ex]
GARCH-M model          & 147.63 & 13.39 & 14.24 & 0.19 & 0.44          &  75.29 & 11.51 & 12.76 & 0.18 & 0.50 \\[1.0ex]
MS model               &  25.65 &  8.73 & 11.77 & 0.12 & 0.41          &  54.04 & 10.69 & 12.78 & 0.16 & 0.44 \\[1.0ex]
Copula-based approach  & {}     & {}    & {}    & {}                  & {}   & {}     & {}    & {}    & {}   & {}   \\
\quad Gaussian         & 165.23 & 13.67 & 14.22 & 0.20\st{*}          & 0.44 & 178.45 & 13.61 & 12.98 & 0.22\st{*} & 0.44 \\
\quad Frank            & 165.23 & 13.67 & 14.22 & 0.20\st{*}          & 0.44 & 176.61 & 13.66 & 13.05 & 0.22\st{*} & 0.44 \\
\quad Clayton          & 117.55 & 12.91 & 14.79 & 0.18               & 0.50 & 178.48 & 13.66 & 13.04 & 0.22\st{*} & 0.44 \\
\quad FGM              & 165.23 & 13.67 & 14.22 & 0.20\st{*}          & 0.44 & 176.61 & 13.66 & 13.05 & 0.22\st{*} & 0.44 \\[1.0ex]
CSM approach           & {}     & {}    & {}    & {}                  & {}   & {}     & {}    & {}    & {}   & {}   \\
\quad Baseline         & 181.68 & 13.89 & 14.13 & 0.21\st{**}         & 0.44 & 168.05 & 13.53 & 12.96 & 0.22 & 0.44 \\
\quad Poly             & 181.68 & 13.89 & 14.13 & 0.21\st{**}         & 0.50 & 149.16 & 13.24 & 12.99 & 0.21 & 0.44 \\
\bottomrule
\end{tabular}}
\begin{minipage}{\textwidth}
\vspace{0.25cm}
\footnotesize
{\textit{Notes:} This table presents out-of-sample performance statistics for the various trading strategies. TW represents the terminal wealth in dollars at the end of the trading period. AV is the annualized average return in percent, SD is the annualized standard deviation of returns in percent, SR is the Sharpe ratio, and MDD is the maximum drawdown in percent.
Stars indicate statistical significance of the SR differential compared to the benchmark buy-and-hold strategy: $^{\ast}$ and $^{\ast\ast}$ denote significance at the 10\% and 5\% levels, respectively.}
\end{minipage}
\end{table}

\newpage

\begin{table}
\vspace*{-1.3 cm}
\begin{center}
\caption{CER gains against historical average for $\gamma = 5$}
\label{CER_k1237}

\begin{tabular*}{\textwidth}{@{\extracolsep{\fill}} l *{4}{S[table-format=1.3]}}
\toprule
& \multicolumn{1}{c}{$k=1$} & \multicolumn{1}{c}{$k=2$} & \multicolumn{1}{c}{$k=3$} & \multicolumn{1}{c}{$k=7$} \\
\midrule

\multicolumn{5}{l}{Panel A: Mean-variance preferences} \\[1.0ex]
Momentum                                  & {}     & {}     & {}     & {}     \\
\quad 3-month momentum                    & -0.706 & -0.706 & -0.706 & -0.706 \\
\quad 6-month momentum                    &  0.042 &  0.042 &  0.042 &  0.042 \\
\quad 12-month momentum                   &  1.436 &  1.436 &  1.436 &  1.436 \\[1.0ex]

Linear model                              &  0.127 &  0.403 &  0.854 &  1.384 \\[1.0ex]
CSR approach                              &  0.000 &  0.596 &  0.239 &  0.207 \\[1.0ex]

GARCH-M model                             &  0.526 &  1.075 &  1.365 &  0.412 \\[1.0ex]
MS model                                  &  1.616 & -0.067 & -1.607 & -0.322 \\[1.0ex]

Copula-based approach                     & {}     & {}     & {}     & {}     \\
\quad Gaussian                            &  0.603 &  0.618 &  1.590 &  2.442 \\
\quad Frank                               &  0.649 &  1.122 &  1.590 &  2.380 \\
\quad Clayton                             &  0.636 &  0.320 &  0.414 &  2.412 \\
\quad FGM                                 &  0.649 &  0.759 &  1.590 &  2.380 \\[1.0ex]

CSM approach                              & {}     & {}     & {}     & {}     \\
\quad Baseline                            &  0.690 &  1.122 &  1.878 &  2.302 \\
\quad Poly                                &  0.649 &  0.546 &  1.878 &  1.993 \\
\midrule

\multicolumn{5}{l}{Panel B: CRRA preferences} \\[1.0ex]
Momentum                                  & {}     & {}     & {}     & {}     \\
\quad 3-month momentum                    & -0.555 & -0.555 & -0.555 & -0.555 \\
\quad 6-month momentum                    &  0.142 &  0.142 &  0.142 &  0.142 \\
\quad 12-month momentum                   &  1.557 &  1.557 &  1.557 &  1.557 \\[1.0ex]

Linear model                              &  0.090 & -0.259 &  0.098 &  0.558 \\[1.0ex]
CSR approach                              &  0.303 &  0.303 &  0.303 &  0.544 \\[1.0ex]

GARCH-M model                             &  0.485 &  1.062 &  1.365 &  0.675 \\[1.0ex]
MS model                                  &  1.714 &  0.085 & -1.597 & -0.187 \\[1.0ex]

Copula-based approach                     & {}     & {}     & {}     & {}     \\
\quad Gaussian                            &  0.573 &  0.658 &  1.643 &  2.748 \\
\quad Frank                               &  0.618 &  1.108 &  1.643 &  2.686 \\
\quad Clayton                             &  0.595 &  0.350 &  0.407 &  2.719 \\
\quad FGM                                 &  0.618 &  0.798 &  1.643 &  2.686 \\[1.0ex]

CSM approach                              & {}     & {}     & {}     & {}     \\
\quad Baseline                            &  0.637 &  1.108 &  1.939 &  2.612 \\
\quad Poly                                &  0.618 &  0.584 &  1.939 &  2.279 \\
\bottomrule
\end{tabular*}
\end{center}
\footnotesize
{\small \textit{Notes:} This table reports the CER gains for each forecasting strategy relative to the historical average benchmark, under two types of investor preferences: mean-variance (Panel A) and constant relative risk aversion (CRRA; Panel B). The parameter $k$ denotes the number of predictors used in each model. The CER gain is interpreted as the annual management fee (in percentage points) an investor would be willing to pay to switch from the benchmark to the corresponding strategy.}
\end{table}

\begin{table}
\begin{center}
\caption{CER gains against CSM (Baseline) for $\gamma = 5$ \label{CER_CSM_k1237}}

\begin{tabular*}{\textwidth}{@{\extracolsep{\fill}} l *{4}{S[table-format=1.3]}}
\toprule
& \multicolumn{1}{c}{$k=1$} & \multicolumn{1}{c}{$k=2$} & \multicolumn{1}{c}{$k=3$} & \multicolumn{1}{c}{$k=7$} \\
\midrule

\multicolumn{5}{l}{Panel A: Mean-variance preferences} \\[1.0ex]
Copula-based approach                     & {}      & {}      & {}      & {}      \\
\quad Gaussian                            & -0.087  & -0.504  & -0.288  &  0.140  \\
\quad Frank                               & -0.041  &  0.000  & -0.288  &  0.078  \\
\quad Clayton                             & -0.054  & -0.802  & -1.464  &  0.110  \\
\quad FGM                                 & -0.041  & -0.363  & -0.288  &  0.078  \\[1.0ex]

CSM approach                              & {}      & {}      & {}      & {}      \\
\quad Poly                                & -0.041  & -0.576  &  0.000  & -0.309  \\
\midrule

\multicolumn{5}{l}{Panel B: CRRA preferences} \\[1.0ex]
Copula-based approach                     & {}      & {}      & {}      & {}      \\
\quad Gaussian                            & -0.064  & -0.450  & -0.296  &  0.136  \\
\quad Frank                               & -0.019  &  0.000  & -0.296  &  0.074  \\
\quad Clayton                             & -0.042  & -0.758  & -1.532  &  0.107  \\
\quad FGM                                 & -0.019  & -0.310  & -0.296  &  0.074  \\[1.0ex]

CSM approach                              & {}      & {}      & {}      & {}      \\
\quad Poly                                & -0.019  & -0.524  &  0.000  & -0.333  \\
\bottomrule
\end{tabular*}
\end{center}
{\small \textit{Notes:} This table reports CER gains relative to the CSM (Baseline) benchmark, using the same setup and preferences as in Table~\ref{CER_k1237}. See the notes to that table for details on interpretation.}
\end{table}

\begin{figure} [hp]
      \caption{Terminal wealth as a function of the number of predictors ($k$) for CSM (Baseline), Clayton, GARCH-M, CSR, and linear models}
      \label{fig:CSM vs. Linear}     
     \centering
     \includegraphics[width= \textwidth , height= \textheight, keepaspectratio]{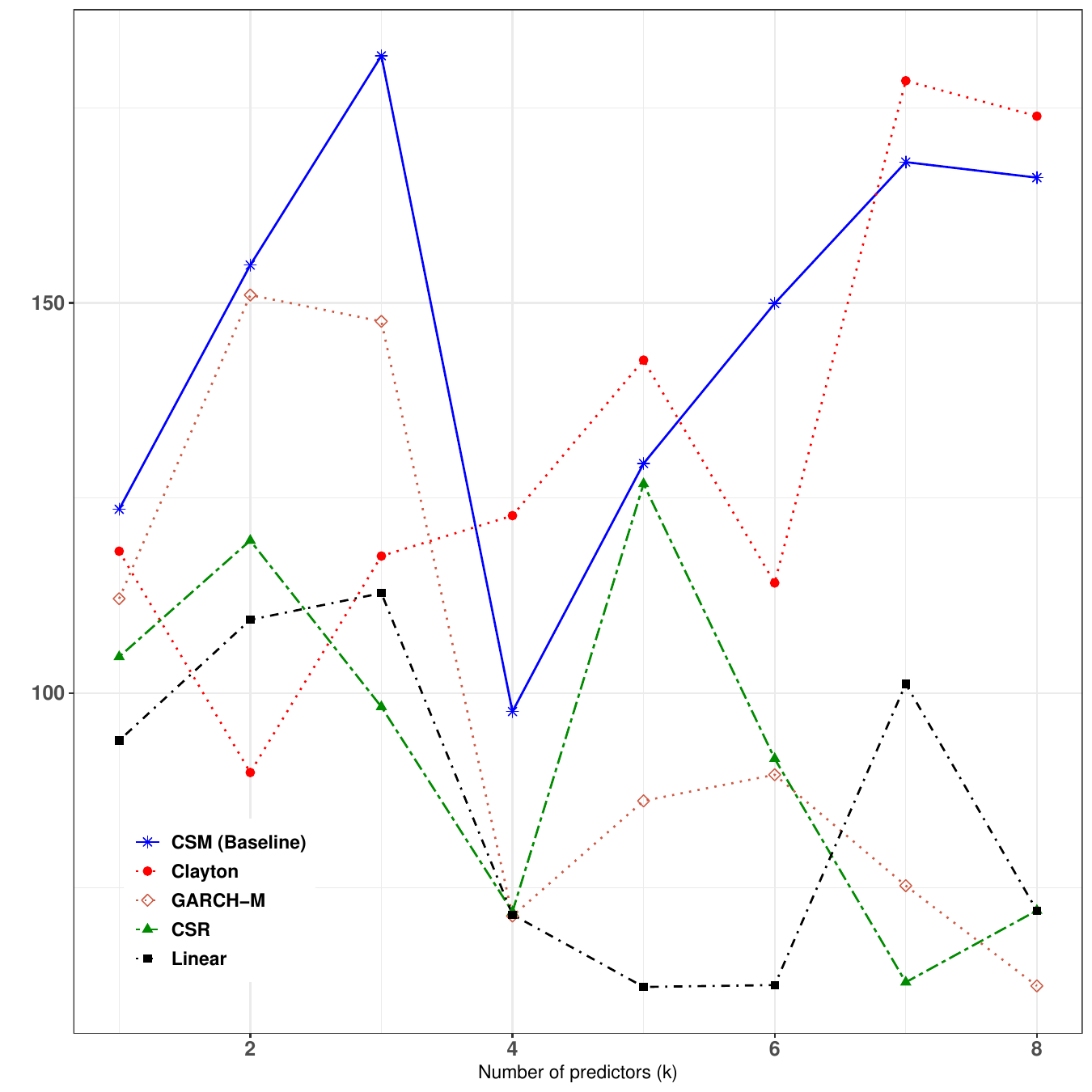}

\end{figure}

\newpage
\clearpage

\appendix
\renewcommand{\thesection}{Appendix}

\counterwithin{table}{section}
\renewcommand{\thetable}{\Alph{section}\arabic{table}}

\section{Copulas and \texorpdfstring{$\varrho_t(z)$}{rho\string_t(z)}}\label{Success_probability}

This appendix presents the copula functions and the corresponding formulas 
for the deformed probability of success, $\varrho_{t}(z)$, for the Gaussian, 
Frank, Clayton, and FGM copulas. In each case, $\varrho_{t}(z)$ is obtained 
by differentiating the copula function with respect to its first argument.

\medskip
\noindent{\textbf{Gaussian copula.}} The Gaussian copula is
\begin{equation}
C(w_{1}, w_{2}) = \Phi_{2}(\Phi^{-1}(w_{1}), \Phi^{-1}(w_{2})),
\end{equation}
where $\Phi_{2}(\cdot, \cdot)$ is the bivariate standard normal CDF with correlation 
parameter $\theta_c \in (-1,+1)$. When $\theta_c=0$, the Gaussian copula 
collapses to the independence copula: $C(w_{1}, w_{2}) = w_1 w_2$. The 
deformed probability function $\varrho_{t}(z)$ appearing in (\ref{copulajoint}) 
is given as
\[
\varrho_{t}(z) = \Phi \bigg( \frac{\Phi^{-1}(p_{t}) + \theta_c \Phi^{-1}(z)}
{\sqrt{1-\theta_c^{2}}} \bigg),
\]
which equals the conditional probability $p_t$ when $\theta_c=0$.

\medskip
\noindent{\textbf{Frank copula.}} The Frank copula is
\begin{equation}
C(w_{1},w_{2}) = - \frac{1}{\theta_c} \log \biggl( 1 + 
\frac{(e^{-\theta_c w_{1}} - 1)(e^{-\theta_c w_{2}} - 1)}{e^{-\theta_c} - 1} 
\biggr),
\end{equation}
where $-\infty < \theta_c < +\infty$, $\theta_c \neq 0$. The Frank copula 
tends to the independence copula as $\theta_c \rightarrow 0$. The deformed 
probability function $\varrho_{t}(z)$ appearing in (\ref{copulajoint}) is 
given by
\[
\varrho_{t}(z) = \biggl( 1 - \frac{1-e^{-\theta_c(1-p_{t})}}{1-e^{\theta_c p_{t}}} 
e^{\theta_c(1-z)} \biggr)^{-1}, \quad \theta_c \neq 0,
\]
and $\varrho_{t}(z) \rightarrow p_t$ as $\theta_c \rightarrow 0$, recovering the marginal success probability under independence.

\medskip
\noindent{\textbf{Clayton copula.}} The Clayton copula is
\begin{equation}
C(w_{1}, w_{2}) = (w_1^{-\theta_c} + w_2^{-\theta_c} - 1)^{-1/\theta_c},
\end{equation}
where $\theta_c > 0$. The Clayton copula tends to the independence copula 
as $\theta_c \rightarrow 0^{+}$. The function $\varrho_{t}(z)$ from 
\eqref{copulajoint} is given by
\[
\varrho_{t}(z) = 1 - \biggl(1 + \frac{(1-p_{t})^{-\theta_c} - 1}{z^{-\theta_c}} 
\biggr)^{-1/\theta_c - 1}, \quad \theta_c > 0,
\]
and $\varrho_{t}(z) \rightarrow p_t$ as $\theta_c \rightarrow 0^{+},$ as expected under independence.

\medskip
\noindent{\textbf{FGM copula.}} The FGM copula is
\begin{equation}
C(w_{1},w_{2}) = w_{1} w_{2} (1 + \theta_c(1-w_{1})(1-w_{2})),
\end{equation}
where $\theta_c \in [-1,+1]$. The FGM copula collapses to the independence 
copula when $\theta_c = 0$. The function $\varrho_{t}(z)$ in 
\eqref{copulajoint} is given by
\[
\varrho_{t}(z) = 1 - (1-p_{t})(1 + \theta_c p_{t}(1-2z)), 
\quad \theta_c \in [-1,+1],
\]
which equals $p_t$ when $\theta_c = 0$.

\clearpage
\setcounter{page}{1}

\begin{bibunit}[chicago]

\renewcommand{\baselinestretch}{1.5}
\normalsize

\renewcommand{\thefootnote}{\fnsymbol{footnote}}

\begin{center}

Supplementary material for:

\noindent{\Large{\bf{A new decomposition approach to modeling financial returns: conditioning sign on magnitude}}}

\smallskip
\bigskip
\bigskip

\renewcommand{\baselinestretch}{1.0}
\normalsize

Ars{\`e}ne Brou and Richard Luger

\end{center}

\renewcommand{\baselinestretch}{1.75}
\normalsize


\setcounter{footnote}{0} 
\renewcommand{\thefootnote}{\arabic{footnote}}

\makeatletter
\renewcommand{\@seccntformat}[1]{\csname the#1\endcsname.\enspace}
\makeatother

\setcounter{section}{0}
\setcounter{table}{0}
\setcounter{figure}{0}

\renewcommand{\thesection}{\Alph{section}}
\renewcommand{\thetable}{A\arabic{table}}
\renewcommand{\thefigure}{A\arabic{figure}}

\section{Results under MSE-based selection}

This section reports additional results when best $k$-predictor subset selection uses the MSE criterion. 
The main text focuses on AUC-based selection; MSE-based selection provides a complementary robustness check.

Table~\ref{Bestsubset2_0LS} reports the selected predictor subsets for the linear model (Panel A), GARCH-M (Panel B), MS (Panel C), and the decomposition-based models (Panel D) across $k=1,\ldots,8$. Selection overlap is substantial across models, even though the subsets are chosen separately for each $k$ and need not be nested. In particular, $dp$ is selected at $k=1$ in all cases, and $dfy$ and $tbl$ appear frequently at intermediate dimensions. The linear and decomposition-based selections coincide for $k=2$--$5$, and differ only at $k=6$--$7$ before converging to the full predictor set at $k=8$. At $k=6$, the linear model selects $infl$ (with $dp$, $dfy$, $tbl$, $ltr$, and $ntis$), whereas the decomposition-based models select $dfr$ instead. At $k=7$, the linear model includes $tms$ and $infl$ but excludes $dfr$, while the decomposition-based models include $dfr$ and $infl$ but exclude $tms$. Importantly, the decomposition-based models select identical subsets across copula specifications (Panel D); GARCH-M and MS broadly converge to similar selections at larger $k$, while differing more at small $k$.

\begin{figure}
\caption{Out-of-sample $R^2$ as a function of the number of predictors ($k$): MSE-based selection}
      \label{fig2:ROOS dynamics}
      \vspace{1.5 cm}
   \begin{minipage}[c]{0.5\linewidth}
     \centering
     (a) Squared errors 
    \vspace{2.5 cm}
     \includegraphics[width=\textwidth, height=\textheight, keepaspectratio]{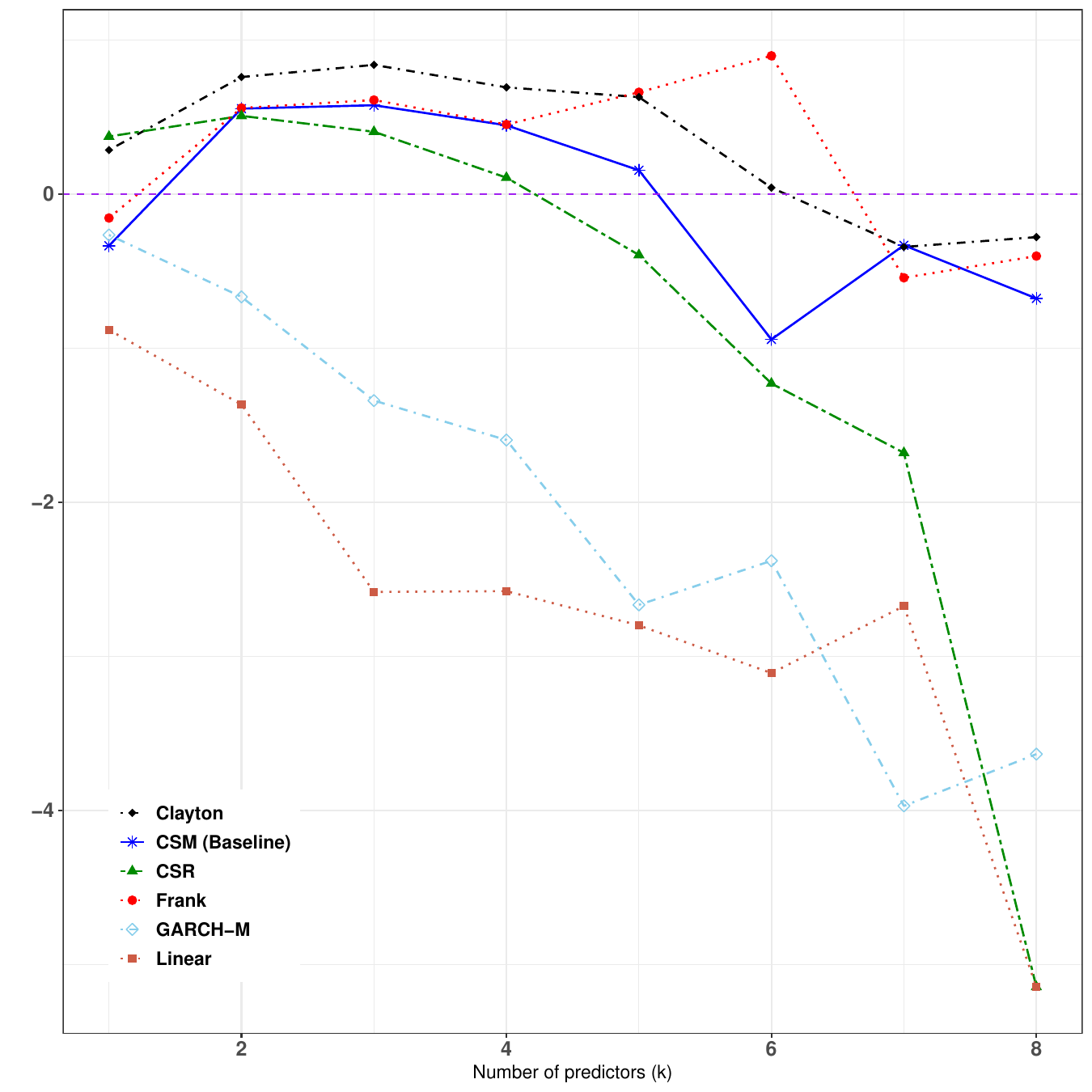}
  \end{minipage}\hspace{1 cm}
  \begin{minipage}[c]{0.5\linewidth}
     \centering
     (b) Absolute errors
    \vspace{2.5 cm}
     \includegraphics[width=\textwidth, height=\textheight, keepaspectratio]{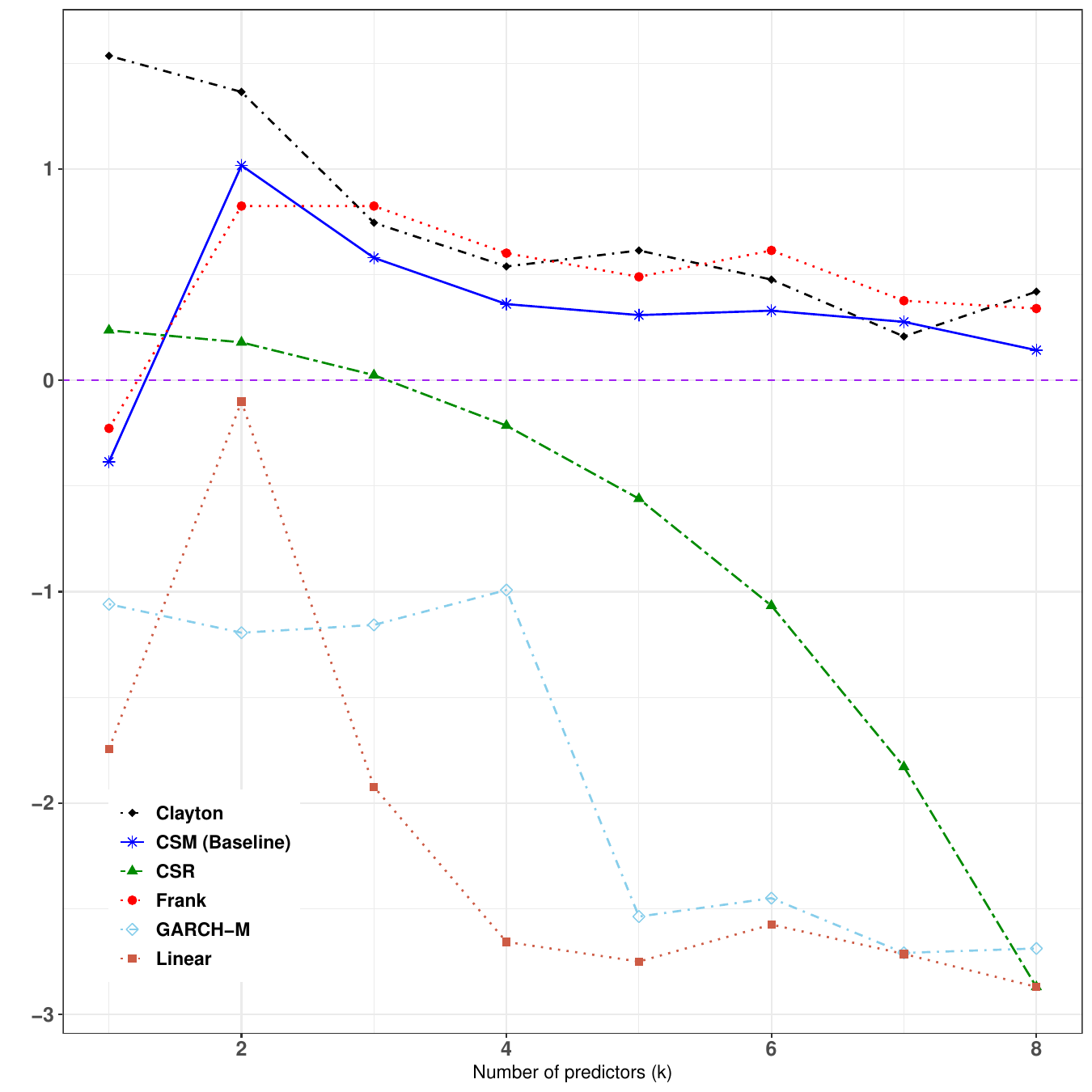}
  \end{minipage}
 \vspace{1.5 cm}
 \begin{minipage}{16 cm}
 \end{minipage}
 \end{figure}

Out-of-sample performance, measured by $R^2_{\text{OOS}}$ (in \%) under squared and absolute loss, is reported in Table~\ref{ROOS_allMSE} and summarized in Figure~\ref{fig2:ROOS dynamics}. The figure shows that the $R^2_{\text{OOS}}$ profiles under squared loss (Panel A) for decomposition-based models peak at intermediate $k$ before declining, whereas conventional benchmarks remain weak and tend to deteriorate at larger $k$. Under absolute loss (Panel B), decomposition-based models deliver uniformly positive $R^2_{\text{OOS}}$ values from $k=2$ onward, while the linear and GARCH-M benchmarks remain negative and CSR turns negative beyond $k=3$.

Overall, MSE-based selection yields weaker performance than AUC-based selection. The deterioration is especially pronounced for conventional benchmarks: the linear model has uniformly negative $R^2_{\text{OOS}}$ values (from $-0.88\%$ to $-5.15\%$ under squared loss and from $-0.10\%$ to $-2.87\%$ under absolute loss), and GARCH-M also produces negative $R^2_{\text{OOS}}$ throughout (e.g., from $-0.27\%$ to $-3.97\%$ under squared loss). CSR achieves only modestly positive $R^2_{\text{OOS}}$ at small $k$ under squared loss, peaking at $0.51\%$ at $k=2$, and then deteriorates, ultimately coinciding with the linear model at $k=8$ by construction. The MS model performs particularly poorly, with squared-loss $R^2_{\text{OOS}}$ values ranging from $-6.23\%$ to $-19.51\%$; frequent DM-test rejections against the historical average indicate statistically significant underperformance.

In contrast, the decomposition-based models are more resilient. Under squared loss, they typically produce positive $R^2_{\text{OOS}}$ values for $k=2$--$5$, with the Clayton copula reaching $0.84\%$ at $k=3$ and CSM (Baseline) attaining $0.58\%$ at $k=3$. These gains attenuate at higher dimensions, with several specifications turning negative for $k \ge 6$, consistent with the declining profiles in Panel (a) of Figure~\ref{fig2:ROOS dynamics}.

Taken together, these results support our reliance on AUC-based subset selection in the main analysis: it delivers stronger and more stable out-of-sample performance across model classes, and it is particularly important for benchmark specifications (linear and GARCH-M) that deteriorate markedly under MSE-based selection.

\begin{table}
\begin{center}
\begin{minipage}{0.95\textwidth} 
\caption{Best subset selection based on MSE criterion}
\label{Bestsubset2_0LS}
\centering
\begin{tabular}{@{\extracolsep{1.5em}}l l *{8}{c}}
\toprule
      & const & dp & dfy & tms & tbl & ltr & dfr & ntis & infl \\
\midrule
\multicolumn{10}{l}{Panel A: Linear model} \\
$k=1$ & *     & *  &     &     &     &     &     &      &      \\
$k=2$ & *     &    & *   &     & *   &     &     &      &      \\
$k=3$ & *     & *  & *   &     & *   &     &     &      &      \\
$k=4$ & *     & *  & *   &     & *   & *   &     &      &      \\
$k=5$ & *     & *  & *   &     & *   & *   &     & *    &      \\
$k=6$ & *     & *  & *   &     & *   & *   &     & *    & *    \\
$k=7$ & *     & *  & *   & *   & *   & *   &     & *    & *    \\
$k=8$ & *     & *  & *   & *   & *   & *   & *   & *    & *    \\
\midrule
\multicolumn{10}{l}{Panel B: GARCH-M model} \\
$k=1$ & *     & *  &     &     &     &     &     &      &      \\
$k=2$ & *     & *  &     & *   &     &     &     &      &      \\
$k=3$ & *     & *  & *   & *   &     &     &     &      &      \\
$k=4$ & *     & *  &     &  *  & *   &     &     & *    &      \\
$k=5$ & *     & *  & *   &     & *   & *   &     &      & *    \\
$k=6$ & *     & *  & *   &     & *   & *   & *   & *    &      \\
$k=7$ & *     & *  & *   & *   & *   & *   &     & *    & *    \\
$k=8$ & *     & *  & *   & *   & *   & *   & *   & *    & *    \\
\midrule
\multicolumn{10}{l}{Panel C: MS model} \\
$k=1$ & *     & *  &     &     &     &     &     &      &      \\
$k=2$ & *     &    & *   &     & *   &     &     &      &      \\
$k=3$ & *     &    & *   &     & *   &     &     & *    &      \\
$k=4$ & *     & *  & *   &  *  &     &     & *   &      &      \\
$k=5$ & *     & *  & *   &  *  &     &     & *   & *    &      \\
$k=6$ & *     & *  & *   &  *  & *   & *   & *   &      &      \\
$k=7$ & *     & *  & *   & *   & *   & *   &     & *    & *    \\
$k=8$ & *     & *  & *   & *   & *   & *   & *   & *    & *    \\
\midrule
\multicolumn{10}{l}{Panel D: Decomposition-based models (CSM and copula-based)} \\
$k=1$ & *     & *  &     &     &     &     &     &      &      \\
$k=2$ & *     &    & *   &     & *   &     &     &      &      \\
$k=3$ & *     & *  & *   &     & *   &     &     &      &      \\
$k=4$ & *     & *  & *   &     & *   & *   &     &      &      \\
$k=5$ & *     & *  & *   &     & *   & *   &     & *    &      \\
$k=6$ & *     & *  & *   &     & *   & *   & *   & *    &      \\
$k=7$ & *     & *  & *   &     & *   & *   & *   & *    & *    \\
$k=8$ & *     & *  & *   & *   & *   & *   & *   & *    & *    \\
\bottomrule
\end{tabular}
\vspace{0.2cm}

\small
\parbox{1.0\textwidth}{\textit{Notes:} A star ($\ast$) indicates that the corresponding variable is included in the best subset of size $k$. The subsets are selected based on the MSE criterion using one-step-ahead in-sample forecasts. An empty cell indicates exclusion of the variable from the selected subset.}
\end{minipage}
\end{center}
\end{table}

\clearpage
\newpage

\begin{landscape}
\begin{table}
\vspace*{-1.3 cm}
\begin{center}
\caption{Out-of-sample $R^2$ values (in \%) for different models, loss functions, and subset sizes ($k$)}
\label{ROOS_allMSE}
\begin{minipage}{1.22\textwidth}

\begin{tabular}{l *{8}{S[table-format=-2.2]}}
\toprule
& \multicolumn{2}{c}{$k=1$} & \multicolumn{2}{c}{$k=2$} & \multicolumn{2}{c}{$k=3$} & \multicolumn{2}{c}{$k=4$} \\
\cmidrule(lr){2-3} \cmidrule(lr){4-5} \cmidrule(lr){6-7} \cmidrule(lr){8-9}
& {Squared} & {Absolute} & {Squared} & {Absolute} & {Squared} & {Absolute} & {Squared} & {Absolute} \\
\midrule
Linear model           & -0.88 & -1.74 & -1.37 & -0.10 & -2.58 & -1.92 & -2.58 & -2.66 \\
CSR approach           &  0.37 &  0.24 &  0.51 &  0.18 &  0.40 &  0.02 &  0.11 & -0.21 \\
MS                     & -13.66 & -6.52 & -6.23 & -2.40 & -7.31 & -4.06 & -9.61\st{***} & -4.26\st{***} \\
GARCH-M                & -0.27 & -1.06 & -0.67 & -1.19 & -1.34 & -1.16 & -1.60 & -0.99 \\
Copula-based approach  & \multicolumn{8}{c}{} \\
\quad Gaussian         & -0.22 & -0.11 &  0.58 &  0.75 &  0.60 &  0.75 &  0.49 &  0.54 \\
\quad Frank            & -0.16 & -0.23 &  0.56 &  0.82 &  0.61 &  0.82 &  0.45 &  0.60 \\
\quad Clayton          &  0.29 &  1.54 &  0.76 &  0.93 &  0.84 &  0.99 &  0.69 &  0.69 \\
\quad FGM              &  0.17 &  1.46 &  0.55 &  0.80 &  0.60 &  0.80 &  0.42 &  0.59 \\
CSM approach           & \multicolumn{8}{c}{} \\
\quad Baseline         & -0.34 & -0.38 &  0.55 &  0.58 &  0.58 &  0.58 &  0.45 &  0.36 \\
\quad Poly             & -0.30 & -0.25 &  0.38 &  0.75 &  0.58 &  0.75 &  0.26 &  0.46 \\
\midrule
\end{tabular}

\vspace{1.5ex}

\begin{tabular}{l *{8}{S[table-format=-2.2]}}
& \multicolumn{2}{c}{$k=5$} & \multicolumn{2}{c}{$k=6$} & \multicolumn{2}{c}{$k=7$} & \multicolumn{2}{c}{$k=8$} \\
\cmidrule(lr){2-3} \cmidrule(lr){4-5} \cmidrule(lr){6-7} \cmidrule(lr){8-9}
& {Squared} & {Absolute} & {Squared} & {Absolute} & {Squared} & {Absolute} & {Squared} & {Absolute} \\
\midrule
Linear model           & -2.80 & -2.75 & -3.11 & -2.57 & -2.67 & -2.71 & -5.15 & -2.87 \\
CSR approach           & -0.40 & -0.56 & -1.23 & -1.07 & -2.68 & -1.83 & -5.15 & -2.87 \\
MS                     & -10.69\st{***} & -5.08\st{***} & -14.71\st{***} & -6.38\st{**} & -19.51\st{***} & -9.84\st{***} & -16.93\st{***} & -7.21\st{***} \\
GARCH-M                & -2.67 & -2.54 & -2.38 & -2.45 & -3.97 & -2.71 & -3.64 & -2.69 \\
Copula-based approach  & \multicolumn{8}{c}{} \\
\quad Gaussian         &  0.70 &  0.43 & -0.02 &  0.48 & -0.52 &  0.21 & -0.39 &  0.35 \\
\quad Frank            &  0.66 &  0.49 &  0.90 &  0.62 & -0.54 &  0.38 & -0.40 &  0.34 \\
\quad Clayton          &  0.63 &  0.50 &  0.04 &  0.41 & -0.34 &  0.28 & -0.28 &  0.42 \\
\quad FGM              &  0.64 &  0.31 &  0.30 &  0.35 & -1.12 &  0.30 & -0.49 &  0.32 \\
CSM approach           & \multicolumn{8}{c}{} \\
\quad Baseline         &  0.15 &  0.33 & -0.94 &  0.36 & -0.33 &  0.04 & -0.68 &  0.14 \\
\quad Poly             & -1.12 &  0.35 & -1.12 &  0.30 & -1.12 &  0.30 & -0.99 &  0.35 \\
\bottomrule
\end{tabular}
\vspace{0.2cm}

\parbox{1.0\textwidth}{\textit{Notes:} This table reports $R^2_{\text{OOS}}$ statistics (in \%) for squared and absolute loss functions, relative to the historical-average benchmark.
Positive values indicate improvement over the historical average, while negative values indicate deterioration.
Stars indicate rejection of equal predictive accuracy (DM test) against the historical average benchmark: $^{\ast}$, $^{\ast\ast}$, and $^{\ast\ast\ast}$ denote significance at the 10\%, 5\%, and 1\% levels, respectively.}

\end{minipage}
\end{center}
\end{table}
\end{landscape}

\clearpage
\newpage

\setcounter{section}{0}
\renewcommand{\thesection}{B}

\setcounter{table}{0}
\setcounter{figure}{0}
\renewcommand{\thetable}{B\arabic{table}}
\renewcommand{\thefigure}{B\arabic{figure}}

\section{Additional economic value results}

This appendix complements the main text by reporting results for additional predictor subset sizes ($k = 4, 5, 6, 8$), extending the analysis beyond the main text's focus on $k = 1, 2, 3, 7$.
Table~\ref{tab:performance_matrix2} reports out-of-sample performance statistics for these subset sizes.
The CSM (Poly) specification performs strongly at intermediate dimensions, achieving terminal wealth of \$139.78 at $k=4$, \$148.04 at $k=5$, and \$137.85 at $k=6$, which is substantially above the linear model (\$71.58, \$62.31, and \$65.92) and also above CSR at these $k$ values.
At larger $k$, the two CSM variants become very similar (TW \$166.06 for CSM Baseline and \$166.86 for CSM Poly at $k=8$).
Copula-based strategies are highly competitive at larger subset sizes, with the Clayton specification delivering the highest terminal wealth among the copulas at $k=8$ (TW \$173.94).
Overall, decomposition-based strategies dominate the conventional benchmarks (linear, GARCH-M, and MS) across these subset sizes, while the MS strategy remains comparatively weak.

Table~\ref{CER_k4568} reports certainty equivalent return (CER) gains relative to the historical average benchmark for an investor with moderate risk aversion ($\gamma = 5$), under mean-variance preferences (Panel A) and CRRA preferences (Panel B).
Across $k$, decomposition-based strategies deliver positive and economically meaningful CER gains, with particularly large gains at $k=8$.
At this subset size, both CSM variants exceed $2\%$ CER under each preference (e.g., 2.267--2.324 in Panel A and 2.579--2.615 in Panel B), and the copula-based strategies are also strongly positive, with Clayton among the top performers (especially under CRRA).
In contrast, the MS benchmark produces negative CER gains throughout, while GARCH-M is negative at $k=4$ but becomes positive for $k=5,6,8$, albeit with gains that remain modest relative to the decomposition-based strategies.

Finally, Table~\ref{CER_CSM_k4568} reports CER differentials computed relative to the 
CSM (Baseline) benchmark, which can be interpreted as the additional annual fee (in 
percentage points) an investor would be willing to pay to switch from the baseline to 
an alternative decomposition-based strategy.
The CSM (Poly) specification delivers sizable incremental gains at $k=4$ and $k=6$, 
and smaller but still positive improvements at $k=5$ and $k=8$.
Among the copula-based alternatives, Clayton delivers positive differentials across all 
subset sizes, whereas Gaussian provides smaller and at times negative differentials, 
particularly at $k=5$ and $k=8$, while Frank turns negative only at $k=8$.

Overall, these supplementary results confirm that the economic value of the decomposition framework remains strong across predictor dimensions.
While copula-based strategies can be highly competitive at larger subset sizes, the CSM specifications provide a robust alternative, and the polynomial variant delivers clear incremental utility at intermediate dimensions.

\clearpage
\newpage

\begin{table}[htbp]
\vspace*{-1 cm}
\centering
\caption{Out-of-sample performance statistics for the trading strategies for $k = 4,5,6,8$}
\label{tab:performance_matrix2}
\resizebox{\textwidth}{!}{%
\begin{tabular}{l
S[table-format=3.2] S[table-format=2.2] S[table-format=2.2] S[table-format=1.2] S[table-format=1.2]
S[table-format=3.2] S[table-format=2.2] S[table-format=2.2] S[table-format=1.2] S[table-format=1.2]}
\toprule
& \multicolumn{5}{c}{$k=4$} & \multicolumn{5}{c}{$k=5$} \\
\cmidrule(lr){2-6} \cmidrule(lr){7-11}
& \multicolumn{1}{c}{TW} & \multicolumn{1}{c}{AV} & \multicolumn{1}{c}{SD} & \multicolumn{1}{c}{SR} & \multicolumn{1}{c}{MDD}
& \multicolumn{1}{c}{TW} & \multicolumn{1}{c}{AV} & \multicolumn{1}{c}{SD} & \multicolumn{1}{c}{SR} & \multicolumn{1}{c}{MDD} \\
\midrule
Buy and hold           & 104.63 & 12.65 & 15.00 & 0.17 & 0.50 & 104.63 & 12.65 & 15.00 & 0.17 & 0.50 \\[1.0ex]
Momentum               & {}     & {}    & {}    & {}   & {}   & {}     & {}    & {}    & {}   & {}   \\
\quad 3-month          &  32.39 &  9.18 & 10.69 & 0.15 & 0.23 &  32.39 &  9.18 & 10.69 & 0.15 & 0.23 \\
\quad 6-month          &  48.71 & 10.26 & 11.30 & 0.17 & 0.23 &  48.71 & 10.26 & 11.30 & 0.17 & 0.23 \\
\quad 12-month         & 100.21 & 12.15 & 12.15 & 0.20 & 0.30 & 100.21 & 12.15 & 12.15 & 0.20 & 0.30 \\[1.0ex]
Linear model           &  71.58 & 11.54 & 13.86 & 0.16 & 0.46 &  62.31 & 11.03 & 12.71 & 0.17 & 0.46 \\[1.0ex]
CSR approach           &  71.94 & 11.57 & 13.91 & 0.16 & 0.45 & 126.80 & 12.81 & 12.87 & 0.20 & 0.44 \\[1.0ex]
GARCH-M model          &  71.40 & 11.63 & 14.47 & 0.16 & 0.50 &  86.16 & 12.01 & 13.96 & 0.17 & 0.46 \\[1.0ex]
MS model               &  40.81 & 10.03 & 13.01 & 0.14 & 0.44 &  32.36 &  9.48 & 11.73 & 0.13 & 0.44 \\[1.0ex]
Copula-based approach  & {}     & {}    & {}    & {}   & {}   & {}     & {}    & {}    & {}   & {}   \\
\quad Gaussian         & 110.36 & 12.70 & 14.46 & 0.18 & 0.45 & 127.65 & 12.96 & 13.82 & 0.19 & 0.50 \\
\quad Frank            & 110.66 & 12.74 & 14.67 & 0.18 & 0.51 & 142.66 & 13.18 & 13.41 & 0.20 & 0.45 \\
\quad Clayton          & 122.73 & 12.70 & 14.55 & 0.18 & 0.45 & 142.67 & 13.26 & 13.93 & 0.20 & 0.50 \\
\quad FGM              & 110.66 & 12.74 & 14.67 & 0.18 & 0.51 & 148.34 & 13.35 & 13.94 & 0.20 & 0.50 \\[1.0ex]
CSM approach           & {}     & {}    & {}    & {}   & {}   & {}     & {}    & {}    & {}   & {}   \\
\quad Baseline         &  97.61 & 12.39 & 14.42 & 0.17 & 0.45 & 129.41 & 12.98 & 13.72 & 0.20 & 0.50 \\
\quad Poly             & 139.78 & 13.28 & 14.41 & 0.19\st{*} & 0.50 & 148.04 & 13.36 & 14.03 & 0.20 & 0.50 \\[1.0ex]
\midrule
& \multicolumn{5}{c}{$k=6$} & \multicolumn{5}{c}{$k=8$} \\
\cmidrule(lr){2-6} \cmidrule(lr){7-11}
& \multicolumn{1}{c}{TW} & \multicolumn{1}{c}{AV} & \multicolumn{1}{c}{SD} & \multicolumn{1}{c}{SR} & \multicolumn{1}{c}{MDD}
& \multicolumn{1}{c}{TW} & \multicolumn{1}{c}{AV} & \multicolumn{1}{c}{SD} & \multicolumn{1}{c}{SR} & \multicolumn{1}{c}{MDD} \\
\midrule
Buy and hold           & 104.63 & 12.65 & 15.00 & 0.17 & 0.50 & 104.63 & 12.65 & 15.00 & 0.17 & 0.50 \\[1.0ex]
Momentum               & {}     & {}    & {}    & {}   & {}   & {}     & {}    & {}    & {}   & {}   \\
\quad 3-month          &  32.39 &  9.18 & 10.69 & 0.15 & 0.23 &  32.39 &  9.18 & 10.69 & 0.15 & 0.23 \\
\quad 6-month          &  48.71 & 10.26 & 11.30 & 0.17 & 0.23 &  48.71 & 10.26 & 11.30 & 0.17 & 0.23 \\
\quad 12-month         & 100.21 & 12.15 & 12.15 & 0.20 & 0.30 & 100.21 & 12.15 & 12.15 & 0.20 & 0.30 \\[1.0ex]
Linear model           &  65.92 & 11.26 & 13.34 & 0.16 & 0.44 &  72.10 & 11.29 & 11.90 & 0.18 & 0.45 \\[1.0ex]
CSR approach           &  91.57 & 11.99 & 12.72 & 0.19 & 0.45 &  72.10 & 11.29 & 11.90 & 0.18 & 0.45 \\[1.0ex]
GARCH-M model          &  89.51 & 12.11 & 13.99 & 0.17 & 0.44 &  62.44 & 10.95 & 12.08 & 0.17 & 0.42 \\[1.0ex]
MS model               &  68.57 & 11.31 & 12.93 & 0.17 & 0.45 &  42.51 & 10.02 & 12.18 & 0.15 & 0.43 \\[1.0ex]
Copula-based approach  & {}     & {}    & {}    & {}   & {}   & {}     & {}    & {}    & {}   & {}   \\
\quad Gaussian         & 102.82 & 12.38 & 13.47 & 0.19 & 0.45 & 167.11 & 13.52 & 13.01 & 0.22 & 0.44 \\
\quad Frank            & 105.97 & 12.47 & 13.56 & 0.19 & 0.45 & 168.24 & 13.54 & 13.04 & 0.22 & 0.44 \\
\quad Clayton          & 114.10 & 12.65 & 13.53 & 0.19 & 0.45 & 173.94 & 13.63 & 13.01 & 0.22 & 0.44 \\
\quad FGM              & 102.81 & 12.39 & 13.57 & 0.18 & 0.45 & 168.24 & 13.54 & 13.04 & 0.22 & 0.44 \\[1.0ex]
CSM approach           & {}     & {}    & {}    & {}   & {}   & {}     & {}    & {}    & {}   & {}   \\
\quad Baseline         &  91.52 & 12.09 & 13.46 & 0.18 & 0.45 & 166.06 & 13.50 & 12.97 & 0.22 & 0.44 \\
\quad Poly             & 137.85 & 13.18 & 13.97 & 0.20 & 0.50 & 166.86 & 13.50 & 12.89 & 0.22 & 0.50 \\
\bottomrule
\end{tabular}}

\begin{minipage}{\textwidth}
\vspace{0.25cm}
\footnotesize
{\textit{Notes:} This table presents out-of-sample performance statistics for the various trading strategies. TW represents the terminal wealth in dollars at the end of the trading period. AV is the annualized average return in percent, SD is the annualized standard deviation of returns in percent, SR is the Sharpe ratio, and MDD is the maximum drawdown in percent.
Stars indicate statistical significance of the SR differential compared to the benchmark buy-and-hold strategy: $^{\ast}$ and $^{\ast\ast}$ denote significance at the 10\% and 5\% levels, respectively.}
\end{minipage}
\end{table}

\begin{table}
\vspace*{-1.3 cm}
\begin{center}
\caption{CER gains against historical average for $\gamma = 5$ \label{CER_k4568}}

\begin{tabular*}{\textwidth}{@{\extracolsep{\fill}} l *{4}{S[table-format=1.3]}}
\toprule
& \multicolumn{1}{c}{$k=4$} & \multicolumn{1}{c}{$k=5$} & \multicolumn{1}{c}{$k=6$} & \multicolumn{1}{c}{$k=8$} \\
\midrule

\multicolumn{5}{l}{Panel A: Mean-variance preferences} \\[1.0ex]
Momentum                                  & {}     & {}     & {}     & {}     \\
\quad 3-month momentum                    & -0.706 & -0.706 & -0.706 & -0.706 \\
\quad 6-month momentum                    &  0.042 &  0.042 &  0.042 &  0.042 \\
\quad 12-month momentum                   &  1.436 &  1.436 &  1.436 &  1.436 \\[1.0ex]

Linear model                              & -0.286 & -0.408 & -0.209 &  0.726 \\[1.0ex]
CSR approach                              & -0.306 &  1.649 &  0.916 &  0.726 \\[1.0ex]

GARCH-M model                             & -0.640 &  0.112 &  0.199 &  0.278 \\[1.0ex]
MS model                                  & -1.232 & -1.850 & -0.330 & -0.670 \\[1.0ex]

Copula-based approach                     & {}     & {}     & {}     & {}     \\
\quad Gaussian                            &  0.448 &  1.163 &  0.818 &  2.262 \\
\quad Frank                               &  0.334 &  1.470 &  0.843 &  2.264 \\
\quad Clayton                             &  0.659 &  1.377 &  1.048 &  2.360 \\
\quad FGM                                 &  0.334 &  1.470 &  0.762 &  2.264 \\[1.0ex]

CSM approach                              & {}     & {}     & {}     & {}     \\
\quad Baseline                            &  0.170 &  1.252 &  0.530 &  2.267 \\
\quad Poly                                &  1.064 &  1.414 &  1.270 &  2.324 \\
\midrule

\multicolumn{5}{l}{Panel B: CRRA preferences} \\[1.0ex]
Momentum                                  & {}     & {}     & {}     & {}     \\
\quad 3-month momentum                    & -0.555 & -0.555 & -0.555 & -0.555 \\
\quad 6-month momentum                    &  0.142 &  0.142 &  0.142 &  0.142 \\
\quad 12-month momentum                   &  1.557 &  1.557 &  1.557 &  1.557 \\[1.0ex]

Linear model                              & -0.220 &  0.240 & -0.131 &  1.073 \\[1.0ex]
CSR approach                              &  0.987 &  0.987 &  1.263 &  1.073 \\[1.0ex]

GARCH-M model                             & -0.636 &  0.189 &  0.274 &  0.631 \\[1.0ex]
MS model                                  & -1.112 & -1.788 & -0.251 & -0.324 \\[1.0ex]

Copula-based approach                     & {}     & {}     & {}     & {}     \\
\quad Gaussian                            &  0.445 &  1.365 &  1.029 &  2.570 \\
\quad Frank                               &  0.325 &  1.672 &  1.057 &  2.572 \\
\quad Clayton                             &  0.663 &  1.581 &  1.255 &  2.667 \\
\quad FGM                                 &  0.325 &  1.672 &  0.977 &  2.572 \\[1.0ex]

CSM approach                              & {}     & {}     & {}     & {}     \\
\quad Baseline                            &  0.172 &  1.462 &  0.747 &  2.579 \\
\quad Poly                                &  1.077 &  1.614 &  1.473 &  2.615 \\
\bottomrule
\end{tabular*}
\end{center}
\footnotesize
{\small \textit{Notes:} This table reports the CER gains for each forecasting strategy relative to the historical average benchmark, under two types of investor preferences: mean-variance (Panel A) and constant relative risk aversion (CRRA; Panel B). The parameter $k$ denotes the number of predictors used in each model. The CER gain is interpreted as the annual management fee (in percentage points) an investor would be willing to pay to switch from the benchmark to the corresponding strategy.}
\end{table}

\begin{table}
\begin{center}
\caption{CER gains against CSM (Baseline) for $\gamma = 5$ \label{CER_CSM_k4568}}

\begin{tabular*}{\textwidth}{@{\extracolsep{\fill}} l *{4}{S[table-format=1.3]}}
\toprule
& \multicolumn{1}{c}{$k=4$} & \multicolumn{1}{c}{$k=5$} & \multicolumn{1}{c}{$k=6$} & \multicolumn{1}{c}{$k=8$} \\
\midrule

\multicolumn{5}{l}{Panel A: Mean-variance preferences} \\[1.0ex]
Copula-based approach                     & {}      & {}      & {}      & {}      \\
\quad Gaussian                            &  0.278  & -0.089  &  0.288  & -0.005  \\
\quad Frank                               &  0.164  &  0.218  &  0.313  & -0.003  \\
\quad Clayton                             &  0.489  &  0.125  &  0.518  &  0.093  \\
\quad FGM                                 &  0.164  &  0.218  &  0.232  & -0.003  \\[1.0ex]

CSM approach                              & {}      & {}      & {}      & {}      \\
\quad Poly                                &  0.894  &  0.162  &  0.740  &  0.057  \\
\midrule

\multicolumn{5}{l}{Panel B: CRRA preferences} \\[1.0ex]
Copula-based approach                     & {}      & {}      & {}      & {}      \\
\quad Gaussian                            &  0.273  & -0.097  &  0.282  & -0.009  \\
\quad Frank                               &  0.153  &  0.210  &  0.310  & -0.007  \\
\quad Clayton                             &  0.491  &  0.119  &  0.508  &  0.088  \\
\quad FGM                                 &  0.153  &  0.210  &  0.230  & -0.007  \\[1.0ex]

CSM approach                              & {}      & {}      & {}      & {}      \\
\quad Poly                                &  0.905  &  0.152  &  0.726  &  0.036  \\
\bottomrule
\end{tabular*}
\end{center}
{\small \textit{Notes:} This table reports CER gains relative to the CSM (Baseline) benchmark, using the same setup and preferences as in Table~\ref{CER_k4568}. See the notes to that table for details on interpretation.}
\end{table}

\clearpage
\newpage

\setcounter{section}{0}
\renewcommand{\thesection}{C}

\setcounter{table}{0}
\setcounter{figure}{0}
\renewcommand{\thetable}{C\arabic{table}}
\renewcommand{\thefigure}{C\arabic{figure}}

\section{Economic performance during crisis periods}

\begin{figure}[hp]
\caption{Wealth growth during the Global Financial Crisis: 3-month momentum, CSR, GARCH-M, and buy-and-hold, together with the CSM (Baseline) and FGM copula strategies. The CSM (Baseline) and FGM wealth paths coincide and therefore overlap.}
  \label{fig:Wealth_dynamics_GFC}
  \centering
  \includegraphics[width=\textwidth, height=\textheight, keepaspectratio]{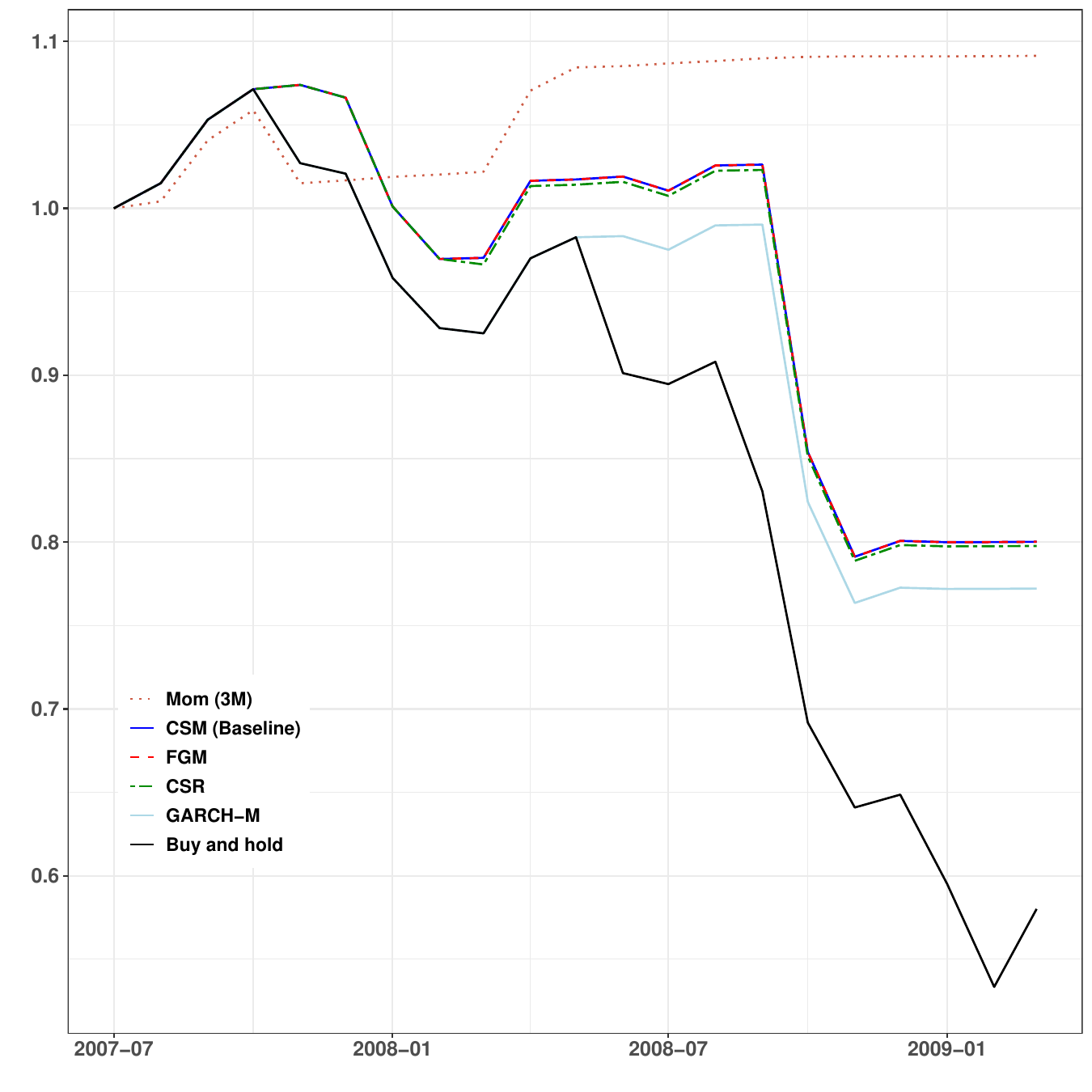}
\end{figure}

\begin{figure}[hp]

\caption{Wealth growth during the COVID crisis period: 3-month momentum, CSM (Poly), FGM copula, CSR, GARCH-M, and buy-and-hold. The FGM and buy-and-hold wealth paths coincide, as do the CSR and CSM (Poly) paths, and the corresponding curves therefore overlap.}

  \label{fig:Wealth_dynamics_COVID}
  \centering
  \includegraphics[width=\textwidth, height=\textheight, keepaspectratio]{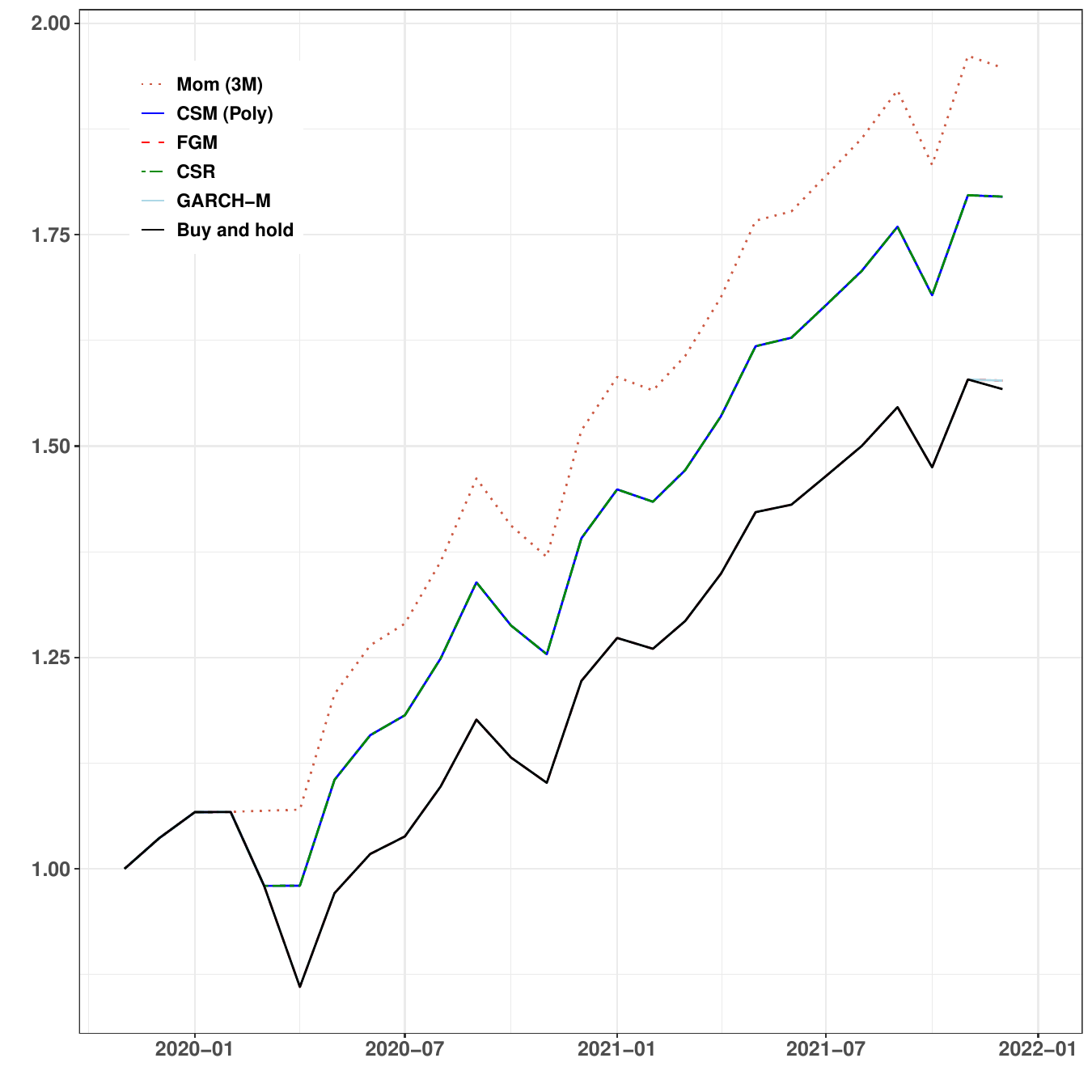}
\end{figure}

This appendix evaluates the economic performance of the forecasting strategies during two major crisis episodes: the Global Financial Crisis (GFC) and the COVID-19 crisis period. The GFC is defined as August 2007 to March 2009, while the COVID-19 crisis period spans December 2019 to December 2021 (the end of our sample). At the beginning of each crisis, models are estimated using the most recent 400 available observations, and portfolio performance is then evaluated from the start to the end of the crisis using a rolling window of 400 observations, as in the main analysis. We normalize initial wealth to \$1 (so TW is a gross wealth multiple). To ensure a consistent comparison across approaches, all models are estimated using the full set of eight predictors (i.e., $k=8$ throughout this appendix).\footnote{Under $k=8$, the CSR and linear strategies coincide by construction. In addition, several decomposition-based strategies yield identical switching decisions over these short windows, which is reflected by overlapping wealth paths in Figures~\ref{fig:Wealth_dynamics_GFC} and \ref{fig:Wealth_dynamics_COVID}.} Table~\ref{Crisisperformance} reports performance statistics over both periods, and Figures~\ref{fig:Wealth_dynamics_GFC} and \ref{fig:Wealth_dynamics_COVID} plot the corresponding cumulative wealth paths.

During the GFC, most model-based switching strategies deliver negative average returns and Sharpe ratios (Table~\ref{Crisisperformance}), although they generally attenuate losses relative to buy-and-hold. Buy-and-hold exhibits the weakest performance (TW $=0.59$ and MDD $=0.50$). Momentum strategies are comparatively resilient during this window, preserving capital and achieving modest gains (TW $\approx 1.04$--$1.09$). The decomposition-based strategies (CSM and copula-based specifications) deliver very similar outcomes during the GFC (TW $=0.80$), close to the linear/MS/CSR benchmarks (TW between $0.79$ and $0.82$). Figure~\ref{fig:Wealth_dynamics_GFC} highlights that CSM (Baseline) and the FGM copula strategy follow essentially identical wealth paths throughout the crisis, declining substantially less than buy-and-hold but underperforming short-horizon momentum.

The COVID-19 crisis period shows a different pattern. All strategies deliver positive average returns and Sharpe ratios, with 3-month momentum achieving the strongest performance (TW $=1.95$). A second group of strategies---Linear/CSR/MS and CSM (Poly)---coincides in this episode and attains strong performance (TW $=1.79$). Buy-and-hold, GARCH-M, and the remaining decomposition-based specifications exhibit more moderate but still positive outcomes (TW $\approx 1.57$--$1.58$). Figure~\ref{fig:Wealth_dynamics_COVID} illustrates the associated overlaps in wealth paths: the FGM copula and buy-and-hold coincide (TW $=1.57$), as do CSR and CSM (Poly) (TW $=1.79$), while 3-month momentum maintains a clear lead over the full window.

Overall, within these two crisis windows, momentum---particularly at short horizons---is the most consistently resilient strategy. Among the model-based switching approaches, performance is broadly similar during the GFC, whereas during COVID the strongest non-momentum performance is achieved by the Linear/CSR/MS and CSM (Poly) strategies, with buy-and-hold performing worst during the GFC.

\begin{table}
\centering
\caption{Out-of-sample performance statistics for the trading strategies during crisis periods}
\label{Crisisperformance}
\resizebox{\textwidth}{!}{%
\begin{tabular}{l
S[table-format=1.2] S[table-format=-2.2] S[table-format=2.2] S[table-format=-1.2] S[table-format=1.2]
S[table-format=1.2] S[table-format=2.2]  S[table-format=2.2] S[table-format=1.2]  S[table-format=1.2]}
\toprule
& \multicolumn{5}{c}{Global Financial Crisis} & \multicolumn{5}{c}{COVID-19 Crisis} \\
\cmidrule(lr){2-6} \cmidrule(lr){7-11}
& \multicolumn{1}{c}{TW} & \multicolumn{1}{c}{AV} & \multicolumn{1}{c}{SD} & \multicolumn{1}{c}{SR} & \multicolumn{1}{c}{MDD}
& \multicolumn{1}{c}{TW} & \multicolumn{1}{c}{AV} & \multicolumn{1}{c}{SD} & \multicolumn{1}{c}{SR} & \multicolumn{1}{c}{MDD} \\
\midrule
Buy and hold
& 0.59 & -29.98 & 21.26 & -0.43 & 0.50
& 1.57 & 23.51  & 19.05 & 0.35  & 0.19 \\

Momentum & {} & {} & {} & {} & {} & {} & {} & {} & {} & {} \\
\quad 3-month
& 1.09 & 5.42 & 5.84 & 0.17 & 0.04
& 1.95 & 33.37 & 14.32 & 0.66 & 0.06 \\
\quad 6-month
& 1.09 & 5.47 & 3.14 & 0.32 & 0.05
& 1.64 & 24.84 & 12.44 & 0.57 & 0.06 \\
\quad 12-month
& 1.04 & 2.25 & 4.79 & 0.02 & 0.05
& 1.70 & 27.41 & 17.70 & 0.44 & 0.12 \\

Linear model
& 0.79 & -12.15 & 16.18 & -0.25 & 0.27
& 1.79 & 29.63  & 16.08 & 0.52  & 0.08 \\

CSR approach
& 0.79 & -12.15 & 16.18 & -0.25 & 0.27
& 1.79 & 29.63  & 16.08 & 0.52  & 0.08 \\

GARCH-M model
& 0.77 & -14.04 & 16.32 & -0.28 & 0.29
& 1.58 & 23.82  & 19.01 & 0.35  & 0.19 \\

MS model
& 0.82 & -10.31 & 16.62 & -0.22 & 0.24
& 1.79 & 29.63  & 16.08 & 0.52  & 0.08 \\

Copula-based approach & {} & {} & {} & {} & {} & {} & {} & {} & {} & {} \\
\quad Gaussian
& 0.80 & -11.97 & 16.09 & -0.25 & 0.26
& 1.57 & 23.51  & 19.05 & 0.35  & 0.19 \\
\quad Frank
& 0.80 & -11.97 & 16.09 & -0.25 & 0.26
& 1.57 & 23.51  & 19.05 & 0.35  & 0.19 \\
\quad Clayton
& 0.80 & -11.97 & 16.09 & -0.25 & 0.26
& 1.57 & 23.51  & 19.05 & 0.35  & 0.19 \\
\quad FGM
& 0.80 & -11.97 & 16.09 & -0.25 & 0.26
& 1.57 & 23.51  & 19.05 & 0.35  & 0.19 \\

CSM approach & {} & {} & {} & {} & {} & {} & {} & {} & {} & {} \\
\quad Baseline
& 0.80 & -11.97 & 16.09 & -0.25 & 0.26
& 1.57 & 23.51  & 19.05 & 0.35  & 0.19 \\
\quad Poly
& 0.80 & -11.97 & 16.09 & -0.25 & 0.26
& 1.79 & 29.63  & 16.08 & 0.52  & 0.08 \\
\bottomrule
\end{tabular}}

\vspace{0.5cm}
\begin{minipage}{\textwidth}
\footnotesize
\textit{Notes:} See Table \ref{tab:performance_matrix2} for details on notation.
\end{minipage}
\end{table}

\clearpage
\newpage

\setcounter{section}{0}
\renewcommand{\thesection}{D}

\setcounter{table}{0}
\setcounter{figure}{0}
\renewcommand{\thetable}{D\arabic{table}}
\renewcommand{\thefigure}{D\arabic{figure}}

\section{Additional results with momentum variable}

This appendix reports results from an augmented predictor set that adds a standard time-series momentum indicator to the eight predictors used in the main analysis. We construct the signal following \citet{Moskowitz-Ooi-Pedersen:2012}. Using the monthly S\&P 500 returns, the momentum variable at time $t-1$ is defined as the cumulative return from month $t-12$ to $t-2$, excluding the most recent month to mitigate short-term reversal effects \citep{Jegadeesh-Titman:1993,Moskowitz-Ooi-Pedersen:2012}:
\[
\text{Mom}_{t-1} = \prod_{j=t-12}^{t-2} (1 + r_j) - 1,
\]
where $r_j$ denotes the simple monthly return.

\begin{figure}[h]
  \caption{Wealth growth for CSM (Baseline), copula-based (Clayton, Gaussian), CSR, and MS strategies with nine predictors (including momentum).}
  \label{fig:Wealth_dynamics_k9}
  \centering
  \includegraphics[width=\textwidth, height=\textheight, keepaspectratio]{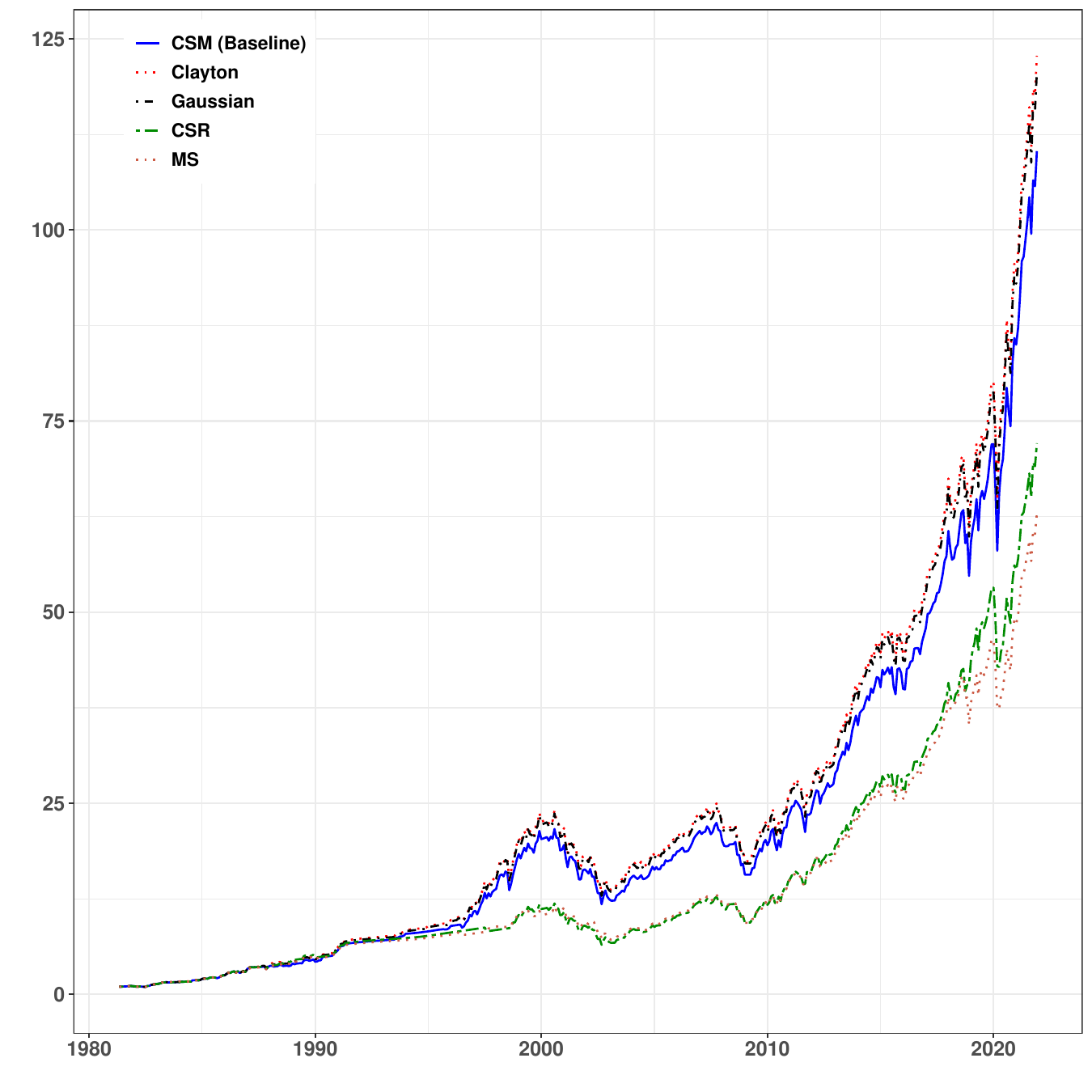}
\end{figure}

We then re-estimate all predictive models using the full set of nine predictors. Table~\ref{FullwithMom} reports out-of-sample $R^2$ statistics and the corresponding portfolio performance, while Figure~\ref{fig:Wealth_dynamics_k9} plots the associated cumulative wealth paths for a representative subset of strategies. Under squared loss, all models deliver negative $R^2_{\text{OOS}}$. The deterioration is severe for the linear/CSR and regime-switching benchmarks (e.g., $-5.15\%$ for the linear model and CSR, and $-21.97\%$ for MS), whereas the decomposition-based models are substantially closer to the historical-average benchmark (ranging from $-0.42\%$ to $-1.12\%$). Under absolute loss, the decomposition-based approaches again provide the strongest relative statistical performance: the copula-based specifications attain small but positive $R^2_{\text{OOS}}$ values (0.15--0.27\%), while CSM with the polynomial sign specification performs comparably (0.23\%); the baseline CSM specification is close to zero ($-0.01\%$). Since the augmented specification uses all nine predictors, CSR equals the linear model here, explaining the identical $R^2_{\text{OOS}}$ and portfolio statistics.

Figure~\ref{fig:Wealth_dynamics_k9} illustrates the temporal wealth evolution for selected strategies. \sloppy The decomposition-based approaches (CSM Baseline, Clayton, Gaussian) track similar paths throughout the sample period and end at markedly higher wealth levels than the benchmarks. CSR follows a broadly similar trajectory early in the sample but begins to lag behind the decomposition-based strategies around the late-1990s/early-2000s, with further relative slippage during subsequent turbulent episodes. The MS strategy exhibits the weakest overall wealth accumulation among the plotted benchmarks, consistent with its very poor $R^2_{\text{OOS}}$ values.

The economic evaluation leads to a similar conclusion. Adding momentum does not materially improve the performance of the linear/CSR, GARCH-in-mean, or regime-switching strategies, which remain dominated in this exercise. Among decomposition-based strategies, the copula-based specifications deliver the 
highest terminal wealth (TW $\approx \$118$--\$123, with Clayton achieving \$122.78) 
with Sharpe ratios around 0.20, while CSM remains competitive 
(TW $\approx \$110$--\$111) and substantially improves on the non-decomposition 
benchmarks (Linear/CSR TW = \$72.10; GARCH-M TW = \$69.87; MS TW = \$62.86).

Overall, incorporating a momentum indicator does not change the main message: modeling returns through sign-magnitude decompositions remains the most reliable approach in this application, while linear and regime-switching benchmarks remain weak when evaluated out-of-sample.

\begin{table}[h]
\begin{center}
\caption{Out-of-sample $R^2$ and economic performance across models}
\label{FullwithMom}
\begin{tabular*}{\textwidth}{@{\extracolsep{\fill}}l
S[table-format=-2.2] S[table-format=-2.2]
S[table-format=3.2]  S[table-format=2.2]  S[table-format=2.2]  S[table-format=1.2]  S[table-format=1.2]}
\toprule
& \multicolumn{2}{c}{$R^2_{\text{OOS}}$} & \multicolumn{5}{c}{Portfolio performance} \\
\cmidrule(lr){2-3} \cmidrule(lr){4-8}
& \multicolumn{1}{c}{Squared} & \multicolumn{1}{c}{Absolute}
& \multicolumn{1}{c}{TW} & \multicolumn{1}{c}{AV} & \multicolumn{1}{c}{SD} & \multicolumn{1}{c}{SR} & \multicolumn{1}{c}{MDD} \\
\midrule
Linear model                  & -5.15  & -2.87  &  72.10 & 11.29 & 11.90 & 0.18 & 0.45 \\
CSR approach                  & -5.15  & -2.87  &  72.10 & 11.29 & 11.90 & 0.18 & 0.45 \\
GARCH-M model                 & -4.36  & -2.77  &  69.87 & 11.44 & 12.19 & 0.18 & 0.42 \\
MS model                      & -21.97 & -10.52 &  62.86 & 10.94 & 11.83 & 0.18 & 0.38 \\
Copula-based approach         & {}     & {}     & {}     & {}    & {}    & {}   & {}   \\
\quad Gaussian                & -0.60  &  0.15  & 120.66 & 12.72 & 13.10 & 0.20 & 0.46 \\
\quad Frank                   & -0.42  &  0.27  & 118.29 & 12.67 & 13.06 & 0.20 & 0.46 \\
\quad Clayton                 & -0.46  &  0.26  & 122.78 & 12.77 & 13.10 & 0.20 & 0.46 \\
\quad FGM                     & -0.59  &  0.16  & 118.29 & 12.67 & 13.06 & 0.20 & 0.46 \\
CSM approach                  & {}     & {}     & {}     & {}    & {}    & {}   & {}   \\
\quad Baseline                & -0.91  & -0.01  & 110.32 & 12.50 & 13.07 & 0.19 & 0.46 \\
\quad Poly                    & -1.12  &  0.23  & 111.12 & 12.49 & 12.86 & 0.20 & 0.46 \\
\bottomrule
\end{tabular*}
\end{center}
{\small \textit{Notes:} $R^2_{\text{OOS}}$ is computed relative to the historical-average benchmark.
TW denotes terminal wealth (initial wealth \$1), AV and SD are annualized in percent, SR is the Sharpe ratio, and MDD is maximum drawdown in percent.}
\end{table}

\putbib[CSM-References]

\end{bibunit}

\end{document}